\documentclass[11pt]{article}
\usepackage{geometry}                
\usepackage{graphicx}
\usepackage{amssymb}
\usepackage{epstopdf}

\usepackage{amsmath} 
\usepackage{latexsym}
\usepackage{curves,epic}

\def\Z{\mathbb Z}
\def\R{\mathbb R}
\def\Q{\mathbb Q}
\def\N{\mathbb N}
\def\A{\mathcal A}
\def\B{\mathcal B}
\def\C{\mathcal C}
\def\D{\,{\rm d}}
\def\LR{L^{2}\mathbb{(R)}}
\def\LA{\Lambda}
\def\pf{\textbf{Proof}}
\def\pfk{\ \hfill \rule{0.5em}{0.5em}}

 \newtheorem{de}{Definition}[section]
 \newtheorem{pozn}[de]{Remark}
 \newtheorem{ex}[de]{Example}

 \newtheorem{prop}[de]{Proposition}
 \newtheorem{thm}[de]{Theorem}
 
  \newtheorem{vlast}[de]{Property}

\title{Nested quasicrystalline discretisations of the line}

\author{J.P. Gazeau\footnote{
UMR 7164, (CNRS, Universit\'e Paris 7, CEA, Observatoire de Paris)}, Z. Mas\'akov\'a and
E. Pelantov\'a}

\date{
APC, Boite 7020, Universit\'e Paris 7-Denis Diderot\\
2 place Jussieu 75251  Paris Cedex 05, France\\
email:\,{\footnotesize Gazeau@ccr.jussieu.fr}
\\[4pt]
Department of Mathematics, FNSPE,
Czech Technical University\\
Trojanova 13, 120 00 Praha 2, Czech Republic\\
emails:\,{\footnotesize Masakova@km1.fjfi.cvut.cz,\,Pelantova@km1.fjfi.cvut.cz} }

\begin{document}
\maketitle

\begin{abstract}
One-dimensional cut-and-project point sets obtained from the square lattice in the plane
are considered  from a  unifying point of view and in the perspective of aperiodic
wavelet constructions.  We successively examine their geometrical aspects, combinatorial
properties from the point of view of the theory of languages,  and self-similarity with
algebraic scaling factor $\theta$. We explain the relation of the cut-and-project sets to
non-standard numeration systems based on $\theta$.  We finally examine the
substitutivity, a weakened version of substitution invariance, which provides us with an
algorithm for symbolic generation of cut-and-project sequences.
\end{abstract}

\subsubsection*{Classification}
52C23, 42C40, 68R15, 11Z05

\subsubsection*{Keywords}
Multiresolution, wavelet, Pisot number, cut-and-project set, quasicrystal,
self-similarity, substitution, combinatorics on words


\section{Introduction}

Initially introduced by Y. Meyer \cite{meyer1,meyer2} in the context of Harmonic Analysis
and more  specifically of \emph{harmonious sets}, the cut-and-project sets or model sets
have become during the two last decades a kind of  geometrical paradigm in
quasicrystalline studies. Quasicrystals are those  alloys whose   first sample was
discovered in 1982 by Shechtman,  Blech,   Gratias, and  Cahn  \cite{shech}, namely the
alloy $Al_{0.86}Mn_{0.14}$, characterized by \begin{enumerate} \item[i)]
  a diffraction pattern like a dense constellation of more-or-less
bright spots, which is an indication of a long-range order,
\item[ii)]
a spatial organisation of those Bragg peaks obeying five- or ten-fold symmetries, at
least locally, which indicates a sort of icosahedral organisation in real space with
five-fold symmetries, \item[iii)] a spatial organisation of those Bragg peaks obeying
specific scale inva\-riance, more precisely invariance under dilations by a factor equal
to some  power of the golden mean $\tau = \frac{1+\sqrt{5}}{2}$ and manifestly consistent
with the five-fold symmetry since $\tau=2\cos{\frac{2\pi}{10}}$. \end{enumerate}

  One-dimensional  examples which are usually presented as  toy
geometrical models of quasicrystals \cite{lest} appertain to the  so-called Fibonacci
chain family. They are  discrete quasiperiodic
  subsets of the real line and are often presented as an illustration of
the cut-and-project method, mainly developed in this  context by
\cite{kadu,kalugin}. Consider a semi-open band $\mathcal{B}$
obtained by translating the unit square through the square lattice
$\Z^2$ along the straight line $D_1$ of slope $\varepsilon$. $D_1$
is referred to as a ``cut'' or ``parallel'' space or ``physical
space''. Then project on $D_1$ and along a straight line $D_2$ the
lattice points lying  in $\mathcal{B}$. Note that the latter
points belong to a unique path made of horizontal  segments $(A)$
and vertical segments $(B)$. The resulting sequence of points
lying in  $D_1$ are the
  nodes of a specific  Fibonacci chain if $\varepsilon =
\frac{1}{\tau}$ and $D_2 = D_1^{\perp}$. Let us denote this  set  of nodes by
$\mathcal{F}$.  The chain itself is made of the projected paths and reads $\dots
ABAABABAABAA\dots$. Note that a short link $B$ is never adjacent to another $B$ whereas
two adjacent long links $A$ can occur. $D_2$ is called the ``internal'' space, and
$\mathcal{B} \cap D_2$ is the ``window'' or ``acceptance zone'', or also ``atomic
surface''.

The set of Fibonacci nodes is equivalently obtained through a purely algebraic filtering
procedure. Let us first consider  the so-called extension ring of the algebraic integer
$\tau$: $$ \Z \lbrack \tau \rbrack = \lbrace x = m + n\tau\  \vert \  m, n \in \Z \rbrace
= \Z + \Z \tau. $$ It can be obtained as the projection onto $D_1$ and along $D_2$ of the
whole square lattice $\Z^2$. There exists in this type of a ring an algebraic
conjugation, called Galois automorphism, and defined by:
 $$ 
 x = m + n\tau \quad \mapsto \quad
 x^{\prime} = m + n\tau^{\prime},
 $$ 
where $\ \tau^{\prime} = - \frac{1}{\tau} = \frac{1-\sqrt{5}}{2}$ is  the other root of
the golden mean equation $x^2 = x + 1$. Then define  the point set $\Sigma(\Omega) $
using an internal sieving rule in the ring $\Z\lbrack \tau \rbrack$ itself \cite{Mopa}:
 $$ 
\Sigma(\Omega) = \left\lbrace x = m
+ n\tau \in \Z \lbrack \tau \rbrack \ \Big |\ x^{\prime} = m - n\frac{1}{\tau} \in \Omega
\right\rbrace \
  =  (\Z\lbrack \tau \rbrack \cap \Omega)^{\prime}.
 $$ 
The Fibonacci point set $\mathcal{F}$ in the above, with link lengths $  A = \tau^2$, $ B
= \tau $, is precisely that set $\Sigma(\Omega)$ with $\Omega = \lbrack 0, 1)$.

As was previously mentioned, self-similarity plays a fundamental structural role in the
existence of quasicrystals. That property is perfectly illustrated by the Fibonacci point
set $\mathcal{F} = \Sigma\lbrack 0, 1)$ since we check from the algebraic definition that
  $\tau^2\,\Sigma\lbrack 0, 1) = \Sigma\lbrack 0, 1/\tau^2) \subset
\Sigma\lbrack 0, 1)$ and so $\tau^2\,\mathcal{F} \subset \mathcal{F}$: this particular
Fibonacci point set is self-similar with scaling factor equal to $\tau^2$. Immediately we
get the infinite nested sequence
 \begin{equation} \label{fibnest}
 \cdots \ \subset\ \mathcal{F}/\tau^{2j-2}\ \subset \ \mathcal{F}_j := \mathcal{F}/\tau^{2j} \
 \subset\ \mathcal{F}/\tau^{2j+2} \ \subset \ \cdots \, ,
 \end{equation}
as increasing aperiodic discretizations of $\mathbb{R}$. Since the distance between two
adjacent points of $\mathcal{F}_j$ is equal to  $1/\tau^{2j}$ or $1/\tau^{2j + 1}$, it is
clear that the inductive  limit $\mathcal{F}_{\infty} := \underset{j \to
\infty}{\lim}\mathcal{F}_j$ densely fills the real line. It is  precisely this property
which led us to examine in recent works (see \cite{gazpat,abug,anbuga} and references
therein) the problem of constructing  wavelets by following discretization schemes of the
real line like (\ref{fibnest}). In \cite{anbuga}, the construction was based on
   multiresolution analysis  with spline wavelets earlier elaborated by
Lemari\'e-Rieusset \cite{lem} and  Bernuau \cite{bern1,bern2} (more details will be given
below). Our aim  was  to eventually apply these wavelets to the analysis of aperiodic
structures, like diffraction spectra of Fibonacci chain or  some other related spectral
problem, and to compare our results with more standard wavelet analysis (e.g.\ dyadic
wavelets) \cite{anav}.

   Let us recall the main  features of wavelet analysis in the framework
of the Hilbert space $\LR$. More complete information can be found in comprehensive
textbooks, like \cite{mallat1}. We shall just  outline here the essential of that field
whose development since  the beginning of the eighties amazingly parallels that one of
quasicrystals. Under the name \emph{wavelet} is commonly understood a  function
$\psi(x)\in\LR$ such that the family of functions $\psi_{j,k}(x):=2^{j/2}\psi(2^j x-k)$
for $j,k\in\Z$ forms an orthonormal (in a restrictive sense) or at least  a \emph{Riesz
basis}  for $\LR$. A family of vectors $(v_n)$ in a separable Hilbert space $V$ is a
Riesz basis if and only if each $v \in V$ can be  expressed uniquely as $ v = \sum_n a_n
v_n $ and there exist positive constants $K_2$ and $K_2$, $0<K_1\leq K_2$, such that
 $$
 K_1 \sum_n \vert a_n \vert^2  \leq \Big\Vert\sum_n a_n v_n \Big\Vert^2
 \leq K_2 \sum_n \vert a_n \vert^2
 $$
for all sequence of scalars $a_n$.  We can say that the $v_n$'s are  strongly linearly
independent and, if $K_1 = 1 = K_2$, then the basis is orthonormal.  A  function (or a
set of functions) generating through dilations and translations an  orthonormal basis
for $\LR$ can be found using a \emph{multiresolution analysis} of $\LR$  (shortly MRA), a
method settled by S.~Mallat \cite{mallat} and precisely based on an increasing sequence
of periodic discretizations of $\R$. The genuine MRA ingredients are:
\begin{enumerate}
\item[(i)] one  \emph{scaling} function $\varphi (x) \in L^2(\R) $ such  that
  $\{ \varphi (x- k) \}_{k\in \Z}$ is an orthonormal  system,

  \item[(ii)] the Hilbert subspace $V_0$ which is the linear span of $\{
\varphi (x - k)\, | \, k\in \Z \}$ and which corresponds to the ``central'' element of
the sequence of discretizations of $\R$,

  \item[(iii)] the increasing sequence of nested Hilbert subspaces
$\cdots V_{j-1} \subset V_j \subset  V_{j+1} \cdots $ which are defined by $f(x) \in V_0
\Leftrightarrow f(2^j x) \in V_j $ and are  such that $\bigcap_j V_j = \{0\} $  and
$\bigcup_j V_j$   is dense in $L^2(\R)$,

\item[(iv)] one  wavelet, i.e.\ a function $\psi(x)$ such  that
  $\{ \psi (x - k) \, | \, k\in \Z\}$ spans the orthogonal complement
$W_0$ of $V_0$ in  $V_1 = V_0 \oplus W_0$. \end{enumerate} Note that the orthonormality
of the basis $\{ \varphi (x- k) \}_{k\in  \Z}$ of the subspace $V_0$ is a strong
constraint. This condition is usually  weakened by just imposing that the system $\{
\varphi (x- k) \}_{k\in  \Z}$ be a Riesz basis of $V_0$.

We thus note that the dilatation factor is genuinely $\theta=2$. Indeed, the construction
of a wavelet basis within the MRA framework relies on the fact that the lattices
$2^{-j}\Z$ are increasing for the inclusion. This property is  preserved only when
$\theta$ is an integer. Then, what about choosing another number  $\theta$ as a scaling
factor? Auscher \cite{auscher} considered the  following  problem: given a real number
$\theta>1$, does there exist a finite set $\{\psi_1,\psi_2,\dots,\psi_\ell\}$ of
functions in $\LR$ such that the family $\theta^{j/2}\psi_i(\theta^j  x-k),\ j,k\in\Z,\ 1
\leq i\leq \ell,$ is an orthonormal basis for $\LR$?  Then he proved  that a basis of
this type exists if $\theta$ is a rational number. More precisely, for $\theta=p/q>1$,
$p$ and $q$ being relatively prime integers, there exists a set of $p-q$  wavelet
functions satisfying the previous condition. Now, what about an irrational  scaling
factor? We already mentioned the works \cite{bern1,bern2} and \cite{anbuga}, the  latter
being mainly devoted to the cases in which $\theta$ is encountered in quasicrystallography, like $\tau = (1 + \sqrt{5})/2$ or $\tau^2 = (3 +
\sqrt{5})/2$. Other ``quasicrystallographic'' numbers have been observed: $ 1 + \sqrt2$, $2 + \sqrt3$. All of them belong to the class of 
quadratic Pisot-Vijayaraghavan units.  To such numbers are associated discretization sequences like in
(\ref{fibnest}),
 \begin{equation} \label{simdelnest}
 \cdots \ \subset\
 \Lambda_{j-1}\ \subset\ \Lambda_j := \Lambda/\theta^{j}\ \subset\
 \Lambda_{j+1}\ \subset\ \cdots \, ,
 \end{equation}
where $\Lambda = \Lambda_0$ is a selfsimilar Delone set with scaling  factor $\theta$,
$\theta\, \Lambda \subset \Lambda$ and is such that the inductive limit $\Lambda_{\infty}
:= \underset{j \to \infty}{\lim}\Lambda_j $  densely fills the real line. Recall that by
Delone set we mean that $\Lambda$ is {\it    uniformly discrete} (the distances between
any pair of points in $\Lambda$ are greater than  a fixed $r >0$) and {\it
   relatively dense} (there exists $R>0$ such that $\R$ is covered by
intervals of length $2R$ centered at points of $\Lambda$).

In 1992 Buhmann and Micchelli \cite{bumi} proposed a construction of a wavelet spline
basis corresponding to non-uniform and non-self-similar knot sequences, which are
actually nested sequences of Delone sets
 \begin{equation} \label{nonsimdelnest}
 \cdots \subset \Lambda_{j-1}\subset \Lambda_j \subset \Lambda_{j+1}\subset\cdots \, .
 \end{equation}
They consider only two successive elements, say $\Lambda_0$  and $\Lambda_1 \supset
\Lambda_0$ for their purpose of proving the existence of what they call prewavelets, with
minimal support, which span the orthogonal complement  $W_0$ of $V_0$ in $V_1 = V_0
\oplus W_0$. Here, $V_0$ and $V_1$ are the spaces of linear combinations of
\emph{$B$-splines} on $\Lambda_0$ and $\Lambda_1$ respectively. They first suppose that
the ``refining'' of the ``coarse'' knot sequence $\Lambda_0$ leading to  the ``finer''
$\Lambda_1$ consists in adding a new knot between each two adjacent knots of $\Lambda_0$.
The case of multiple insertions is eventually examined.  Let us now give some insight on
these spline functions, so intimately linked to the notion of a Delone set.

Any Delone set $\Lambda$ determines a space of splines of order $s$, $s\geq 2$, in the
following way.

\begin{de} \label{spline} Let $s\geq 2$.  Then ${V_0^{(s)}(\Lambda)}$ is the closed subspace
   of $\LR$ defined by
   $$
   {V_0^{(s)}(\Lambda)} := \left\{f(x)\in \LR \, \Big
     | \,\cfrac[l]{\D^s}{\D x^s}f(x)=\sum_{\lambda\in\Lambda}
     a_{\lambda} \delta_{\lambda}\right\}.
   $$
\end{de}

An equivalent definition is given in terms of the restriction of functions to intervals
determined by consecutive elements of $\Lambda$. Suppose $\Lambda=\{ \lambda_n \mid
n\in\Z\}$, where $\lambda_n < \lambda_{n +1}$ for all $n\in\Z$. There results from
Definition~\ref{spline} that
 $$
 {V_0^{(s)}(\Lambda)}=\left\{f\in C^{s-2}\cap\LR\,\big|\,
 f_{\mid_{[\lambda_n,\lambda_{n+1}]}} \mbox{ is a
 polynomial of degree }\leq s\! -\! 1 \right\}.
 $$
Therefore, ${V_0^{(s)}(\Lambda)}$ is the space of splines of order $s$ with nodes in
$\Lambda$. Let us now give a classical result about the existence of a  Riesz basis for
${V_0^{(s)}(\Lambda)}$ \cite{schum}.

\begin{thm}
For all Delone  sets $\Lambda=\{\lambda_n\mid n\in\Z\}\subset\R$  and for all
$s\geq 2$, there exists a Riesz basis $\{ B_{\lambda}^{(s)}\mid\lambda\in\Lambda\}$ of
${V_0^{(s)}(\Lambda)}$.  The function $B_{\lambda}^{(s)}$ (called B-spline) is the unique
function in ${V_0^{(s)}(\Lambda)}$ satisfying the following conditions:
    \begin{enumerate}
    \item[(i)]{$\mathrm{supp}\ B_{\lambda}^{(s)}=[\lambda,\lambda']$,
        where $\lambda'\in\Lambda.$}
    \item[(ii)]{The interval $(\lambda,\lambda')$ contains exactly
        $s-1$ points of $\Lambda$.}
    \item[(iii)]{$\displaystyle\int_{\R}B_{\lambda}^{(s)}=\frac{\lambda'-
\lambda}{s}.$}
    \end{enumerate}
\end{thm}

See \cite{schum} for proof.  Note that (i) and (ii) give precise information on the
(compact) support of $B_{\lambda}^{(s)}$ whilst (iii) is a normalization condition.

The construction of $B^{(s)}_{\lambda}(x)$ can be carried out in various ways, by
recurrence, by using the condition of minimal support, or by inverse Fourier transform.
In the latter case, one can prove that the Fourier transform of $B_{\lambda_n}^{(s)} (x +
\lambda_n)$ depends on the $s$-tuple $(\lambda_{n + 1} - \lambda_n, \dots, \lambda_{n +
s} - \lambda_n ) $ only. Now suppose that the Delone set $\Lambda$ is of \emph{finite
local complexity}, which means \cite{LaPle,lagarias-finite} that, for all $R > 0$, the point set
 $$
 \bigcup_{\lambda \in \Lambda}\bigl\{(\Lambda - \lambda) \cap (-R,R) \bigr\}
 $$
is finite, i.e.\ local environments of points in $\Lambda$ are not different in infinite
fashions.  Typically, such sets $\Lambda$ are mathematical models for one-dimensional
structures having a long-range order, like quasicrystals. We can then assert the
following:

\begin{prop} \label{flcdel}
   Let $\Lambda\subset\R$ be a Delone set of finite local complexity. Then
   the B-splines of order $s$ based on $\Lambda$ are of the form
   $B^{(s)}_{\lambda}(x)=\phi_{\lambda}(x-\lambda),\ \lambda\in\Lambda$,
   where the set $\{\phi_{\lambda}(x)\mid \lambda\in\Lambda\}$ is a finite
   set of functions with compact support.
\end{prop}

Therefore, in the finite local complexity case, it is possible to partition the indexing
set $\Z$ for $\Lambda$ into a finite set of equivalence classes $\bar{0}$, $\bar{1}$,
\dots, $\bar{q}$, where the equivalence between $k$ and $n$ is given by
$$
 B_{\lambda_{k}}^{(s)} (x +\lambda_{k}) = B_{\lambda_n}^{(s)} (x + \lambda_n)\,,\qquad
 \hbox{ for all } x\,.
$$
Correspondingly, for a given $s$, the point set $\Lambda$ is partitioned into $\Lambda =
\overset{\bar{q}}{\underset{\bar{n} = \bar{0}}\cup} \Lambda_{\bar{n}}$ with
$\Lambda_{\bar{n}} = \left\{\lambda_k \in \LA \, | \, k\in \bar{n} \right\} $. The
equivalence between $k$ and $n$ means that $\lambda_k$ and $\lambda_n$ are left-hand ends
of identical $s$-letter words if we identify each interval $\left( \lambda_k ,
\lambda_{k+1}\right)$ with a letter of the allowed alphabet. To each class $\bar{n}$ is
biunivocally associated the function $\phi_{\bar{n}} (x) \equiv \phi_{\lambda_k} (x) =
B_{\lambda_k}^{(s)} (x + \lambda_k), \ k \in \bar{n} $.  In this way, the space
$V_0^{(s)}(\Lambda)$ decomposes into the direct sum
 $$
 V_0^{(s)}(\Lambda) = \overset{\bar{q}}{\underset{\bar{n} =
 \bar{0}}\bigoplus} V_{0, \bar{n}},
 $$
where $V_{0, \bar{n}} $ is the closure of the linear span of the functions
$\phi_{\bar{n}} (x - \lambda_k), \ k \in \bar{n}$.

Let us now go back to the case  in which there is self-similarity (like in
(\ref{simdelnest})) and  finite local complexity and let us
 see which issue holds in term of MRA and existence and properties of
wavelets. More  concretely, let $\Lambda$ be a Delone set of finite local complexity and
self-similar  with inflation factor $\theta > 1$, $\theta\Lambda\subset\Lambda$.
Changing the  scale allows us to define subspaces ${V_j^{(s)}(\Lambda)},\ j\in\Z$, as

   $${V_j^{(s)}(\Lambda)}=\left\{f(x)\in \LR\ \mid \ \frac{\D^s}{\D
       x^s}f(x)=\sum_{\lambda\in\Lambda} a_{\lambda}
     \delta_{\theta^{-j}\lambda} \right\}.$$
 Therefore, ${V_j^{(s)}(\Lambda)}$ is the space of splines of order $s$ with nodes in
the $j$th scaled version of $\Lambda$. 

We now have at our disposal an inductive chain of spaces allowing analysis at any scale.
More precisely, with the above notations, we have the following statement.

\begin{prop}
   The sequence of subspaces $(V_j^{(s)}(\Lambda))_{j\in \mathbb{Z}}$ is a
   $\theta$-multi\-reso\-lution analysis of $L^{2}\mathbb{(R})$,
i.e.
\begin{enumerate}\vspace{-2mm}\label{974.multiresolution}
\item[(i)]{for any $j\in\Z, \ V_j^{(s)}(\Lambda)$ is a closed subspace
     of $L^{2}\mathbb{(R})$,}
\item[(ii)]{$\cdots\subset V_{-1}^{(s)}(\Lambda)\subset
     V_0^{(s)}(\Lambda)\subset V_1^{(s)}(\Lambda)
     \subset\cdots$,}
\item[(iii)]{$\bigcup_{j\in\Z}V_j^{(s)}(\Lambda)$ is dense in
     $L^{2}\mathbb{(R})$,}
\item[(iv)]{$\bigcap_{j\in\Z}V_j^{(s)}(\Lambda)=\{0\}$,} 
\item[(v)]{ $f(x)\in V_j^{(s)}(\Lambda)$ if and only if $
     f(\theta^{-j}x) \in V_{0}^{(s)}(\Lambda)$,}
\item[(vi)]{there exists a finite number of functions
     $\phi_{\bar{n}}(x)\in V_0^{(s)}$, called {scaling}
     {functions} such that
     $\left\{\phi_{\bar{n}}(x-\lambda_k)\right\}_{k\in \bar{n},\,
       {0} \leq {n} \leq {q}}$ is a Riesz basis in
     $V_0^{(s)}$.}
\end{enumerate}
\end{prop}

The proof is straightforward from definitions and Proposition \ref{flcdel}.

As a consequence of the the above statements, we have the  important result obtained by
Bernuau  \cite{bern1,bern2}:

\begin{thm}
   Let $\Lambda = \{\lambda_n \mid n\in \Z\} \subset \R$ be a
   Delone set of finite local complexity, self-similar with factor
   $\theta>1.$ Let us denote the elements of $\theta^{-1}\Lambda$ by
   $\kappa_n = \theta^{-1}\lambda_n$, $n\in \Z$. Then for all
   $s>1$ there exists a Riesz basis of $\LR$ of the form:
 $$
 \left\{\theta^{j/2}\psi_{\kappa_n}^{(s)}(\theta^{j}x-\kappa_n),\ \kappa_n \in
 \theta^{-1} \Lambda, \ \kappa_{n+1} \notin \Lambda,\ j\in\Z \right\},
 $$
 where $\left\{\psi_{\kappa_n}^{(s)} \right\}$ is a finite set of compactly supported
 functions of order $C^{s-2}$.
\end{thm}

Such a result offers the possibility of constructing explicit (spline) wavelet basis for
a large class of  self-similar Delone set of finite local complexity, like in particular
those ones obtained through the cut-and-project method. It was done for some very
specific cases in \cite{anbuga}. Nevertheless, we think that a systematic study of the
properties of (not necessarily self-similar) cut-and-project  sets is still lacking.

The aim of the present paper is to fill this gap. Its content is devoted to the study of
one-dimensional cut-and-project sets built in a rather generic way, without supposing a
priori any algebraic nature for the respective slopes of $D_1$ and $D_2$ and by using the
freedom of making the window $\Omega$ vary in a continuous way. Hence, our results can be
viewed as paving the way to further investigations concerning explicit wavelet
constructions for arbitrary cut-and-project sets. Indeed, what we learn from previous
works is the importance of knowing in a precise way the  environment of each point in the
Delone set, i.e.\ its complexity, in order to build the scaling functions and the
associated wavelets. Moreover, we should not underestimate the structural importance of
the nested sequence (\ref{nonsimdelnest}) in five of its features,  namely

\begin{itemize}
   \item[(i)] the way in which  the new points (i.e.\ the
\emph{details})
   intertwine the old ones at each step $\Lambda_j \subset
\Lambda_{j+1}$ of the increasing sequence (\ref{nonsimdelnest}), or, equivalently, the
characteristics of the detail sets $\Lambda_{j+1}\setminus \Lambda_j$,
   \item[(ii)] the way in which the sequence behaves at the limit $j \to
\infty$, i.e. the structure of its inductive limit which should  be dense in $\R$,
   \item[(iii)] the specific advantages brought by the self-similarity
hypothesis,
     \item[(iv)] related to (i) and (iii), the specific advantages brought by  the
\emph{substitivity} hypothesis, a notion which is introduced   in (\ref{subst}).
        \item[(v)] the relevance of a specific choice of a nested
sequence  with regard to the domain of application of the corresponding wavelet analysis.
\end{itemize}

Note the crucial importance of the first point in connection with the so-called
\emph{scaling} or \emph{refinement} equations which couple with the  inclusion $V_0
\subset V_1 = V_0 \oplus W_0$. Suppose there exist spline bases
$\left(\varphi_{0,k}\right)_{k \in \mathcal{I}_0 \subset \Z}$,
$\left(\varphi_{1,k}\right)_{k \in \mathcal{I}_1 \subset \Z}$ for subspaces $V_0$ and
$V_1$ respectively, and a wavelet basis $\left(\psi_{0,k}\right)_{k \in  \mathcal{J}_0
\subset \Z}$ for $W_0$. Then the following hilbertian decomposition  should hold:
 \begin{equation} \label{refin}
 \varphi_{1,j}(x) = \sum_{k \in \mathcal{I}_0} c_{1,jk}
 \varphi_{0,k}(x)  + \sum_{k \in \mathcal{J}_0} d_{1,jk} \psi_{0,k}(x).
 \end{equation}
The way the  \emph{tendency} coefficients $c_{1,jk}$ and the  \emph{detail} coefficients
$d_{1,jk}$ behave for large $k$ is a crucial question in wavelet analysis, and so the way  this
question depends on the set inclusion  $\Lambda_0 \subset \Lambda_{1}$ deserves special
attention.

In consequence, we have organized the paper by following the hierarchy  of questions
(\emph{i})-(\emph{v}), and furthermore including in the scheme other  aspects of possible
interest, like some considerations on numeration systems related to  cut-and-project
scheme.
  Section 2 is devoted to the geometrical aspects of those point sets in
the line issued from the square lattice in the plane through ``cut''  and ``projection'':
definition, study of distances between adjacent  points, properties of invariance or
covariance under the group $SL(2,  \Z) \times \{-1,1\}$ acting on the square lattice. The
material  presented  there is not specifically new. However it represents an original
overview, in which one focuses on the universality of many features of  these
cut-and-project sets, independently of specific algebraic or  substitutional
characteristics.

  In Section 3 we examine the combinatorial aspects of the cut-and-project
sequences from the point of view of the theory of languages: subword complexity, Rauzy
graphs, and occurrence of specific classes of finite words (or {\it factors}) in the
bidirectional word biunivocally associated to cut-and-project sequences. One can find
there original results (Propositions~\ref{pripady},~\ref{factor}, and \ref{rauzy}).  The
 important case of sturmian words is also considered and we establish
three properties, \ref{vl:1},  \ref{vl:2},  \ref{vl:3}, describing their factors.

Self-similar cut-and-project sets  are the object of Section 4 in which we solve (Theorem
\ref{thmselfs}) the precise relation between self-similarity and algebraic properties of
the irrational scaling factor(s) on one hand and the irrational numbers involved in the
cut-and-project scheme on the other hand.

In Section 5 we revisit the  important notion of $\beta$-integers, i.e.\ those real
numbers which do not have ``$\beta$-fractional'' part when expanded in ``basis'' $\beta >
1$, in the light of their possible or not relation with cut and projection, and we give a
necessary and sufficient condition for $\beta$ (Proposition~\ref{betapisot}) under which
the positive  $\beta$-integers coincide with  the positive part of a cut-and-project set.

Another original part of the paper is found in Section 6.  It is well known that infinite
words associated to many cut-and-project sets present the so-called  substitution
invariance, and this property can be crucial for understanding or even  for creating the
relation $\Lambda_j \to \Lambda_{j+1}$ in a multiresolution sequence of  sets and the
scaling equations (\ref{refin}) issued from the companion inclusion   $V_j \subset
V_{j+1}$. Now, even though the substitution invariance is absent for a given
bidirectional infinite word $u = \cdots u_{-2} u_{-1} | u_0 u_1 u_2 \cdots $, where $u_k$ belongs to some alphabet $\mathcal{A}$, there exist cases in which it could  be ``hidden''
behind the weaker notion of \emph{substitutivity}, which means that  there exists another
infinite word $v = \cdots v_{-2} v_{-1} | v_0 v_1 v_2 \cdots$ over an  alphabet
$\mathcal{B}$ which {has} substitution invariance and a  letter projection $\psi :
\mathcal{B} \to \mathcal{A}$ such that $u = \cdots u_{-2} u_{-1} | u_0 u_1  u_2 \cdots =
\cdots \psi(v_{-2})  \psi(v_{-1}) | \psi(v_0) \psi(v_1) \psi(v_2) \cdots $.

With regards to this property, we shall give in Section 6 an algorithm (Theorem
\ref{algor}) allowing to ``pull back'', a given word $u$ pertaining to the algebraic
cut-and-project  scheme to the  word $v$ mentioned in the above. In order to illustrate
this result, the  algorithm is  carried out on an example of cut-and-project set defined
by the algebraic (Sturm) number $1/\sqrt{2} $.

Eventually, we shall give in the conclusion some hints about possible applications of our
results, mainly  in direction  of  wavelet constructions, of mathematical diffraction,
and of design of aperiodic pseudo-random number generators.


\section{Cut-and-project sequences}

In this section we define cut-and-project sequences arising by a projection of a 2-dimensional
lattice. We also describe their basic properties, including the invariance under
 certain transformations. We further show that cut-and-project sequences
 are geometric representations of a three or two interval exchange.
 Codings of two interval exchanges are in one-to-one correspondence
 with mechanical words $\underline{s}_{\alpha,\beta}$, $\overline{s}_{\alpha,\beta}$, (see
 definition by~\eqref{eq:dolnimech} and~\eqref{eq:hornimech}), which are in fact sturmian words
 (Definition~\ref{de:sturm}).

\subsection{Definition and properties}

The construction of a cut-and-project sequence starts with a choice of a 2-dimensional
lattice $L$ and two straight lines $D_1$, $D_2$. One of the lines plays the role of the
space onto which the lattice $L$ is projected, the other line determines the direction of
the projection. If $A$ is an arbitrary non-singular linear map on $\R^2$, then the
cut-and-project sequence constructed using a lattice $AL$ and straight lines $AD_1$,
$AD_2$ is the same as the cut-and-project sequence constructed using a lattice $L$ and
straight lines $D_1$, $D_2$. Therefore it is not necessary to consider general $L$ and
$D_1,D_2$. Some authors allow arbitrary lattice $L$ and for $D_1,D_2$ take mutually
orthogonal straight lines. Others, including us, prefer to fix the lattice $\Z^2$ and
consider arbitrary straight lines $D_1$, $D_2$.

Let us take two distinct irrational numbers $\varepsilon$, $\eta$ and let us consider
straight lines $D_1: y=\varepsilon x$, $D_2:y=\eta x$. If we choose vectors
$$
\vec{x}_1=\frac{1}{\varepsilon-\eta}(1,\varepsilon)\quad\hbox{ and }\quad
\vec{x}_2=\frac{1}{\eta-\varepsilon}(1,\eta)
$$
in the subspaces $D_1$, $D_2$ of $\R^2$, then for every lattice point $(a,b)\in\Z^2$ we
have
$$
(a,b)=(b-a\eta)\vec{x}_1 + (b-a\varepsilon)\vec{x}_2\,.
$$
Obviously, the projection of $\Z^2$ on $D_1$ along $D_2$ is the set
$$
\Z[\eta]\vec{x}_1\,,
$$
where $\Z[\eta]$ is the abelian group
$$
\Z[\eta]:=\{a+b\varepsilon\mid a,b\in\Z\}\,.
$$
Similarly, the projection of $\Z^2$ on $D_2$ along $D_1$ is the set $\Z[\varepsilon]\vec{x}_2$,
Since numbers $\varepsilon$, $\eta$ are irrational, the mappings $(a,b)\mapsto a+b\eta$,
$(a,b)\mapsto a+b\varepsilon$ are bijections between $\Z^2$ and $\Z[\eta]$, resp. $\Z^2$
and $\Z[\varepsilon]$. Therefore there exists also a bijection
$$
\star: \Z[\eta] \to \Z[\varepsilon]
$$
defined by the prescription
$$
x=a+b\eta \quad\mapsto\quad x^\star=a+b\varepsilon\,,
$$
which is called the star map. Directly from its definition we obtain
$$
(x+y)^\star = x^\star+ y^\star \qquad\hbox{ for every }\ x,y\in\Z[\eta]\,.
$$

Let us now introduce the definition of cut-and-project sets. It is easy to observe that,
as a consequence of irrationality of $\varepsilon$ and $\eta$, the sets
$\Z[\varepsilon]$, $\Z[\eta]$ are dense in $\R$. However, if instead of all the lattice
$\Z^2$, we project only those points in $\Z^2$ that belong to a chosen strip parallel to
$D_1$, the resulting set in $D_1$ has no limit points. The width and position of the
projected strip is determined by an interval in $D_2$. Formally, we have the following
definition.

\begin{de}
Let $\varepsilon$, $\eta$ be distinct irrational numbers and let $\Omega$ be a bounded
interval. The set
$$
\Sigma_{\varepsilon,\eta}(\Omega)=\{a+b\eta\mid a,b\in\Z,\ a+b\varepsilon\in\Omega\} =
\{x\in\Z[\eta] \mid x^\star\in\Omega\}
$$
is called a cut-and-project sets, or C\&P set. The interval $\Omega$ is called the
acceptance window of $\Sigma_{\varepsilon,\eta}(\Omega)$.
\end{de}

The above definition is a special case of the very general `model
sets'. Important contributions to the study of model sets as
mathematical models of quasicrystals are due
to~\cite{lagarias-finite,LaPle,meyer2,moody,Mopa,MoPaDensity}.

Let us mention some of the properties of cut-and-project sequences
that follow directly from the definition or were derived by cited
authors.

\begin{pozn}\label{poznamka}~\
\begin{enumerate}

\item
Trivially from the definition we have
$$
\Sigma_{\varepsilon,\eta}(\Omega_1)\subset\Sigma_{\varepsilon,\eta}(\Omega_2)
\quad\hbox{ for }\quad \Omega_1\subset\Omega_2\,.
$$
More generally, if $(\Omega_i)_{i\in\Z}$ is a sequence of nested bounded intervals such
that
$$
\cdots\subset \Omega_{i-1}\subset\Omega_i\subset\Omega_{i+1}\subset \cdots \qquad\hbox{ and }\qquad
\bigcup_{i\in\Z}\Omega_i = \R\,,
$$
then
$$
\cdots\subset \Sigma_{\varepsilon,\eta}(\Omega_{i-1})\subset\Sigma_{\varepsilon,\eta}(\Omega_i)
\subset\Sigma_{\varepsilon,\eta}(\Omega_{i+1})\subset \cdots
$$
and
$$
\bigcup_{i\in\Z}\Sigma_{\varepsilon,\eta}(\Omega_i) = \Z[\eta]\,.
$$

\item
Since $\Z[\eta]$ and $\Z[\varepsilon]$ are additive groups, the C\&P sequence satisfies
$$
x+\Sigma_{\varepsilon,\eta}(\Omega) = \Sigma_{\varepsilon,\eta}(\Omega+x^\star)\qquad
\hbox{ for every }\ x\in\Z[\eta].
$$
This property further implies that
$\Sigma_{\varepsilon,\eta}(\Omega)$ is not invariant under any
translation, i.e.\ is aperiodic.

\item
Any model set is Delone, see~\cite{moody}. In our one-dimensional
case it implies that there exists a sequence $(x_n)_{n\in\Z}$, and
two positive numbers $r_1,r_2$ such that $r_1<x_{n+1}-x_n<r_2$,
for all $n\in\Z$, and $\Sigma_{\varepsilon,\eta}(\Omega)=\{x_n\mid
n\in\Z\}$.

\item
The density of points of $\Sigma_{\varepsilon,\eta}(\Omega)$, defined as
$$
\varrho\left(\Sigma_{\varepsilon,\eta}(\Omega)\right) := \lim_{N\to+\infty} \frac{\#\
\bigl([-N,N]\cap\Sigma_{\varepsilon,\eta}(\Omega)\bigr)}{2N+1}
$$
is proportional to the length of the interval $\Omega$,
see~\cite{MoPaDensity}.

\item
Since $\Omega$ is an interval, it is easy to see that there exists a finite set
$F^\star\subset\Z[\varepsilon]$ such that $\Omega-\Omega\subset\Omega+F^\star$. Hence
there also exists a finite set $F$ such that
$$
\Sigma_{\varepsilon,\eta}(\Omega)-\Sigma_{\varepsilon,\eta}(\Omega)\subset
\Sigma_{\varepsilon,\eta}(\Omega)+F\,.
$$
Thus $\Sigma_{\varepsilon,\eta}(\Omega)$ satisfies the so-called
Meyer property. In fact, every model set $\Lambda\subset\R^{n}$ is
a Meyer set, i.e.\ is Delone and satisfies $\Lambda-\Lambda\subset
\Lambda+F$ for a finite set $F$, cf.~\cite{moody}. Note that in
this sense a cut-and-project set is a generalization of a lattice,
because a lattice satisfies the above property with $F=\{0\}$.

\item
Since $\Sigma_{\varepsilon,\eta}(\Omega)$ is a Meyer set, it is of
finite local complexity, i.e.\ it has only a finite number of
local configurations of a fixed size~\cite{lagarias-finite}. More
precisely, for $\varrho>0$ we define the $\varrho$-neighbourhood
of a point $x\in\Sigma_{\varepsilon,\eta}(\Omega)$ as
$$
N_\varrho(x)=\{ y\in \Sigma_{\varepsilon,\eta}(\Omega) \mid |x-y|<\varrho\}\,.
$$
The family of $\varrho$-neighbourhoods $\{N_\varrho(x)-x \mid x\in \Sigma_{\varepsilon,\eta}(\Omega)\}$
is finite for any positive $\varrho$. In particular, there is only finitely many distances between
adjacent points of any C\&P sequence, i.e. the set $\{x_{n+1}-x_n \mid n\in\Z\}$ is finite.

\item
The boundary of the acceptance interval $\Omega$ influences the structure of
the C\&P sequence only trivially. The sets $\Sigma_{\varepsilon,\eta}[c,c+\ell)$,
$\Sigma_{\varepsilon,\eta}[c,c+\ell]$, $\Sigma_{\varepsilon,\eta}(c,c+\ell)$, and
$\Sigma_{\varepsilon,\eta}(c,c+\ell]$ differ at most in two points. If
$c,c+\ell\notin\Z[\varepsilon]$, then all these sets coincide.

\item
If the acceptance window $\Omega$ is chosen to be a semi-closed interval, then
the number and shape of $\varrho$-neighbourhoods of a C\&P sequence does not depend on the
position $\Omega$, but only on its length $|\Omega|$.

\item
The set $\Sigma_{\varepsilon,\eta}(\Omega)$, where $\Omega=[c,c+\ell)$ or
$\Omega=(c,c+\ell]$ contains every finite configuration infinitely many times. More
precisely, for every $\varrho>0$ and every $x\in\Sigma_{\varepsilon,\eta}(\Omega)$
there exists infinitely many points $y\in\Sigma_{\varepsilon,\eta}(\Omega)$,
such that
$$
N_\varrho(x)-x =N_\varrho(y)-y\,.
$$
We say that such C\&P sequences are repetitive.

\end{enumerate}
\end{pozn}

\subsection{Distances}
\label{subs:dist}

As it was mentioned in (6) of Remark~\ref{poznamka}, any cut-and-project sequence has
only a finite number of distances between adjacent points. It turns out that the number
of distances does not exceed 3. We quote the result of~\cite{kombi}, which is a
generalization of the famous 3-distance theorem, and provide algorithms for determining
the distances for any particular acceptance interval.

\begin{thm}\label{mezery}
Let $\Omega$ be a semi-closed interval. For every $\Sigma_{\varepsilon,\eta}(\Omega)$
there exist positive numbers $\Delta_1$, $\Delta_2\in\Z[\eta]$ such that the distances
between adjacent points in $\Sigma_{\varepsilon,\eta}(\Omega)$ take values in
$\{\Delta_1,\Delta_2,\Delta_1+\Delta_2\}$. The numbers $\Delta_1$, $\Delta_2$ depend only
on the parameters $\varepsilon,\eta$ and on the length $|\Omega|$ of the interval
$\Omega$. They are linearly independent over $\Q$ and satisfy $\Delta_1^\star>0$,
$\Delta_2^\star<0$, and $\Delta_1^\star-\Delta_2^\star\geq |\Omega|$.
\end{thm}

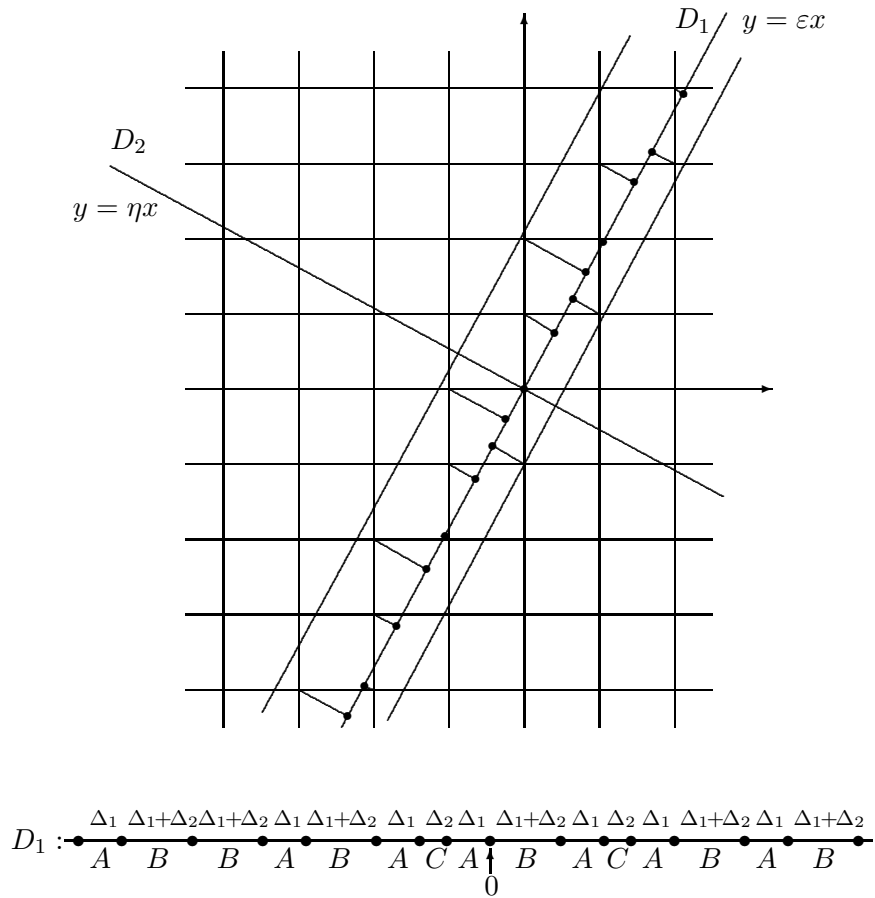
\begin{figure}[hbt]\begin{center}
\setlength{\unitlength}{1mm}
\begin{picture}(80,95)
\put(-5, 82){$D_2$}
\put(79, 98){$y=\varepsilon x$}
\put(-10, 73){$y=\eta x$}
\put(70,98){$D_1$}
\put(5,10){\line(1,0){70}}
\put(5,20){\line(1,0){70}}
\put(5,30){\line(1,0){70}}
\put(5,40){\line(1,0){70}}
\put(5,50){\vector(1,0){78}}
\put(5,60){\line(1,0){70}}
\put(5,70){\line(1,0){70}}
\put(5,80){\line(1,0){70}}
\put(5,90){\line(1,0){70}}
\put(10,5){\line(0,1){90}}
\put(20,5){\line(0,1){90}}
\put(30,5){\line(0,1){90}}
\put(40,5){\line(0,1){90}}
\put(50,5){\vector(0,1){95}}
\put(60,5){\line(0,1){90}}
\put(70,5){\line(0,1){90}}
\qbezier(15.2,7)(42,56)(64.1,97)
\qbezier(25.75,5)(50,50)(76.9,100)
\qbezier(-5,79.65)(50,50)(76.5,35.69)
\qbezier(31.9,6)(57,53)(78.8,94)
\qbezier(50,40)(47.9,41.25)(45.8,42.5)
\put(45.8,42.5){\circle*{1.2}}
\put(50,50){\circle*{1.2}}
\qbezier(50,60)(52,58.75)(54,57.5)
\put(54,57.5){\circle*{1.2}}
\qbezier(50,70)(54.25,67.65)(58.5,65.3)
\put(58.2,65.5){\circle*{1.2}}
\qbezier(60,60)(58.25,61)(56.5,62)
\put(56.5,62){\circle*{1.2}}
\qbezier(60,70)(60.55,69.75)(60.9,69.5)
\put(60.5,69.6){\circle*{1.2}}
\qbezier(60,80)(62.3,78.75)(64.6,77.5)
\put(64.6,77.5){\circle*{1.2}}
\qbezier(70,80)(68.5,80.75)(67,81.5)
\put(67,81.5){\circle*{1.2}}
\qbezier(70,90)(70.75,89.5)(71.5,89)
\put(71.15,89.25){\circle*{1.2}}
\qbezier(40,30)(39.5,30.25)(39,30.5) \put(39.5,30.5){\circle*{1.2}}
\qbezier(40,40)(41.75,39)(43.5,38) \put(43.5,38){\circle*{1.2}}
\qbezier(40,50)(43.75,48)(47.5,46) \put(47.5,46){\circle*{1.2}}
\qbezier(30,30)(33.5,28)(37,26) \put(37,26){\circle*{1.2}}
\qbezier(30,20)(31.5,19.25)(33,18.5) \put(33,18.5){\circle*{1.2}}
\qbezier(30,10)(29.25,10.25)(28.5,10.5) \put(28.8,10.5){\circle*{1.2}}
\qbezier(20,10)(23.25,8.25)(26.5,6.5) \put(26.5,6.5){\circle*{1.2}}
\end{picture}
{\setlength{\unitlength}{0.36mm}
\begin{picture}(310,47)
\put(-5,17){$D_1:$}
\put(15,20){\line(1,0){298}}
\put(20,20){\circle*{4}}
\put(36,20){\circle*{4}}
\put(62,20){\circle*{4}}
\put(88,20){\circle*{4}}
\put(104,20){\circle*{4}}
\put(130,20){\circle*{4}}
\put(146,20){\circle*{4}}
\put(156,20){\circle*{4}}
\put(172,20){\circle*{4}}
\put(170,0){0}
\put(172,8){\vector(0,1){9}}
\put(198,20){\circle*{4}}
\put(214,20){\circle*{4}}
\put(224,20){\circle*{4}}
\put(240,20){\circle*{4}}
\put(266,20){\circle*{4}}
\put(282,20){\circle*{4}}
\put(308,20){\circle*{4}}
\put(24,26){\scriptsize $\Delta_1$}
\put(38,26){\scriptsize $\Delta_1\!\!+\!\!\Delta_2$}
\put(64,26){\scriptsize $\Delta_1\!\!+\!\!\Delta_2$}
\put(92,26){\scriptsize $\Delta_1$}
\put(104,26){\scriptsize $\Delta_1\!\!+\!\!\Delta_2$}
\put(134,26){\scriptsize $\Delta_1$}
\put(148,26){\scriptsize $\Delta_2$}
\put(160,26){\scriptsize $\Delta_1$}
\put(174,26){\scriptsize $\Delta_1\!\!+\!\!\Delta_2$}
\put(202,26){\scriptsize $\Delta_1$}
\put(215,26){\scriptsize $\Delta_2$}
\put(228,26){\scriptsize $\Delta_1$}
\put(242,26){\scriptsize $\Delta_1\!\!+\!\!\Delta_2$}
\put(270,26){\scriptsize $\Delta_1$}
\put(284,26){\scriptsize $\Delta_1\!\!+\!\!\Delta_2$}
\put(24,10){$A$}
\put(45,10){$B$}
\put(71,10){$B$}
\put(92,10){$A$}
\put(112,10){$B$}
\put(134,10){$A$}
\put(148,10){$C$}
\put(160,10){$A$}
\put(181,10){$B$}
\put(202,10){$A$}
\put(215,10){$C$}
\put(228,10){$A$}
\put(249,10){$B$}
\put(270,10){$A$}
\put(291,10){$B$}
\end{picture}
}
\caption{Construction of a cut-and-project sequence and assignment of the infinite word
which codes the order of distances in the cut-and-project sequence. }
\label{f:cap}
\end{center}\end{figure}

More precisely, every C\&P sequence $\Sigma_{\varepsilon,\eta}(\Omega)=\{x_n\mid
n\in\Z\}$ has always two or three type of distances between adjacent points, namely
\begin{equation}\label{e:tri}
\{x_{n+1}-x_n\mid n\in\Z\}=\left\{\begin{array}{cl}
\{\Delta_1,\Delta_2,\Delta_1+\Delta_2\} &\hbox{ if } \Delta_1^\star-\Delta_2^\star>|\Omega|\,,\\[2mm]
\{\Delta_1,\Delta_2\} &\hbox{ if } \Delta_1^\star-\Delta_2^\star=|\Omega|\,.
\end{array}\right.
\end{equation}
Therefore one can naturally assign to it a binary or ternary bidirectional infinite word
$u_{\varepsilon,\eta}(\Omega)=(u_n)_{n\in\Z}$, for example in the alphabet $\{A,B,C\}$,
by
\begin{equation}\label{eq:slovo}
u_n=\left\{\begin{array}{cl}
A&\hbox{ if }\ x_{n+1}-x_n=\Delta_1\,, \\[2mm]
B&\hbox{ if }\ x_{n+1}-x_n=\Delta_1+\Delta_2\,,\\[2mm]
C&\hbox{ if }\ x_{n+1}-x_n=\Delta_2\,.
\end{array}\right.
\end{equation}

An example of construction of a cut-and-project sequence together with the assignment of
the infinite word is shown in Figure~\ref{f:cap}.

The successor of a point $x$ in the C\&P sequence is determined using its star-map image
$x^\star$. If $\Omega=[c,c+\ell)$, then the inequality $\Delta^\star_1-\Delta^\star_2\geq
\ell$ from Theorem~\ref{thmx} ensures that the nearest right neighbour of the point $x$
in $\Sigma_{\varepsilon,\eta}(\Omega)$ is equal to $x+\Delta_1$ if
$x^\star\in[c,c+\ell-\Delta_1^\star)$, to $x+\Delta_2$ if
$x^\star\in[c-\Delta_2^\star,c+\ell)$, or to $x+\Delta_1+\Delta_2$ if
$x^\star\in[c+\ell-\Delta_1^\star,c-\Delta_2^\star)$. Thus we can define a piecewise
linear map $f:[c,c+\ell)\to[c,c+\ell)$ which satisfies $f(x^\star)=y^\star$ if $y$ is the
nearest right neighbour of $x$. This mapping plays an important role in our
considerations.

\begin{de}\label{de:steppingf}
Let $c\in\R$, $\ell>0$. The stepping function of the interval $[c,c+\ell)$ is a mapping
$f:[c,c+\ell)\to[c,c+\ell)$ defined by
$$
f(y)=\left\{\begin{array}{ll}
y+\Delta^\star_1 &\hbox{ if } \ y \in[c,c+\ell-\Delta_1^\star)\,,\\[2mm]
y+\Delta^\star_1+\Delta^\star_2 &\hbox{ if } \ y \in[c+\ell-\Delta_1^\star,c-\Delta_2^\star)\,,\\[2mm]
y+\Delta^\star_2&\hbox{ if } \ y \in[c-\Delta_2^\star,c+\ell) \,.
\end{array}\right.
$$
\end{de}

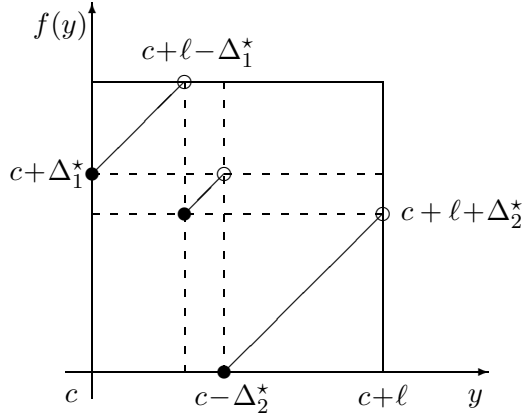
\begin{figure}[hbt]
\begin{center}
\begin{picture}(175,146)
\put(20,10){\vector(1,0){160}}
\put(30,0){\vector(0,1){150}}
 \put(18,141){\makebox(0,0){$f(y)$}}
 \put(175,0){\makebox(0,0){$y$}}
\put(140,10){\line(0,1){110}}
\put(30,120){\line(1,0){110}}
\put(80,10){\line(1,1){60}}
\put(65,70){\line(1,1){15}}
\put(30,85){\line(1,1){35}}
 \put(22,1){\makebox(0,0){$c$}}
 \put(70,131){\makebox(0,0){$c\!+\!\ell\!-\!\Delta_1^\star$}}
 \put(83,0){\makebox(0,0){$c\!-\!\Delta_2^\star$}}
 \put(140,1){\makebox(0,0){$c\!+\!\ell$}}
 \put(13,85){\makebox(0,0){$c\!+\!\Delta_1^\star$}}
 \put(170,70){\makebox(0,0){$c+\ell\!+\!\Delta_2^\star$}}
\put(80,10){\circle*{5}}
 \put(65,70){\circle*{5}}
 \put(30,85){\circle*{5}}
 \put(140,70){\circle{5}}
 \put(80,85){\circle{5}}
 \put(65,120){\circle{5}}
\dashline{3}(80,10)(80,120)
 \dashline{3}(65,10)(65,120)
 \dashline{3}(30,70)(140,70)
\dashline{3}(30,85)(140,85)
\end{picture}
\caption{Stepping function $f$ associated to $\Sigma_{\varepsilon,\eta}[c,c+\ell)$. If
the distances between neighbours in the C\&P set $\Sigma_{\varepsilon,\eta}[c,c+\ell)$
take only two values $\Delta_1$, $\Delta_2$, i.e. $\ell=\Delta_1^\star-\Delta_2^\star$,
then the discontinuity points $c+\ell-\Delta_1^\star$ and $c-\Delta_2^\star$ coincide.}
\label{f}
\end{center}
\end{figure}

The graph of the map $f$ is illustrated on Figure~\ref{f}.

The stepping function $f$ of Figure~\ref{f} has been studied in the field of dynamical
systems under the name of three interval exchange (in case that $f$ has two discontinuity
points) or two interval exchange (if $f$ has only one discontinuity point). In the former
situation, the two discontinuity points divide the acceptance interval $\Omega$ into
three disjoint intervals, say $\Omega_A$, $\Omega_B$, $\Omega_C$, from left to right. The
image $f(\Omega)=\Omega$ is again divided into three disjoint intervals $f(\Omega_C)$,
$f(\Omega_B)$, $f(\Omega_A)$, in the order from left to right. Therefore one sometimes
uses the graphical notation shown in the following scheme.

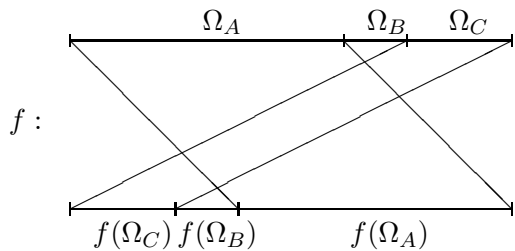
\begin{center}
{\setlength{\unitlength}{0.28mm}
\begin{picture}(270,130)
\put(30,30){\line(1,0){210}} \put(30,27){\line(0,1){6}} \put(240,27){\line(0,1){6}}
\put(110,27){\line(0,1){6}} \put(80,27){\line(0,1){6}} \put(30,110){\line(1,0){210}}
\put(30,107){\line(0,1){6}} \put(240,107){\line(0,1){6}} \put(190,107){\line(0,1){6}}
\put(160,107){\line(0,1){6}} \put(30,110){\line(1,-1){80}} \put(160,110){\line(1,-1){80}}
\put(30,30){\line(2,1){160}} \put(80,30){\line(2,1){160}} \put(1,68){$f:$}
\put(41,14){$f(\Omega_C)$} \put(81,14){$f(\Omega_B)$} \put(163,14){$f(\Omega_A)$}
\put(93,115){$\Omega_A$} \put(171,115){$\Omega_B$} \put(210,115){$\Omega_C$}
\end{picture}
}
\end{center}

The relation of three interval exchange to simultaneous
approximation of a pair of irrational numbers is treated
in~\cite{adam,3ExchI,rauzy}. In the theory of symbolic dynamical
systems one studies the orbit of a point $x\in\Omega$, under the
mapping $f$, i.e. the sequence
$(f^{(n)})_{n\in\N_0}$\footnote{Note that we use the notation
$\N=\{1,2,3,\dots\}$ and $\N_0=\{0\}\cup\N$.}. Therefore C\&P
sequences can be viewed as geometric representations of three
interval exchange transformations.

\smallskip
Changing continuously the length $\ell$ of the acceptance interval $\Omega=[c,c+\ell)$
causes discrete changes of the triplet of distances ($\Delta_1$, $\Delta_2$,
$\Delta_1+\Delta_2$). Recall that the triplet does not depend on $c$, therefore we
consider $c=0$. Let $\Sigma_{\varepsilon,\eta}[0,\ell)$ be a C\&P sequence with three
distances between its neighbours, and let $\Delta^\star_1>0$, $\Delta^\star_2<0$,
$\Delta^\star_1+\Delta^\star_2$, be the star map images of these distances. According
to~\eqref{e:tri}, we must have $\ell<\Delta_1^\star-\Delta_2^\star$. Growing $\ell$ up to
the value $\Delta_1^\star-\Delta_2^\star$ causes appearance of new points in the C\&P
sequence, which split the large distance $\Delta_1+\Delta_2$ into two distances
$\Delta_1$ and $\Delta_2$. When $\ell$ reaches the value $\Delta_1^\star-\Delta_2^\star$,
the large distance $\Delta_1+\Delta_2$ disappears completely.

On the other hand, diminishing the length $\ell$ of the acceptance interval causes that
the frequency of the distance $\Delta_1+\Delta_2$ grows to the detriment of occurrences
of the distances $\Delta_1$ and $\Delta_2$. This happens until $\ell$ reaches a certain
limit value for which one of the distances $\Delta_1$ or $\Delta_2$ disappears.

Starting from a given initial value $\ell_0$, for which the set
$\Sigma_{\varepsilon,\eta}[0,\ell_0)$ has two distances between
adjacent points, we can determine by recurrence the increasing sequence of lengths
$\ell_n$, $n\in\Z$, of the acceptance windows for which
$\Sigma_{\varepsilon,\eta}[0,\ell_n)$ has only two distances.
The initial value $\ell_0$ is determined below (Remark~\ref{pozn:init}).

\smallskip Let $\Delta_{n1}^\star>0$ and $\Delta_{n2}^\star<0$ be the star images of
distances occurring in the sequence $\Sigma_{\varepsilon,\eta}[0,\ell_n)$, i.e.,
according to~\eqref{e:tri}, $\ell_n=\Delta_{n1}^\star-\Delta_{n2}^\star$.
 \begin{equation}\label{eq:11}
 \begin{array}{l}
 \hbox{If } \Delta_{n1}^\star+\Delta_{n2}^\star > 0\,
 \hbox{ then } \\[2mm]
 \hspace*{1.3cm}\ell_{n-1}:= \Delta_{n1}^\star,\quad
 \Delta_{(n-1)1}^\star:= \Delta_{n1}^\star+\Delta_{n2}^\star,\quad
 \Delta_{(n-1)2}^\star:= \Delta_{n2}^\star.\\[2mm]
 \hbox{If }  \Delta_{n1}^\star+\Delta_{n2}^\star < 0\,
 \hbox{ then } \\[2mm]
 \hspace*{1.3cm}\ell_{n-1}:= -\Delta_{n2}^\star,\quad
 \Delta_{(n-1)1}^\star:= \Delta_{n1}^\star,\quad
 \Delta_{(n-1)2}^\star:= \Delta_{n1}^\star+\Delta_{n2}^\star.
 \end{array}
 \end{equation}

\noindent Similarly, the algorithm which determines the triple
$\ell_{n+1},\Delta_{(n+1)1}, \Delta_{(n+1)2}$ from the triple
$\ell_n,\Delta_{n1}, \Delta_{n2}$ has the inverse form
 \begin{equation}\label{eq:11+}
 \begin{array}{l}
 \hbox{If } \Delta_{n1}>\Delta_{n2}\,
 \hbox{ then } \\[2mm]
 \hspace*{0.4cm}\ell_{n+1}:= \Delta_{n1}^\star-2\Delta_{n2}^\star,\quad
 \Delta_{(n+1)1}^\star:= \Delta_{n1}^\star-\Delta_{n2}^\star,\quad
 \Delta_{(n+1)2}^\star:= \Delta_{n2}^\star.\\[2mm]
 \hbox{If }  \Delta_{n1}<\Delta_{n2}\,
 \hbox{ then } \\[2mm]
 \hspace*{0.4cm}\ell_{n+1}:= 2\Delta_{n1}^\star-\Delta_{n2}^\star,\quad
 \Delta_{(n+1)1}^\star:= \Delta_{n1}^\star,\quad
 \Delta_{(n+1)2}^\star:= \Delta_{n2}^\star-\Delta_{n1}^\star.
 \end{array}
 \end{equation}

From this algorithm it can be seen that the C\&P sequences
$\Sigma_{\varepsilon,\eta}[0,\ell_n)$ and $\Sigma_{\varepsilon,\eta}[0,\ell_{n-1})$ have
exactly one type of distances in common. It is the shorter one among $\Delta_{(n-1)1}$,
$\Delta_{(n-1)2}$. Moreover, the distances in $\Sigma_{\varepsilon,\eta}[0,\ell)$, for
$\ell_{n-1}<\ell<\ell_{n}$, are three, and they are given by the union of the sets of
distances for $\Sigma_{\varepsilon,\eta}[0,\ell_n)$ and
$\Sigma_{\varepsilon,\eta}[0,\ell_{n-1})$.

\subsection{Transformations}

We would like to identify those parameters $\varepsilon,\eta,\Omega$ which provide
essentially the same cut-and-project sequences. For example, we have
\begin{equation}\label{eq:cisloc}
a+b\eta+\Sigma_{\varepsilon,\eta}(\Omega) =
\Sigma_{\varepsilon,\eta}(\Omega+a+b\varepsilon)\,,\qquad \hbox{ for }\ a,b\in\Z\,.
\end{equation}
Such translation of the C\&P sequence corresponds to a translation of the lattice $\Z^2$.
The group of all linear transformations of the lattice $\Z^2$ onto itself is
$$
G=\bigl\{{\mathbb A}\in M_2(\Z) \,\bigm|\, \det{\mathbb
A}=\pm1\bigr\}\,.
$$
Consider the matrix ${\mathbb A}=\binom{a\ b}{c\ d}$. For arbitrary irrational numbers
$\varepsilon$, $\eta$ and arbitrary interval $\Omega$ it holds that
\begin{eqnarray*}
\Sigma_{\varepsilon,\eta}(\Omega)
  &=& \left\{p+q\eta \;\bigm|\; p,q\in\Z,\, p+
q\varepsilon\in\Omega\right\}
    =\nonumber\\[1mm]
  &=& \left\{(1,\eta)\bigl(\begin{smallmatrix}p\\
q\end{smallmatrix}\bigr)
      \;\bigm|\; p,q\in\Z,\, (1,\varepsilon)
\bigl(\begin{smallmatrix}p\\
       q\end{smallmatrix}\bigr)\in\Omega\right\} =\nonumber\\[1mm]
  &=& \left\{(1,\eta){\mathbb A}\bigl(\begin{smallmatrix}p\\
q\end{smallmatrix}\bigr)
      \;\bigm|\; p,q\in\Z,\, (1,\varepsilon){\mathbb A}
\bigl(\begin{smallmatrix}p\\
      q\end{smallmatrix}\bigr)\in\Omega\right\} =\nonumber\\[1mm]
  &=& \left\{(a+c\eta,b+d\eta)\bigl(\begin{smallmatrix}p\\
q\end{smallmatrix}\bigr)
      \;\bigm|\; p,q\in\Z,\, (a+c\varepsilon,b+d\varepsilon)
      \bigl(\begin{smallmatrix}p\\
q\end{smallmatrix}\bigr)\in\Omega\right\}
     =\nonumber\\[1mm]
  &=& (a+c\eta)
       \left\{\bigl(1,\tfrac{b+d\eta}{a+c\eta}\bigr)
    \bigl(\begin{smallmatrix}p\\ q\end{smallmatrix}\bigr)
      \;\bigm|\; p,q\in\Z,\, \bigl(1,\tfrac{b+d\varepsilon}{a+
c\varepsilon}\bigr)
      \bigl(\begin{smallmatrix}p\\ q\end{smallmatrix}\bigr)
      \in\tfrac{1}{a+c\varepsilon}\Omega\right\} =\nonumber\\[1mm]
   &=& (a+c\eta) \Sigma_{\tfrac{b+d\varepsilon}{a+c\varepsilon},
       \tfrac{b+d\eta}{a+c\eta}}\left(\tfrac1{a+
c\varepsilon}\Omega\right)\,.
\end{eqnarray*}
Let us study the consequences of the above relation if we choose for the matrix
${\mathbb A}$ one of the three generators ${\mathbb A}_1=\binom{1\ 1}{0\ 1}$,
${\mathbb A}_2=\binom{1\ \ 0}{0\ -\!1}$ , ${\mathbb A}_3=\binom{0\ 1}{1\ 0}$ of the group $G$,
\begin{eqnarray}
\Sigma_{\varepsilon,\eta}(\Omega)
    &=& \Sigma_{1+\varepsilon,\ 1+\eta}(\Omega)\,, \label{e:trans1}
    \\[1mm]
\Sigma_{\varepsilon,\eta}(\Omega)
    &=& \Sigma_{-\varepsilon,-\eta}(-\Omega)\,, \label{e:trans2}
    \\[1mm]
\Sigma_{\varepsilon,\eta}(\Omega)
    &=&\eta\ \Sigma_{\tfrac1\varepsilon,\tfrac1\eta}
    (\tfrac{1}{\varepsilon}\Omega)\,. 
\end{eqnarray}

The mentioned transformations were used in~\cite{kombi} for the proof of the following
theorem.

\begin{thm}\label{thmx}
For every irrational numbers $\varepsilon,\eta$, $\varepsilon\neq\eta$ and every bounded
interval $\Omega$, there exist $\tilde{\varepsilon}\in(-1,0)$, $\tilde{\eta}>0$ and an
interval $\tilde{\Omega}$, satisfying $\max(1+\tilde{\varepsilon},-\tilde{\varepsilon}) <
|\tilde{\Omega}| \leq 1$, such that
$$
\Sigma_{\varepsilon,\eta}(\Omega) =
s\Sigma_{\tilde{\varepsilon},\tilde{\eta}}(\tilde{\Omega})\,,\qquad \hbox{ for some }\ s\in\R\,.
$$
Moreover, if $|\tilde{\Omega}| \neq 1$, then the distances between adjacent points
in $\Sigma_{\tilde{\varepsilon},\tilde{\eta}} (\tilde{\Omega})$ are $\tilde{\eta}$,
$1+\tilde{\eta}$, and $1+2\tilde{\eta}$. The distances take only two values
$\tilde{\eta}$, $1+\tilde{\eta}$, only if $|\tilde{\Omega}| = 1$.
\end{thm}

According to the above theorem, every C\&P sequence is geometrically similar to another
C\&P sequence whose parameters satisfy certain restricted conditions. In particular,
without loss of generality we can consider $\varepsilon\in(-1,0)$, $\eta>0$ and
the length of the acceptance interval $\Omega$ in the range
$(\max(1+{\varepsilon},-{\varepsilon}),1]$.
If moreover we are interested only in the ordering of the distances in the C\&P sequence
and not on their actual lengths, i.e. we consider only the infinite word $u_{\varepsilon,\eta}(\Omega)$
we can choose any fixed $\eta>0$. The words $u_{\varepsilon,\eta_1}(\Omega)$,
$u_{\varepsilon,\eta_2}(\Omega)$ for $\eta_1\neq\eta_2$ coincide. Therefore choosing
$\eta=-\frac1\varepsilon$, which corresponds to a cut-and-project scheme with orthogonal projection,
causes no loss of generality when studying only combinatorial properties of C\&P sequences. The choice
of $\eta$ however influences geometry of the sequences, such as existence of self-similarity factor,
etc.~(cf. Section~\ref{s:selfs}).

\begin{pozn}\label{pozn:init}
Note that according to Theorem~\ref{thmx}, the length $|\Omega|$ of the interval $\Omega$
being equal to 1 is the only case among
$\bigl(\max(1+\tilde{\varepsilon},-\tilde{\varepsilon}),1\bigr]$
for which the C\&P set has only two distances between neighbours.
We shall thus take it as the initial case for the algorithm given
in~\eqref{eq:11},~\eqref{eq:11+}. We have
\begin{equation}\label{e:init}
\ell_0=1,\quad \Delta^\star_{01}=1+\varepsilon,\quad \Delta^\star_{02}=\varepsilon\,.
\end{equation}
\end{pozn}

\begin{ex}\label{ex}
As an example, let us study the case $|{\Omega}|=1$, which gives a C\&P sequence with two distances
between adjacent points.
Set $\alpha=-\varepsilon\in(0,1)$ and put  $\Omega=(\beta-1,\beta]$ for some
$\beta\in\R$ as the acceptance window. Since the condition $a+b\varepsilon\in\Omega$ rewrites
as $\beta-1<a-b\alpha\leq\beta$, we obtain $a=\lfloor b\alpha+\beta\rfloor$ and the C\&P sequence
is of the form
\begin{equation}\label{e:expl}
\Sigma_{-\alpha,\eta}(\beta-1,\beta] = \{\lfloor b\alpha+\beta\rfloor + b\eta \mid b\in\Z\}\,.
\end{equation}
Since $\alpha,\eta>0$, the sequence $x_n:=\lfloor n\alpha+\beta\rfloor + n\eta$ is strictly increasing
and thus the distances between adjacent points of the C\&P set $\Sigma_{-\alpha,\eta}(\beta-1,\beta]$
are of the form
\begin{equation}\label{e:mezery}
x_{n+1}-x_n = \eta + \lfloor (n+1)\alpha+\beta\rfloor - \lfloor n\alpha+\beta\rfloor =\left\{
\begin{array}{c}
\eta+1\,,\\
\eta\,.
\end{array}\right.
\end{equation}
From this expression it can be seen that the distances in the C\&P sequence are arranged in the same
order as 0's and 1's in the so-called lower and upper mechanical word.
Recall that the lower mechanical word $\underline{s}_{\alpha,\beta}:\Z\to\{0,1\}$
is defined by the prescription
\begin{equation}\label{eq:dolnimech}
\underline{s}_{\alpha,\beta}(n) = \lfloor (n+1)\alpha+\beta\rfloor - \lfloor n\alpha+\beta\rfloor\,,
\end{equation}
where $\alpha$ is called the slope and $\beta$ the intercept of the word $\underline{s}_{\alpha,\beta}$.
Similarly, upper mechanical word $\overline{s}_{\alpha,\beta}:\Z\to\{0,1\}$ is defined by the prescription
\begin{equation}\label{eq:hornimech}
\overline{s}_{\alpha,\beta}(n) = \lceil (n+1)\alpha+\beta\rceil - \lceil n\alpha+\beta\rceil\,.
\end{equation}
The infinite word $u_{\varepsilon,\eta}(\beta-1,\beta]$ with
parameters $\eta>0$, $\varepsilon=-\alpha$ is in fact the lower
mechanical word $\underline{s}_{\alpha,\beta}$. Similarly, the
choice $[\beta,\beta+1)$ for the acceptance window provides the
upper mechanical word $\overline{s}_{\alpha,\beta}$. The
mechanical words are in fact related to the well-known sturmian
words, see~Definition~\ref{de:sturm} and
Remark~\ref{pozn:mechkom}.
\end{ex}

\section{Combinatorial properties of C\&P sequences}

Ordering of the distances in the C\&P sequence $\Sigma_{\varepsilon,\eta}(\Omega)$ on the
real line defines naturally an infinite binary or ternary word
$u_{\varepsilon,\eta}(\Omega)$ (cf. equation~\eqref{eq:slovo}). In this section we
describe some combinatorial properties of these infinite words. Some of the results
derived here can be found in~\cite{3ExchII}. Nevertheless, the geometric approach to
three interval exchange makes the proof simpler.

Obviously, geometrically similar C\&P sequences correspond to the same infinite words.
Therefore according to Theorem~\ref{thmx} we can consider only
\begin{align}\label{eq:buno}
&\varepsilon\in(-1,0),\ \eta>0\quad \hbox{ and  }\quad \Omega=[c,c+\ell),\\
&\nonumber \quad \hbox{ where  }\ \max(1+{\varepsilon},-{\varepsilon})\ <\ \ell\ \leq\ 1.
\end{align}
In this case the stepping function has the form
\begin{equation}\label{eq:stepf}
f(y)=\left\{\begin{array}{lll}
y+1+\varepsilon &\hbox{ if } \ y\ \in\ [c,c+\ell-1-\varepsilon)&=:\ \Omega_A\,,\\[2mm]
y+1+2\varepsilon&\hbox{ if } \ y\ \in\ [c+\ell-1-\varepsilon,c-\varepsilon)&=:\ \Omega_B\,,\\[2mm]
y+\varepsilon &\hbox{ if }\ y\ \in\ [c-\varepsilon,c+\ell)&=:\ \Omega_C\,.
\end{array}\right.
\end{equation}
For simplicity, we denote the discontinuity points of the stepping function
$$
\delta_1:=c+\ell-1-\varepsilon\,,\qquad
\delta_2:=c-\varepsilon\,.
$$
As was already mentioned, the infinite word $u_{\varepsilon,\eta}(\Omega)$ is defined
over a binary alphabet if and only if the length of the acceptance window is $\ell=1$,
because in that case $\delta_1=\delta_2$ and thus $\Omega_B=\emptyset$. Otherwise the
alphabet of $u_{\varepsilon,\eta}(\Omega)$ has three letters.

Let us recall some basic notions of combinatorics on words. An alphabet ${\mathcal A}$ is
a finite set of symbols - letters. A finite concatenation $w$ of letters is called a
finite word. The set of all finite words (including the empty word $\epsilon$) over the
alphabet $\A$ is denoted by $\A^*$. The concatenation of $n$ letters $a$ is denoted by
$a^n$. The length of a word $w$ is the number of letters concatenated in $w$, it is
denoted by $|w|$. One considers also one-directional infinite words
$$
u=u_0u_1u_2u_3\cdots
$$
and bidirectional infinite words
$$
u=\cdots u_{-2}u_{-1}u_0u_1u_2\cdots\,.
$$
In relation to C\&P sequences, mainly bidirectional infinite words are important. We
denote the set of such words by $\A^\Z$. A word $w=w_0w_1\cdots w_{k-1}$ is called a {\em
factor} of a word $u\in\A^\Z$ if $w=u_iu_{i+1}\cdots u_{i+k-1}$ for some $i$. Note that
such $i$ is called the {\em occurrence} of $w$ in $u$. The set of factors of a word
$u\in\A^\Z$ with the length $n$ is denoted by
$$
{\mathcal L}_n=\{u_iu_{i+1}\cdots u _{i+n-1}\mid i\in\Z\}\,.
$$
The set of all factors of the word $u$ (the {\em language} of $u$) is denoted by
$$
{\mathcal L} = \bigcup_{n\in\N} {\mathcal L}_n\,.
$$
The number of different $n$-tuples that appear in the infinite
word is given by the so-called complexity function, see for
example~\cite{allouche}.

\begin{de}
The complexity of a word $u\in\A^\Z$ is a mapping ${\mathcal C}:\N\to\N$ such that
$$
{\mathcal C}(n) = \#\{u_iu_{i+1}\cdots u _{i+n-1}\mid i\in\Z\} = \#{\mathcal L}_n\,.
$$
\end{de}

Obviously, if $u$ is an infinite word over a $k$-letter alphabet, then its complexity satisfies
$$
1\leq {\mathcal C}(n)\leq k^n\,,\qquad\hbox{ for every }n\in\N\,.
$$
It is known~\cite{morse1} that if there exists an $n\in\N$ such
that ${\mathcal C}(n)\leq n$, then the word $u$ is periodic, i.e.\
of the form $u=\cdots wwww\cdots $ for a finite word $w$. An
aperiodic word of minimal complexity thus satisfies ${\mathcal
C}(n)=n+1$ for all $n\in\N$. An example of such a word is the word
$\cdots 0001000\cdots$ on the alphabet $\{0,1\}$. The structure of
such words is little interesting, since the occurrence of the
letter 1 is singular. Obviously, they cannot be obtained by a
cut-and-project scheme. In order to avoid such strange phenomena,
we consider only those words which have reasonable density of
their letters. The density of a letter $a$ in the infinite word
$u=\cdots u_{-2}u_{-1}u_0u_1u_2\cdots$ is defined by
$$
 \varrho_a := \lim_{k\to\infty}
 \frac{\# \bigl\{i\in\Z\cap[-k,k] \,\bigm|\, u_i=a \bigr\}}{2k+1}\,,
$$
if the limit exists.

\begin{de}\label{de:sturm}
An infinite word $u\in\A^\Z$ is called sturmian if ${\mathcal
C}(n)=n+1$ for all $n\in\N$ and the densities of its letters are
irrational.
\end{de}

Such words have been extensively studied. We shall focus on them
later in this section. Let us mention that the condition
${\mathcal C}(n)=n+1$ for all $n\in\N$ in the case of
one-directional infinite words already implies irrationality of
the densities of letters. Our notion of sturmian words
follows~\cite{morse}, however, sturmian words are often considered
only as one-directional. A survey of properties of one-directional
sturmian words can be found in~\cite{lothaire}.


\subsection{Complexity}\label{subs:compl}

For the determination of the factors in the
infinite bidirectional word $u_{\varepsilon,\eta}(\Omega)$ associated with the C\&P
sequence $\Sigma_{\varepsilon,\eta}(\Omega)$ it is essential to study the stepping function $f$.
Its properties imply that the word $w=w_0w_1\cdots w_{k-1}$ in the alphabet $\A=\{A,B,C\}$
is a factor of $u_{\varepsilon,\eta}(\Omega)$ if and only if there exists an
$x\in\Sigma_{\varepsilon,\eta}(\Omega)$ such that
$$
x^\star\in\Omega_{w_0}\,,\quad f(x^\star)\in\Omega_{w_1}\,,\quad \dots\,,\quad
f^{k-1}(x^\star)\in\Omega_{w_{k-1}}\,,
$$
where $\Omega_A$, $\Omega_B$, $\Omega_C$ are defined in~\eqref{eq:stepf}. This means that
$$
w=w_0w_1\cdots w_{k-1}\in{\cal L}_k \quad\iff\quad
\Omega_{w_0}\cap f^{-1}(\Omega_{w_1})\cap\dots\cap f^{-(k-1)}(\Omega_{w_{k-1}})\neq\emptyset\,.
$$
In case that $w=w_0w_1\cdots w_{k-1}\in{\cal L}_k$, we denote
$$
\Omega_w:=\Omega_{w_0}\cap f^{-1}(\Omega_{w_1})\cap\dots\cap f^{-(k-1)}(\Omega_{w_{k-1}})
\,.
$$
Properties of the stepping function $f$ imply that $\Omega_w$ is an interval, closed from the left,
open from the right. Obviously, we have
$$
\Omega=\bigcup_{w\in{\cal L}_k}\Omega_w\,,
$$
where the union is disjoint. In order that points $x^\star,y^\star\in\Omega$ belong to
different intervals $x^\star\in\Omega_{w^{(1)}}$, $y^\star\in\Omega_{w^{(2)}}$, where
$w^{(1)}\neq w^{(2)}$, $w^{(1)},w^{(2)}\in{\cal L}_k$, there must exist $i=0,1,\dots,k-1$
such that at least one discontinuity point of the function $f$ lies between
$f^{i}(x^\star)$ and $f^{i}(y^\star)$. Thus boundaries between intervals $\Omega_w$ for
$w\in{\cal L}_k$ are all points $z$ such that $f^{i}(z)$ is a discontinuity point of the
function $f$, i.e. $f^{i}(z)\in\{\delta_1,\delta_2\}$. This implies that the number of
different factors of the word $u_{\varepsilon,\eta}(\Omega)$ of length $k$ is equal to
the number of elements
\begin{equation}\label{eq:jazyk}
\#{\cal L}_k=\#\Bigl\{c,\delta_1,f^{-1}(\delta_1),\dots,f^{-k+1}(\delta_1),
\delta_2,f^{-1}(\delta_2),\dots,f^{-k+1}(\delta_2)\Bigr\}\,.
\end{equation}

For determination of the cardinality of the set ${\cal L}_k$, i.e. complexity of the infinite word,
we need to use two properties of the stepping function $f$, which follow from the irrationality of
$\varepsilon$.
\begin{equation}\label{eq:xy}
\begin{array}{ll}
1)\ \hbox{Let } x\in\Omega. &\hbox{Then } f^i(x)\neq x \hbox{ for all } i\in\Z,\ i\neq0.\\[3mm]
2)\ \hbox{Let } x,y\in\Omega. &\hbox{Then } \exists\, i\in\Z, \hbox{ such that }
f^i(x)= y \hbox{ iff } x-y\in\Z[\varepsilon].
\end{array}
\end{equation}

Since $c=f(\delta_2)$, the property 1) implies that
$c,\delta_2,f^{-1}(\delta_2),\dots,f^{-k+1}(\delta_2)$ are distinct. Thus
$$
{\cal C}(k)=\#{\cal L}_k\geq k+1\,,\qquad\hbox{ for all }\ k\in\N\,.
$$
Equality holds only in the case that $\delta_1=\delta_2$. Therefore for C\&P sequences
$u_{\varepsilon,\eta}(\Omega)$ with parameters $\varepsilon,\eta,\Omega$ satisfying~\eqref{eq:buno}
it holds that
$$
{\cal C}(k)= k+1 \qquad\iff\qquad |\Omega|=1\,.
$$
We have thus derived the following well known fact.

\begin{pozn}\label{pozn:mechkom}
Every mechanical word~\eqref{eq:dolnimech} or~\eqref{eq:hornimech}
is a sturmian word. The opposite is also true~\cite{CoHe,morse}.
\end{pozn}

The results about the complexity function of all C\&P sequences are summarized
in the following theorem.

\begin{thm}[\cite{kombi}]\label{thmcompl}
Let ${\mathcal C}$ be the complexity function of the infinite word
$u_{\varepsilon,\eta}(\Omega)$ with $\Omega=[c,c+\ell)$, and let $f$ be the corresponding
stepping function.
 \begin{itemize}
 \item If $\ell\notin\Z[\varepsilon]$, then
 $$
 {\mathcal C}(n)=2n+1\,,\qquad\hbox{for }\ n\in\N\,.
 $$
 \item If $\ell\in\Z[\varepsilon]$, then there exists a unique
 $n_0\in\N_0$ such that
 $$
 {\mathcal C}(n)=\left\{
 \begin{array}{cl}
 2n+1 &\ \hbox{for }\ n\leq n_0\,,\\[1mm]
 n+n_0+1 &\ \hbox{for }\ n>n_0\,.
 \end{array}
 \right.
 $$
 \end{itemize}
\end{thm}

Obviously, generic cut-and-project sequences have complexity $2n+1$. In case that the
length of the acceptance window is in $\Z[\varepsilon]$, the cut-and-project sequence has
a specific property which is explained in the following remark.

\begin{pozn}
Theorem~\ref{thmcompl} says that infinite words
$u_{\varepsilon,\eta}(\Omega)$ with $|\Omega|\in\Z[\varepsilon]$
have complexity $\C(n)= n+ {\it const.}$ for sufficiently large
$n$. One-directional words with such complexity are
 called quasisturmian by Cassaigne in~\cite{cassaigne2}. This author shows that
such words have a sturmian structure, i.e.\ up to a finite prefix
they are images under a morphism of a one-direction sturmian word.
Following the same ideas, one can show that the
 bidirectional infinite word
$u_{\varepsilon,\eta}(\Omega)$ with $|\Omega|\in\Z[\varepsilon]$
corresponding to a C\&P sequence satisfies the following: there
exists a sturmian word $v=\cdots v_{-2}v_{-1}|v_0v_1v_2\cdots
\in\{0,1\}^{\Z}$ and finite words $W_0,W_1\in\{A,B,C\}^*$ such
that
$$
u_{\varepsilon,\eta}(\Omega) = \cdots
W_{v_{-2}}W_{v_{-1}}|W_{v_0}W_{v_1}W_{v_2}\cdots\,,
$$
i.e. $u_{\varepsilon,\eta}(\Omega)$ can be obtained by concatenation of words $W_0,W_1$
in the order of 0's and 1's in the sturmian word $v$.
\end{pozn}

\begin{ex}
Consider $\varepsilon=-\frac1\tau$ and $\eta=\tau$, where $\tau=\frac12(1+\sqrt5)$, (see
Introduction). For the acceptance window choose
$\Omega_1=\bigl[-\frac{7}\tau+4,-\frac{17}\tau+11\bigr)$,
$\Omega_2=\Omega_1+2-\frac{3}\tau$. Since the length of the acceptance windows
$\ell=|\Omega_1|=|\Omega_2|$ satisfies
$\ell=7-\frac{10}\tau\in\bigl(\max(1+\varepsilon,\varepsilon),1\bigr]$, according to
Theorem~\ref{thmx}, the distances between adjacent points in the C\&P sequences
$\Sigma_{\varepsilon,\eta}(\Omega_1)$, $\Sigma_{\varepsilon,\eta}(\Omega_2)$ are
$1+\tau$, coded by the letter $A$; $1+2\tau$, coded by the letter $B$; and $\tau$, coded
by the letter $C$.

Using~\eqref{eq:cisloc} the sequences $\Sigma_{\varepsilon,\eta}(\Omega_1)$,
$\Sigma_{\varepsilon,\eta}(\Omega_2)$ are the same, up to a shift by $2+3\tau$. Both of
them contain 0, since $0\in\Omega_1$, $0\in\Omega_2$. Figure~\ref{f:quasisturmian} shows a
segment of the infinite words coding these sequences, where we mark the point 0 in both
of them.

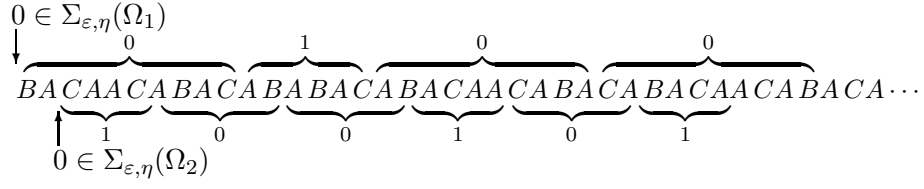
\begin{figure}[ht]
\begin{center}
\begin{picture}(330,75)
\put(8,58){$0\in\Sigma_{\varepsilon,\eta}(\Omega_1)$}
\put(10,56){\vector(0,-1){13}} \put(10,30){\small
$BA\,CAA\,CA\,BA\,CA\,BA\,BA\,CA\,BA\,CAA\,CA\,BA\,CA\,BA\,CAA\,CA\,BA\,CA\cdots$}
\put(24,3){$0\in\Sigma_{\varepsilon,\eta}(\Omega_2)$}
\put(26,12){\vector(0,1){13}}
\put(13,39){$\overbrace{\hspace*{2.8cm}}^0$}
\put(98,39){$\overbrace{\hspace*{1.5cm}}^1$}
\put(146.5,39){$\overbrace{\hspace*{2.8cm}}^0$}
\put(232,39){$\overbrace{\hspace*{2.8cm}}^0$}
\put(27,28){$\underbrace{\hspace*{1.2cm}}_1$}
\put(65,28){$\underbrace{\hspace*{1.55cm}}_0$}
\put(112.5,28){$\underbrace{\hspace*{1.55cm}}_0$}
\put(160,28){$\underbrace{\hspace*{1.2cm}}_1$}
\put(198,28){$\underbrace{\hspace*{1.55cm}}_0$}
\put(246,28){$\underbrace{\hspace*{1.2cm}}_1$}
\end{picture}
\end{center}
\caption{Block structure of quasisturmian words.}\label{f:quasisturmian}
\end{figure}

It can be shown that the word $u_{\varepsilon,\eta}(\Omega_1)$ can be obtained from the
upper mechanical sequence $\overline{s}_{\frac1\tau,-\frac1{\tau^2}}$ by substituting for
0 the word $w_0=BACAACA$ and for 1 the word $w_1=BABACA$. Similarly, the infinite word
$u_{\varepsilon,\eta}(\Omega_2)$ can be obtained from the upper mechanical sequence
$\overline{s}_{\frac1\tau,-\frac1{\tau}}$ by substituting for 0 the word $w_0=BACABA$ and
for 1 the word $w_1=CAACA$.

\end{ex}

\subsection{Properties of the language}

Let us study the language of the infinite word $u_{\varepsilon,\eta}(\Omega)$.
Since the stepping function $f_\Omega$ corresponding to the C\&P sequence
$\Sigma_{\varepsilon,\eta}(\Omega)$ and the stepping function $f_{\Omega+z}$
corresponding to the C\&P sequence $\Sigma_{\varepsilon,\eta}(\Omega+z)$ satisfy
the relation
$$
f_{\Omega+z}(x)=z+f_{\Omega}(x-z)\,,
$$
the language $\cal L$ and the complexity $\C$ of a C\&P sequence depend only on the length
of the acceptance interval $\Omega$ and not on its position.

First we determine the density of a given factor. Recall that the density of a particular
factor $w$ in the infinite bidirectional word $u=\cdots u_{-2}u_{-1}u_0u_1u_2\cdots$ is
defined by
\begin{equation}\label{density}
 \varrho_w := \lim_{k\to\infty}
 \frac{\# \bigl\{i\in\Z\cap[-k,k] \,\bigm|\, u_iu_{i+1}\ldots u_{i+n-1}=w \bigr\}}{2k+1}\,,
\end{equation}
if the limit exists. Elements $x\in\Sigma_{\varepsilon,\eta}(\Omega)$ such that the word
corresponding to the $k$-tuple of the right neighbours of $x$ is $w=w_0w_1\cdots
w_{k-1}\in{\cal L}_k$ satisfy $x^\star\in\Omega_w$. Therefore the occurrences of the
factor $w$ in the infinite word $u_{\varepsilon,\eta}(\Omega)$ are given by the set
$$
\{x\in\Z[\eta] \mid x^\star\in\Omega_w\}\,.
$$
Since $\Omega_w$ is a semi-closed interval
$\Omega_w\subset\Omega$, it is a C\&P set, and its density is
proportional to the length of the acceptance window (see fact 4 of
Remark~\ref{poznamka}). This implies that the density of a factor
$w\in {\cal L}$ in $u_{\varepsilon,\eta}(\Omega)$ is given by
$$
\varrho_w = \frac{|\Omega_w|}{|\Omega|}\,.
$$

Another important property of the language of the infinite word
$u_{\varepsilon,\eta}(\Omega)$ is given by the following
proposition.

\begin{prop}\label{p:reverse}
The language ${\cal L}$ of the infinite word $u_{\varepsilon,\eta}[c,c+\ell)$ is stable
under mirror image, i.e.
$$
w=w_0w_1\cdots w_{n-1}\in{\cal L} \quad\implies\quad
\overline{w}=w_{n-1}w_{n-2}\cdots w_{0}\in{\cal L}\,.
$$
Moreover, the densities of the factors $w$ and $\overline{w}$ coincide,
$\varrho_w=\varrho_{\overline{w}}$.
\end{prop}

\pf
Since the language of the infinite word $u_{\varepsilon,\eta}[c,c+\ell)$ depends only on the length
$\ell$ of the acceptance interval and not on its position, it suffices to show the statement for
the infinite word $u_{\varepsilon,\eta}[-\frac{\ell}{2},\frac{\ell}{2})$, which codes the
C\&P sequence $\Sigma_{\varepsilon,\eta}[-\frac{\ell}{2},\frac{\ell}{2})$.
If $-\frac{\ell}{2}\notin\Z[\varepsilon]$, then
$$
\Sigma_{\varepsilon,\eta}[-\tfrac{\ell}{2},\tfrac{\ell}{2})=
\Sigma_{\varepsilon,\eta}(-\tfrac{\ell}{2},\tfrac{\ell}{2})
$$
and thus it is a centrally symmetric set and the statement of the proposition is obvious.
If $-\frac{\ell}{2}=a+b\varepsilon\in\Z[\varepsilon]$, then for the proof it suffices to
realize that the central symmetry of the set
$\Sigma_{\varepsilon,\eta}[-\frac{\ell}{2},\frac{\ell}{2})$ is broken by a unique point,
namely $a+b\eta$. Since every factor $w\in{\cal L}$ occurs in
$u_{\varepsilon,\eta}[-\frac{\ell}{2},\frac{\ell}{2})$ infinitely many times, we can
still use the same argument to justify the proposition. \pfk

For sturmian words the above property is well known, its proof can
be found in~\cite{lothaire}.

For different lengths
$\ell_1,\ell_2\in\bigl(\max(-\varepsilon,1+\varepsilon),1\bigr]$
the languages of the infinite words
$u_{\varepsilon,\eta}[c,c+\ell_1)$,
$u_{\varepsilon,\eta}[c,c+\ell_2)$ are different. However, if we
are interested only in factors of a given length $n$, the sets
${\cal L}_n$ can coincide even for infinite words corresponding to
acceptance intervals of different lengths. Let
$u_{\varepsilon,\eta}(\Omega)$ be an infinite word with the length
of the acceptance window $|\Omega|=\ell$. We denote ${\cal
L}_n(\ell)$ its set of factors of length $n$. For example
${\mathcal L}_1(\ell)$ is equal to the alphabet $\{A,B,C\}$ for
every length $\max(-\varepsilon,1+\varepsilon)<\ell<1$. Let us now
see how much we can change the length $\ell$ of the acceptance
interval $\Omega$ without changing the set ${\mathcal L}_n(\ell)$.

\begin{prop}\label{pripady}
Let $n\in\N$ be fixed. Denote by ${\mathcal C}_\ell$ the complexity function of the
infinite word $u_{\varepsilon,\eta}(\Omega)$ with $\Omega=[c,c+\ell)$.
Define
 $$
 {\mathcal D}_n=\bigl\{\ \ell\ \bigm|\ \max(-\varepsilon,1+\varepsilon)<\ell\leq1,\ {\mathcal
 C}_\ell(n)<2n+1\bigr\}\,.
 $$
Then elements of ${\mathcal D}_n$ divide the interval
$(\max(-\varepsilon,1+\varepsilon),1]$ into a finite disjoint
union of sub-intervals, such that ${\mathcal L}_n(\ell)$ is
constant on the interior of each of these intervals.
\end{prop}

The proof for special case $\varepsilon=-\frac1\tau$, $\eta=\tau$, can be found
in~\cite{zich1}. The demonstration of the general statement follows analogous ideas.

\begin{ex}
Consider again the parameters $\varepsilon=-\frac1\tau$,
$\eta=\tau$. For the sake of illustration  of the previous proposition, let us
choose $n=4$ and find the division of the interval
$\bigl(\max(-\varepsilon,1+\varepsilon),1\bigr]=(\frac1\tau,1]$
into intervals such that the set ${\mathcal L}_4(\ell)$ is
constant on the interior of these intervals. For that, we need to
find $\ell$ so that ${\mathcal
 C}_\ell(n)<2n+1$ for $n=4$. Using~\eqref{eq:jazyk} this happens if
 $$
 f^{(k)}(\delta_1)=\delta_2\quad\hbox{ or }\quad f^{(k)}(\delta_2)=\delta_1\,,\qquad
 \hbox{ for some }  k=0,1,2,3\,.
 $$
For solving these equations, one has to realize that not only the discontinuity points
$\delta_1=c+\ell-1-\varepsilon$, $\delta_2=c-\varepsilon$ depend on $\ell$, but also the
prescription for the function $f$ depends on it. However, since every iteration of $f$ is
piecewise linear, the above equations can be easily solved. We find that
$$
{\mathcal D}_4=\{4-2\tau,-4+3\tau,1\}\,.
$$
The division of the interval $(\frac1\tau,1]$ by the elements of
${\mathcal D}_4$ is illustrated in Figure~\ref{f:mnozinaD}. The
figure also shows the set of factors of length $4$ for each of the
subintervals and for the singular lengths $\ell\in{\mathcal D}_4$.
Note that the set of factors corresponding to the interior of a
subinterval is a union of sets of factors corresponding to the
boundary points of the subinterval, for example
$$
{\mathcal L}_4(\ell) = {\mathcal L}_4(4-2\tau) \cup {\mathcal L}_4(-4+3\tau)\,,
\qquad\hbox{ for all }\ \ell\in(4-2\tau, -4+3\tau)\,.
$$

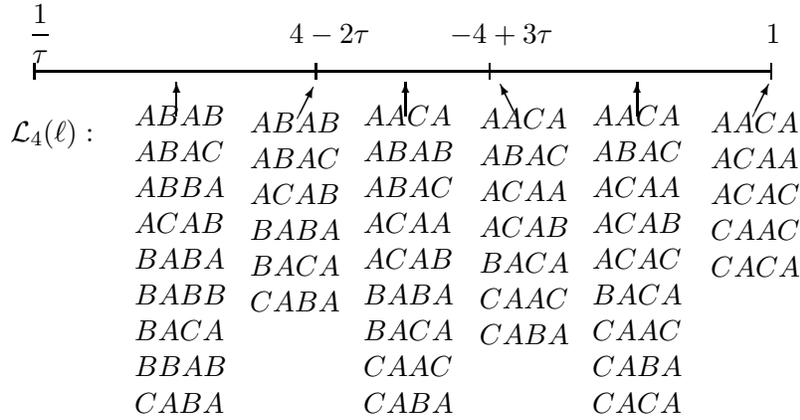
\begin{figure}[ht]
\begin{center}
{\setlength{\unitlength}{0.35mm}
\begin{picture}(310,150)
\put(1,93.5){${\mathcal L}_4(\ell):$}
\put(43,45){$\begin{array}{l}
ABAB\\
ABAC\\
ABBA\\
ACAB\\
BABA\\
BABB\\
BACA\\
BBAB\\
CABA
\end{array}$}
\put(87,62.8){$\begin{array}{l}
ABAB\\
ABAC\\
ACAB\\
BABA\\
BACA\\
CABA
\end{array}$}
\put(130,45){$\begin{array}{l}
AACA\\
ABAB\\
ABAC\\
ACAA\\
ACAB\\
BABA\\
BACA\\
CAAC\\
CABA
\end{array}$}
\put(174,57.2){$\begin{array}{l}
AACA\\
ABAC\\
ACAA\\
ACAB\\
BACA\\
CAAC\\
CABA
\end{array}$}
\put(217,45){$\begin{array}{l}
AACA\\
ABAC\\
ACAA\\
ACAB\\
ACAC\\
BACA\\
CAAC\\
CABA\\
CACA
\end{array}$}
\put(262,69.5){$\begin{array}{l}
AACA\\
ACAA\\
ACAC\\
CAAC\\
CACA
\end{array}$}
\put(10,120){\line(1,0){280}} \put(10,117){\line(0,1){6}} \put(117,117){\line(0,1){6}}
\put(183,117){\line(0,1){6}} \put(290,117){\line(0,1){6}}
\put(8,131){$\displaystyle{\frac1\tau}$} \put(107,131){$4-2\tau$}
\put(168,131){$-4+3\tau$} \put(288,131){$1$}
\put(64,103){\vector(0,1){13}}
\put(151,103){\vector(0,1){13}}
\put(239,103){\vector(0,1){13}}
\put(110,102){\vector(1,2){6}}
\put(193,103){\vector(-1,2){6}}
\put(283,103){\vector(1,2){6}}
\end{picture}
}
\end{center}
\caption{The appearance of factors of length $4$ in
$\Sigma_{-\frac{1}{\tau},\tau}(-\Omega)$ in function of the length of the acceptance
window.} \label{f:mnozinaD}
\end{figure}
\end{ex}

\subsection{Special factors}

Let us introduce some important notions which help us understand
the structure of factors in the language ${\cal L}$. The notions
have been introduced in~\cite{cassaigne}. Consider arbitrary
bidirectional infinite word $v$ in an alphabet $\A$,
$$
v= \cdots v_{-2}v_{-1}v_0v_1v_2\cdots
$$
For every factor $w\in{\cal L}$ of $v$ there exists at least one letter
$a\in\A$ such that $aw\in{\cal L}$. Such letter $a$ is called
a left extension of the factor $w$. The set of left extensions of the factor $w$ is denoted by
${\rm Lext}(w)\subset\A$.

\begin{pozn}\label{pozn:576}
If $\tilde{w}$ is a prefix of the factor $w$, then
$$
{\rm Lext}(\tilde{w})\supseteq {\rm Lext}(w)\,.
$$
\end{pozn}

Since for every factor $\tilde{w}\in{\cal L}_{n+1}$ we have $\tilde{w}=aw$ for some
$w\in{\cal L}_n$ and a letter $a\in{\rm Lext}(w)$, the increment of the complexity function
can be computed as
\begin{equation}\label{eq:314}
\Delta{\cal C}(n) = {\cal C}(n+1)-{\cal C}(n) = \#{\cal L}_{n+1}-\#{\cal L}_n =
\sum_{w\in{\cal L}_n} \bigl(\#{\rm Lext}(w)-1 \bigr)\,.
\end{equation}
Similarly one can define the notion of right extension of a factor and obtain analogical relation
\begin{equation}\label{eq:315}
\Delta{\cal C}(n) = \sum_{w\in{\cal L}_n} \bigl(\#{\rm Rext}(w)-1 \bigr)\,.
\end{equation}
Obviously, for determining the increment of complexity, only such factors $w$ are
interesting that have $\#{\rm Lext}(w)\geq 2$ or $\#{\rm Rext}(w)\geq 2$. Such factors
are called left (resp.\ right) special factor.

Let us study these notions for infinite words corresponding to C\&P sequences.
Proposition~\ref{p:reverse} implies
$$
\begin{gathered}
w \hbox{ is a left special factor of } u_{\varepsilon,\eta}(\Omega)\\
 \Updownarrow \\
\overline{w} \hbox{ is a right special factor of } u_{\varepsilon,\eta}(\Omega)\,.
\end{gathered}
$$
Therefore we can limit our considerations to the study of left special factors.
Theorem~\ref{thmcompl} implies that
$$
1\leq\Delta{\cal C}(n)\leq 2\,.
$$
Thus for every $n\in\N$ there exists at least one and at most two left special factors of length
$n$. Let us explain how one can decide whether a given factor $w\in{\cal L}_n$ is a left special
factor or not. Recall from the beginning of Section~\ref{subs:compl} that every factor
$w\in{\cal L}_n$ is linked with an interval
$\Omega_w\subset\Omega$ of the form $\Omega_w=[a,b)$, where
$$
a,b\in\Bigl\{c,\delta_1,f^{-1}(\delta_1),\dots,f^{-k+1}(\delta_1),
\delta_2,f^{-1}(\delta_2),\dots,f^{-k+1}(\delta_2)\Bigr\}\,,
$$
and $\delta_1,\delta_2$ are the discontinuity points of the stepping function $f$.

For every $x\in\Sigma_{\varepsilon,\eta}(\Omega)$ such that $x^\star\in\Omega_w$ the
$n$-tuple of distances in the right neighbourhood of the point $x$ corresponds to the
word $w$. The nearest left neighbour of the point $x$ is determined by $f^{-1}(x^\star)$.
In order that the word $w$ is a left special factor, the interior $\Omega_w^\circ$ must
contain at least one discontinuity point of $f^{-1}$. If the discontinuity point of
$f^{-1}$ lies only on the boundary of $\Omega_w$, then $w$ has only one left extension
and thus is not a left special factor. The discontinuity points of the function $f^{-1}$
are
$$
c+\ell+\varepsilon = f(\delta_1)\qquad\hbox{ and }\qquad
c+1+\varepsilon=f(c)=f^2(\delta_2)\,.
$$
Properties~\eqref{eq:xy} of the stepping function imply that if $\ell\notin\Z[\varepsilon]$,
then the discontinuity points of $f^{-1}$ do not belong to the set
$$
\Bigl\{c,\delta_1,f^{-1}(\delta_1),\dots,f^{-n+1}(\delta_1),
\delta_2,f^{-1}(\delta_2),\dots,f^{-n+1}(\delta_2)\Bigr\}
$$
for any $n\in\N$, and, therefore, if a discontinuity point of
$f^{-1}$ lies in $\Omega_w$, then it lies in its interior. We can
therefore conclude with the following proposition, which is proved
in a different way in~\cite{3ExchII}.

\begin{prop}\label{factor}
Let $\ell\notin\Z[\varepsilon]$. Consider the one-directional infinite word
$u^{(i)}=u^{(i)}_0u^{(i)}_1u^{(i)}_2u^{(i)}_3\cdots$, $i=1,2$, coding the orbits
$\{f^{n}(c+\ell+\varepsilon)\mid n\in\N_0\}$ and
$\{f^{n}(c+1+\varepsilon)\mid n\in\N_0\}$.
Then a finite word $w$ is a left special factor of $u_{\varepsilon,\eta}(\Omega)$
if and only if it is a prefix of $u^{(1)}$ or $u^{(2)}$.
\end{prop}

\begin{pozn}\label{pozn:638}
Since $c+\ell+\varepsilon$ is the image of $\delta_1$, which is on the boundary between
intervals $\Omega_B,\Omega_C$, then every prefix of the infinite word $u^{(1)}$ has
$\{B,C\}$ as its left extension. Similarly, every prefix of $u^{(2)}$ has in its left
extension letters $A,B$. It can happen that a word $w$ is a prefix of both $u^{(1)}$ and
$u^{(2)}$. Then ${\rm Lext}(w)=\{A,B,C\}$. However, since the infinite words $u^{(1)}$,
$u^{(2)}$ are different, starting from a certain length of the factor $w$ we have $\#{\rm
Lext}(w)=2$.
\end{pozn}

\subsection{Rauzy graphs}

Another important tool for the study of combinatorial properties
of infinite words are the so-called Rauzy graphs,
\cite{rauzygraf,AR}.

\begin{de}
Let $u$ be an infinite word in the alphabet $\A$ and let ${\cal L}_n$ be the set of its factors of
length $n$, $n\in\N$. Rauzy graph $\Gamma_n$ is a directed graph whose set of vertices
is ${\cal L}_n$ and set of directed edges is ${\cal L}_{n+1}$. The edge $e\in{\cal L}_{n+1}$
starts at a vertex $x\in{\cal L}_{n}$ and ends at a vertex $y\in{\cal L}_{n}$, if $x$ is a prefix
of $e$ and $y$ is its suffix, i.e.
\begin{center}
\begin{picture}(220,30)
\put(50,20){\circle*{5}}
\put(210,20){\circle*{5}}
\put(60,20){\vector(1,0){140}}
\put(10,8){$x=w_0w_1\cdots w_{n-1}$}
\put(170,8){$y=w_1\cdots w_{n-1}w_n$}
\put(82,25){$e=w_0w_1\cdots w_{n-1}w_n$}
\end{picture}
\end{center}
The number of edges starting at a vertex $x$ is called the outdegree of $x$ and denoted by $d^+(x)$,
the number of edges ending at $x$ is called the indegree of $x$ and denoted by $d^-(x)$.
\end{de}

From the definition of a Rauzy graph, we have
\begin{equation}\label{eq:413}
d^+(w)=\#{\rm Rext}(w)\qquad\hbox{ and }\qquad d^-(w)=\#{\rm Lext}(w)
\end{equation}

Remark~\ref{pozn:576} implies that if $\Gamma_{n+1}$ contains a vertex with outdegree $K$, then
the graph $\Gamma_n$ contains a vertex with outdegree $\geq K$. Similar statement holds also for
indegrees. Therefore
\begin{equation}\label{eq:577}
\max_{w\in{\cal L}_{n}} d^-(w) \ \geq\ \max_{w\in{\cal L}_{n+1}} d^-(w) \qquad\hbox{ and
}\qquad \max_{w\in{\cal L}_{n}} d^+(w) \ \geq\ \max_{w\in{\cal L}_{n+1}} d^+(w)\,.
\end{equation}

\begin{ex}\label{ex:2}
Let us consider the lower mechanical word
$$
u_n=\left\lfloor\frac{n+1}\tau\right\rfloor -
\left\lfloor\frac{n}\tau\right\rfloor\,,\qquad n\in\Z,\qquad\hbox{where }\tau=\frac{1+\sqrt5}2\,.
$$
Using Example~\ref{ex}, this infinite word in the alphabet $\{0,1\}$ is a coding of the
C\&P sequence $\Sigma_{-\frac1\tau,\eta}(\beta-1,\beta]$. According to
Remark~\ref{pozn:mechkom}, it is a sturmian word, i.e. of complexity ${\cal C}(n)=n+1$.
It can be easily computed that
$$
\begin{array}{rcl}
{\cal L}_3 &=& \{010,011,101,110\}\,,\\
{\cal L}_4 &=& \{0101,0110,1010,1011,1101\}\,,\\
{\cal L}_5 &=& \{01011,01101,10101,10110,11010,11011\}\,.\\
\end{array}
$$
The Rauzy graphs $\Gamma_3$, $\Gamma_4$ are illustrated on Figure~\ref{f:grafysturm}.

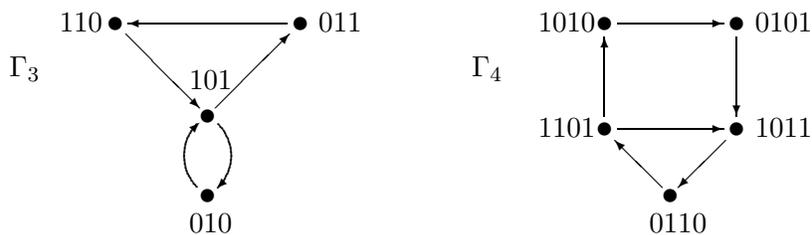
\begin{figure}[ht]
\begin{center}
\begin{picture}(125,90)
\put(0,60){$\Gamma_3$}
\put(75,15){\circle*{5}}
\put(75,45){\circle*{5}}
\put(40,80){\circle*{5}}
\put(110,80){\circle*{5}}
\put(68,1){010}
\put(68,55){101}
\put(19,77){110}
\put(117,77){011}
\put(105,80){\vector(-1,0){60}}
\put(44,76){\vector(1,-1){28}}
\put(78,48){\vector(1,1){28}}
\put(83,30){\arc(-12,12){90}}
\put(67,30){\arc(12,-12){90}}
\put(66,36.75){\vector(1,1){6}}
\put(84.5,23.5){\vector(-1,-1){6}}
\end{picture}
\hspace*{1.5cm}
\begin{picture}(125,90)
\put(0,60){$\Gamma_4$}
\put(75,15){\circle*{5}}
\put(50,40){\circle*{5}}
\put(50,80){\circle*{5}}
\put(100,40){\circle*{5}}
\put(100,80){\circle*{5}}
\put(67,1){0110}
\put(25,37){1101}
\put(25,77){1010}
\put(107,37){1011}
\put(107,77){0101}
\put(72,18){\vector(-1,1){18}}
\put(96,36){\vector(-1,-1){18}}
\put(100,75){\vector(0,-1){30}}
\put(50,45){\vector(0,1){30}}
\put(55,80){\vector(1,0){40}}
\put(55,40){\vector(1,0){40}}
\end{picture}
\caption{Rauzy graphs for the Fibonacci word.}
\label{f:grafysturm}
\end{center}
\end{figure}
\end{ex}

Let us list some of the properties of the Rauzy graph $\Gamma_n$ of the infinite word
$u_{\varepsilon,\eta}[c,c+\ell)$ corresponding to a C\&P sequence.

\begin{enumerate}
\item The graph $\Gamma_n$ is strongly connected for every $n\in\N$. It means that for every
pair of vertices $x,y$ of the graph, there exists a directed path starting at $x$ ending at $y$.
This is a consequence of the repetitivity of the infinite word
$u_{\varepsilon,\eta}[c,c+\ell)$, see (9) of Remark~\ref{poznamka}.

\item For the length of the acceptance window $\ell=1$, the infinite word
$u_{\varepsilon,\eta}[c,c+1)$ is sturmian and thus $\Delta{\cal C}(n)=1$ for every
$n\in\N$. Using~\eqref{eq:314},~\eqref{eq:315} and~\eqref{eq:413} for every $n$ the graph
$\Gamma_n$ contains exactly one vertex $x\in{\cal L}_n$ with outdegree 2 and exactly one
vertex $y\in{\cal L}_n$ with indegree 2. These vertices may or may not coincide, as we
have seen in Example~\ref{ex:2}.

For $\ell\in\Z[\varepsilon]$ the Theorem~\ref{thmcompl} implies $\Delta{\cal C}(n)=1$ for
sufficiently large $n$. Therefore the graphs have the same indegrees and outdegrees as in
the sturmian case.

\item If $\ell\notin\Z[\varepsilon]$, then using Theorem~\ref{thmcompl} we have $\Delta{\cal C}(n)=2$
for all $n\in\N$. Since the language of the infinite word $u_{\varepsilon,\eta}[c,c+\ell)$
is stable under mirror image (Proposition~\ref{p:reverse}), relations~\eqref{eq:314}
and~\eqref{eq:315} imply that in the graph $\Gamma_n$ there is either one vertex with indegree $3$ and
one with outdegree $3$, or there are two vertices with indegree $2$ and two with outdegree $2$.
Remark~\ref{pozn:638} states that a vertex with out or indegree 3 can occur only in a graph $\Gamma_n$
for small $n$.

\item Let us denote by $\bar{\Gamma}_n$ the graph created from $\Gamma_n$ by
the change of the orientation of the edges. Then $\bar{\Gamma}_n$ and $\Gamma_n$ are
isomorphic graphs, i.e. there exists a bijection $\pi$ between the vertices of
$\bar{\Gamma}_n$ and $\Gamma_n$, such that for every two vertices $x,y$ of $\bar{\Gamma}_n$
there is a directed edge from $x$ to $y$ if and only there is a directed edge in the graph
$\Gamma_n$ from $\pi(x)$ to $\pi(y)$. This property follows from Proposition~\ref{p:reverse}.
\end{enumerate}

The last mentioned property can be stated in an even stronger version, if we consider the
densities of factors in ${\cal L}_{n+1}$ as labels of the edges in the graph $\Gamma_n$.

\begin{de}
If the densities of all factors of the infinite word $u$ are well defined, every edge $e$
in the Rauzy graph $\Gamma_n$ can be assigned a non-negative number, namely the density
$\varrho_e$ of the factor $e$. The resulting graph is called a weighted Rauzy graph.
\end{de}

For every vertex $x$ of the weighted Rauzy graph $\Gamma_n$ we have obviously a `conservation law',
\begin{equation}\label{eq:789}
\sum_{\hbox{\scriptsize edge $e$ ending in }x }\varrho_e
\quad=\quad \sum_{\hbox{\scriptsize edge $f$ starting in }x }\varrho_f\,.
\end{equation}

With the mentioned properties we can prove that the factors in a C\&P word take at most 5 values.

\begin{prop}\label{rauzy}
Let ${\cal L}_n$ be the set of factors of length $n$ of the infinite word
$u_{\varepsilon,\eta}[c,c+\ell)$, $\ell\notin\Z[\varepsilon]$. The densities of factors
in ${\cal L}_n$ take at most $5$ values, i.e.
$$
\#\{\varrho_w \mid w\in {\cal L}_{n}\}\leq 5\,.
$$
\end{prop}

\begin{figure}[ht]
\begin{center}
\begin{picture}(300,25)
\put(10,10){\circle*{5}}
\put(15,10){\vector(1,0){30}}
\put(50,10){\circle*{5}}
\put(55,10){\vector(1,0){30}}
\put(90,10){\circle*{5}}
\put(7,0){$y$}
\put(47,0){$x$}
\put(87,0){$z$}
\put(26,15){$\varrho_e$}
\put(66,15){$\varrho_f$}
\put(125,8){is reduced to}
\put(210,10){\circle*{5}}
\put(215,10){\vector(1,0){30}}
\put(250,10){\circle*{5}}
\put(207,0){$y$}
\put(247,0){$z$}
\put(226,15){$\varrho_e$}
\end{picture}
\caption{Reduction of the weighted Rauzy graph.}
\label{f:edges}
\end{center}
\end{figure}
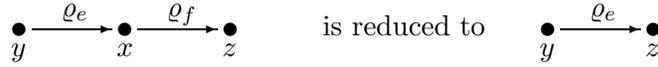

\pf Consider the weighted Rauzy graph $\Gamma_n$ of $u_{\varepsilon,\eta}[c,c+\ell)$. If
for every vertex $x$ of $\Gamma_n$ we have $d^+(x)=d^-(x)=1$, then the
relation~\eqref{eq:789} implies that the density of the edge $e$ ending at $x$ and of the
edge $f$ starting at $x$ coincide. Since the graph is strongly connected, these edges are
different, $e\neq f$. Denote by $y$ the starting vertex of the edge $e$ and by $z$ the
ending vertex of the edge $f$. From the graph $\Gamma_n$ we remove the vertex $x$ and
edges $e,f$ and replace it by a new edge starting at $y$ and ending at $z$. We assign the
new edge with the weight $\varrho_e=\varrho_f$. This reduction of the graph is
illustrated of Figure~\ref{f:edges}.

The reduction of the Rauzy graph $\Gamma_n$ is repeated until there are no vertices with
both outdegree and indegree 1. The resulting graph is called the reduced weighted Rauzy graph
$R\Gamma_n$. The construction implies that also $R\Gamma_n$ is a strongly connected graph,
the weights of its edges satisfy the conservation law and the set of weights of the graph
$R\Gamma_n$ is the same as the set of weight of the graph $\Gamma_n$. Moreover, the graph
$\overline{R\Gamma}_n$ created by reversing the direction of edges in $R\Gamma_n$ is isomorphic
to $R\Gamma_n$.

Using the property 3 of the Rauzy graph $\Gamma_n$ for sufficiently large $n$ there are two
vertices with outdegree 2, the outdegree of the remaining vertices is 1. Similarly, there are
two vertices with indegree 2 and the indegree of other vertices is 1. It may happen that a vertex
with outdegree 2 coincides with a vertex with indegree 2. This implies that the reduced Rauzy
graph has 2, 3 or 4 vertices.

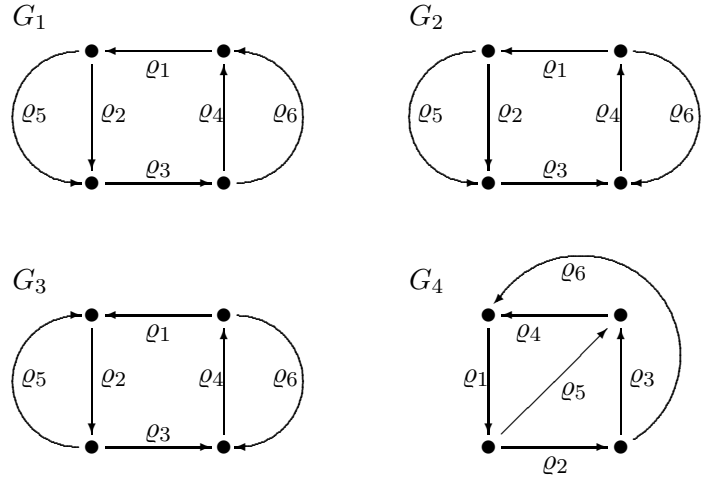
\begin{figure}[ht]
\begin{center}
\begin{picture}(250,170)
\put(0,160){$G_1$}
\put(150,160){$G_2$}
\put(0,60){$G_3$}
\put(150,60){$G_4$}
\put(30,100){\circle*{5}}
\put(35,100){\vector(1,0){40}}
\put(80,100){\circle*{5}}
\put(75,150){\vector(-1,0){40}}
\put(30,150){\circle*{5}}
\put(30,145){\vector(0,-1){40}}
\put(80,150){\circle*{5}}
\put(80,105){\vector(0,1){40}}
\put(25,125){\arc(0,25){180}}
\put(23,100){\vector(1,0){3}}
\put(85,125){\arc(0,-25){180}}
\put(87,150){\vector(-1,0){3}}
\put(33,125){$\varrho_2$}
\put(50,104){$\varrho_3$}
\put(70,125){$\varrho_4$}
\put(50,142){$\varrho_1$}
\put(3,125){$\varrho_5$}
\put(98,125){$\varrho_6$}
\put(180,100){\circle*{5}}
\put(185,100){\vector(1,0){40}}
\put(230,100){\circle*{5}}
\put(225,150){\vector(-1,0){40}}
\put(180,150){\circle*{5}}
\put(180,145){\vector(0,-1){40}}
\put(230,150){\circle*{5}}
\put(230,105){\vector(0,1){40}}
\put(175,125){\arc(0,25){180}}
\put(173,100){\vector(1,0){3}}
\put(235,125){\arc(0,-25){180}}
\put(237,100){\vector(-1,0){3}}
\put(183,125){$\varrho_2$}
\put(200,104){$\varrho_3$}
\put(220,125){$\varrho_4$}
\put(200,142){$\varrho_1$}
\put(153,125){$\varrho_5$}
\put(248,125){$\varrho_6$}
\put(30,0){\circle*{5}}
\put(35,0){\vector(1,0){40}}
\put(80,0){\circle*{5}}
\put(75,50){\vector(-1,0){40}}
\put(30,50){\circle*{5}}
\put(30,45){\vector(0,-1){40}}
\put(80,50){\circle*{5}}
\put(80,5){\vector(0,1){40}}
\put(25,25){\arc(0,25){180}}
\put(23,50){\vector(1,0){3}}
\put(85,25){\arc(0,-25){180}}
\put(87,0){\vector(-1,0){3}}
\put(33,25){$\varrho_2$}
\put(50,4){$\varrho_3$}
\put(70,25){$\varrho_4$}
\put(50,42){$\varrho_1$}
\put(3,25){$\varrho_5$}
\put(98,25){$\varrho_6$}
\put(180,0){\circle*{5}}
\put(185,0){\vector(1,0){40}}
\put(230,0){\circle*{5}}
\put(225,50){\vector(-1,0){40}}
\put(180,50){\circle*{5}}
\put(180,45){\vector(0,-1){40}}
\put(230,50){\circle*{5}}
\put(230,5){\vector(0,1){40}}
\put(185,5){\vector(1,1){40}}
\put(215,35){\arc(20,-32){205}} 
\put(188,60.5){\vector(-1,-1){6}} 
\put(170,25){$\varrho_1$} 
\put(200,-8){$\varrho_2$} 
\put(233,25){$\varrho_3$} 
\put(190,42){$\varrho_4$} 
\put(207,20){$\varrho_5$} 
\put(207,65){$\varrho_6$} 
\end{picture}
\caption{Possible reduces weighted Rauzy graphs.}
\label{f:grafy}
\end{center}
\end{figure}

Let us discuss the case that $R\Gamma_n$ has 4 vertices, i.e. the case when none of the vertices
has in the same time indegree and outdegree $2$. It can be easily derived that the reduced
weighted Rauzy graph has one of the forms illustrated on Figure~\ref{f:grafy}.
Since all the possible reduced graphs have six edges, the original weighted Rauzy graph has
at most six different densities. We can eliminate the sixth value in graphs $G_1$, $G_2$ and $G_3$
using the conservation law. In the graph $G_1$ we have $\varrho_1=\varrho_4+\varrho_6=\varrho_3$.
Similarly in the graph $G_2$ we have $\varrho_1=\varrho_2+\varrho_5=\varrho_3$.
In the graph $G_3$ we have $\varrho_1=\varrho_4-\varrho_6=\varrho_3$.

\begin{figure}[ht]
\begin{center}
\begin{picture}(260,88)
\put(5,55){$G_4$}
\put(40,10){\circle*{5}}
\put(45,10){\vector(1,0){40}}
\put(90,10){\circle*{5}}
\put(85,60){\vector(-1,0){40}}
\put(40,60){\circle*{5}}
\put(40,55){\vector(0,-1){40}}
\put(90,60){\circle*{5}}
\put(90,15){\vector(0,1){40}}
\put(45,15){\vector(1,1){40}}
\put(75,45){\arc(20,-32){205}} 
\put(48,70.5){\vector(-1,-1){6}} 
\put(30,35){$\varrho_1$} 
\put(60,2){$\varrho_2$} 
\put(93,35){$\varrho_3$} 
\put(50,52){$\varrho_4$} 
\put(67,30){$\varrho_5$} 
\put(67,75){$\varrho_6$} 
\put(31,2){$x$}
\put(93,2){$y$}
\put(93,63){$z$}
\put(32,63){$v$}
\put(155,55){$\bar{G}_4$}
\put(190,10){\circle*{5}}
\put(195,60){\vector(1,0){40}}
\put(240,10){\circle*{5}}
\put(235,10){\vector(-1,0){40}}
\put(190,60){\circle*{5}}
\put(240,55){\vector(0,-1){40}}
\put(240,60){\circle*{5}}
\put(190,15){\vector(0,1){40}}
\put(235,55){\vector(-1,-1){40}}
\put(225,45){\arc(20,-32){205}} 
\put(250,18.1){\vector(-1,-1){6}} 
\put(180,35){$\varrho_1$} 
\put(210,2){$\varrho_2$} 
\put(243,35){$\varrho_3$} 
\put(200,52){$\varrho_4$} 
\put(217,30){$\varrho_5$} 
\put(217,75){$\varrho_6$} 
\put(181,2){$x$}
\put(243,2){$y$}
\put(243,63){$z$}
\put(182,63){$v$}
\end{picture}
\caption{Isomorphic reduced graphs $G_4$ and $\bar{G}_4$.}
\label{f:grafy2}
\end{center}
\end{figure}
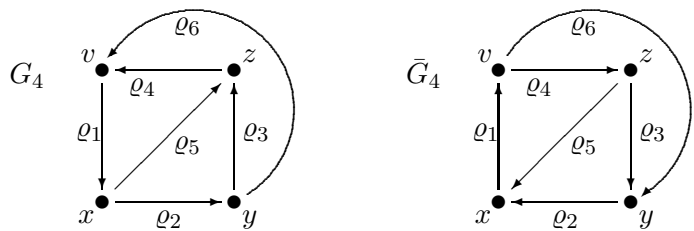

The conservation law is not sufficient for reducing the number of
densities in the graph $G_4$. Here we use the property 4 of the
Rauzy graph of a C\&P, namely that by changing the direction of
the edges in $G_4$ we obtain an isomorphic graph $\bar{G}_4$. The
graphs are illustrated on Figure~\ref{f:grafy2}. The only
permutation $\pi$ of the vertices which realizes the isomorphism
of the graphs $G_4$ and $\bar{G}_4$ is the permutation $\pi(x)=v$,
$\pi(y)=z$, $\pi(z)=y$, $\pi(v)=x$. The isomorphism preserves the
densities, thus $\varrho_2=\varrho_4$, $\varrho_5=\varrho_6$.

We have thus solved the case that the reduced weighted graph $R\Gamma_n$ has 4 vertices.
If $R\Gamma_n$ has 2 or 3 vertices, then such a graph has at most 5
edges. Therefore there are at most 5 values of densities.

For small values of $n$ it can happen that the graph $\Gamma_n$ has one vertex with outdegree 3
and one vertex with indegree 3, the other vertices having both outdegree and indegree 1.
In this case the reduced Rauzy graph $R\Gamma_n$ has 1 or 2 vertices and at most 4 edges, thus
the number of different values of densities is less or equal to 4.
\pfk

\begin{pozn}
In case that $\ell\in\Z[\varepsilon]$, then the densities of
factors of length $n$ of the infinite word
$u_{\varepsilon,\eta}[c,c+\ell)$ take at most $3$ values for
sufficiently large $n$ , because the resulting word is either
sturmian or quasisturmian and the number of densities can be read
from the corresponding reduced Rauzy graph, which has always at
most three edges. Let us mention that the fact that the densities
of factors in sturmian words take at most three values has been
stated in~\cite{berthe2}, in fact, it can be deduced already
from~\cite{sos}.
\end{pozn}

\subsection{Sturmian words}






As we have seen, sturmian words can be defined in several
different equivalent ways, namely as

\begin{itemize}
\item bidirectional infinite words with complexity ${\cal
C}(n)=n+1$and irrational densities of letters,
\item mechanical words $\underline{s}_{\alpha,\beta}$, $\overline{s}_{\alpha,\beta}$ with
irrational slope $\alpha$,
\item codings of cut-and-project sequences $u_{\varepsilon,\eta}(\Omega)$, where $\Omega$
is a semi-closed interval of unit length, and $\varepsilon$, $\eta$ are irrational
numbers satisfying $\varepsilon\in(-1,0)$, $\eta>0$.
\end{itemize}

There exist other equivalent definitions, for example using the
number of palindromes of given length or using the so-called
return words. For a nice overview of these definitions
see~\cite{berstel,lothaire}.

The arithmetical definition of mechanical words allows one to easily derive further
combinatorial properties of sturmian words. In Example~\ref{ex} we have shown that the
upper mechanical word $\overline{s}_{\alpha,\beta}$ corresponds to a cut-and-project
sequence with an acceptance window which is closed from the left and open from the right.
The upper mechanical word $\underline{s}_{\alpha,\beta}$ corresponds to a cut-and-project
sequence with acceptance interval of opposite type. Since
$\Sigma_{\varepsilon,\eta}(-\Omega)=-\Sigma_{\varepsilon,\eta}(\Omega)$ and since the
language of a sturmian sequence is closed under reversal, for the study of the properties
of the language we can limit our considerations to upper mechanical words
$\overline{s}_{\alpha,\beta}$, see~\eqref{eq:hornimech}.

Let us now prove three properties which have been used
in~\cite{abug} for the construction of aperiodic wavelets. Note
that Property~\ref{vl:2} can be found already
in~\cite{cassaigne2}.

\begin{vlast}\label{vl:1}
The number of letters 1 in a factor of length $n$ of the mechanical word
$\overline{s}_{\alpha,\beta}$ is equal to $\lfloor n\alpha\rfloor$ or $\lceil
n\alpha\rceil$.
\end{vlast}

\pf Consider a factor $w$ of length $n$, $w=\overline{s}_{\alpha,\beta}(i)
\overline{s}_{\alpha,\beta}(i+1)\cdots \overline{s}_{\alpha,\beta}(i+n-1)$. Since the
alphabet of the mechanical word is $\{0,1\}$, the number of letters 1 in $w$ is equal to
$$
\begin{aligned}
\sum_{j=0}^{n-1}\overline{s}_{\alpha,\beta}(i+j)& = \lceil(i+n)\alpha+\beta\rceil -
\lceil
i\alpha+\beta\rceil = \\
&=\bigl\lceil n\alpha+ \underbrace{i\alpha +\beta - \lceil
i\alpha+\beta\rceil}_{\in(-1,0)}\bigr\rceil = \left\{
\begin{array}{c}
 \!\lfloor n\alpha\rfloor, \\[1mm]
 \!\lceil n\alpha \rceil.
\end{array}
\right.
\end{aligned}
$$
\pfk

\begin{vlast}\label{vl:2}
All $n+1$ factors of length $n$ of the mechanical word $\overline{s}_{\alpha,\beta}$
appear in the factor $w$ of length $2n$ of the mechanical word
$\overline{s}_{\alpha,-\alpha}$, given by
$$
w=\overline{s}_{\alpha,-\alpha}(-n+1)\overline{s}_{\alpha,-\alpha}(-n+2)\cdots
\overline{s}_{\alpha,-\alpha}(0)\cdots \overline{s}_{\alpha,-\alpha}(n)\,.
$$
\end{vlast}

\pf Example~\ref{ex} says that $\overline{s}_{\alpha,\beta}$ codes the distances in the
cut-and-project sequence $\Sigma_{-\alpha,\eta}[\beta,\beta+1)$ for arbitrary $\eta>0$.
Since the language of a cut-and-project sequence does not change by translation of the
acceptance interval, we can study without loss of generality the language of the
mechanical word $\overline{s}_{\alpha,-\alpha}$, i.e. of the cut-and-project sequence
$\Sigma_{-\alpha,\eta}[-\alpha,1-\alpha)$. The stepping function has a unique
discontinuity point, namely $\delta_1=0$. The same considerations as for determining the
complexity in Section~\ref{subs:compl} lead to the fact that the acceptance window
$\Omega=[-\alpha,1-\alpha)$ is divided by $n$ points $\delta_1$, $f^{-1}(\delta_1)$,
\dots, $f^{-n+1}(\delta_1)$ into $n+1$ disjoint subintervals closed from the left and
open from the right, say $\Omega_{w^{(1)}}$, $\Omega_{w^{(2)}}$, \dots,
$\Omega_{w^{(n+1)}}$, with the following property: if $x,y$ are elements of
$\Sigma_{-\alpha,\eta}[-\alpha,1-\alpha)$, then the $n$-tuples of distances starting from
$x$ and from $y$ coincide if and only if $x^\star$, $y^\star$ belong to the same interval
$\Omega_{w^{(i)}}$ for some $1\leq i \leq n+1$. Since the left boundary points of all the
intervals belong to $\Z[\alpha]$, these boundary points are star map images of points of
$\Sigma_{-\alpha,\eta}[-\alpha,1-\alpha)$. The boundary points of the intervals $w^{(i)}$
are explicitly given by:
$$
-\alpha=f(\delta_1),0=\delta_1, f^{-1}(\delta), \cdots, f^{-n+1}(\delta_1)\,.
$$
Therefore it suffices to consider all $n$-tuples of distances in
$\Sigma_{-\alpha,\eta}[-\alpha,1-\alpha)$, starting at point $0$, at its right neighbour,
and at its $n-1$ left neighbours. From Example~\ref{ex} we know that every element
$x_k\in\Sigma_{-\alpha,\eta}[-\alpha,1-\alpha)$ has the form $x_k=\lceil k\alpha -\alpha
\rceil + k\eta$ for $k\in\Z$. Thus $x_0=0$ and we must study the $n$-tuples of distances
between points $x_{-n+1}$, $x_{-n+2}$, \dots, $x_0$, $x_1$, \dots, $x_{n+1}$. Since
$\overline{s}_{\alpha,-\alpha}(k)$ codes the distance between $x_k$ and $x_{k+1}$, the
proof is finished.
 \pfk

\begin{vlast}\label{vl:3}
The number of factors of length $n$ in the mechanical word $\overline{s}_{\alpha,\beta}$
prefixed by $1$ is equal to $\lceil n\alpha \rceil$.
\end{vlast}

\pf Property~\ref{vl:2} implies that, for the description of the first letter of all
$n+1$ different factors of length $n$, it suffices to focus on letters
$\overline{s}_{\alpha,-\alpha}(-n+1)$, $\overline{s}_{\alpha,-\alpha}(-n+2)$, \dots,
$\overline{s}_{\alpha,-\alpha}(0)$, $\overline{s}_{\alpha,-\alpha}(1)$. The number of
letters 1 among them is
$$
\sum_{k=-n+1}^1 \overline{s}_{\alpha,-\alpha}(k) = \lceil\alpha\rceil-
\lceil-n\alpha\rceil = 1- \lceil-n\alpha\rceil = \lceil n\alpha\rceil\,.
$$
 \pfk

Let us mention an interesting consequence of Property~\ref{vl:1}.
If $w$ and $w'$ are factors of $\overline{s}_{\alpha,\beta}$ of
the same length, then the numbers of letters 1 in $w$ and in $w'$
differ at most by $1$. This obviously implies that also numbers of
letters 0 in $w$ and $w'$ differ at most by $1$. Infinite words
with this property are called balanced. Sturmian words are
balanced. On the other hand, every aperiodic balanced infinite
word is sturmian. We have thus obtained another equivalent
definition of sturmian words. The above implies other properties:
\begin{itemize}
\item
Either 00 or 11 is not a factor of a sturmian word.
\item
If 00 is not a factor of $\overline{s}_{\alpha,\beta}$ or $\underline{s}_{\alpha,\beta}$,
and if $01^x0$ is a factor, then $x=b$ or $x=b+1$, where $b=[\frac{\alpha}{1-\alpha}]$.
\end{itemize}

Other interesting properties of sturmian words concern substitution invariance. This is
the topic of the following section.

\begin{pozn}
Generic cut-and-project sequences with three distances between
adjacent points do not have explicit formula for determining the
$n$-th letter, which exists for sturmian words. Therefore the
study of properties analogous to that mentioned in this subsection
is significantly more difficult~\cite{3ExchII}.
\end{pozn}

\section{Selfsimilarity of C\&P sequences}\label{s:selfs}

We now turn our attention to cut-and-project sequences with self-simi\-larity. We say
that a set $\Lambda\subset\R$ is self-similar if there exists a factor $\gamma>1$ such
that
$$
\gamma\Lambda\subset\Lambda\,.
$$
An infinite bidirectional word corresponding to a self-similar C\&P set may have many
interesting properties, namely under some very general condition it is a fixed point of a
nontrivial morphism, or it is an image of such a fixed point. These properties are
studied in Section~\ref{s:subst}.

In this section we describe the conditions on the parameters $\varepsilon$, $\eta$, and
interval $\Omega$, under which the C\&P set $\Sigma_{\varepsilon,\eta}(\Omega)$ is
self-similar. For that we need to recall some basic number theoretical notions that will
be useful also in studying the invariance of C\&P sets under morphisms. It turns out that
$\varepsilon$ and $\eta$ are different roots of one quadratic equation with integer
coefficients. Therefore, we restrict ourselves to notions connected to quadratic numbers.

For an irrational number $\alpha$ we denote by $\Q(\alpha)$ the minimal number field
containing $\Q$ and $\alpha$. If $\alpha$ is a quadratic number, i.e. an irrational solution of
a quadratic equation with integer coefficients, then
$$
\Q(\alpha)=\{a+b\alpha\mid a,b\in\Q\}\,.
$$
The other root $\alpha'$ of the quadratic equation is the algebraic conjugate of $\alpha$ and
obviously we have $\alpha'\in\Q(\alpha)$. On $\Q(\alpha)$ one defines the mapping
$$
x=a+b\alpha\in\Q(\alpha) \qquad\mapsto\qquad x'=a+b\alpha'\in\alpha\in\Q(\alpha)\,,
$$
which is (the so-called Galois) automorphism on $\Q(\alpha)$. This means that it satisfies
$(x+y)'=x'+y'$ and $(xy)'=x'y'$ for all $x,y\in\Q(\alpha)$.

A root of a monic quadratic polynomial with integer coefficients
is called a quadratic integer. A quadratic integer $\gamma$ is a
quadratic Pisot number, if $\gamma>1$ and its algebraic conjugate
$\gamma'$ satisfies $|\gamma'|<1$. The following result may be
found also in~\cite{balazi}.

\begin{thm}\label{thmselfs}
\
\begin{itemize}
\item[1.] The C\&P sequence $\Sigma_{\varepsilon,\eta}(\Omega)$ is self-similar if and only
if $\varepsilon$ is a quadratic number, $\eta=\varepsilon'$ is its algebraic conjugate,
and the closure $\overline{\Omega}$ of the acceptance $\Omega$ contains the origin. In
that case
$$
\Sigma_{\varepsilon,\eta}(\Omega) = \Sigma_{\varepsilon,\varepsilon'}(\Omega)=
\Sigma_{\eta',\eta}(\Omega)=\{x\in\Z[\eta] \mid x'\in\Omega\}\,.
$$
\item[2.] If $\gamma$ is the self-similarity factor of $\Sigma_{\eta',\eta}(\Omega)$,
then $\gamma$ is a quadratic Pisot number in $\Q[\eta]$.
\end{itemize}
\end{thm}

\pf First let us show that if $\varepsilon$ and $\eta$ are mutually conjugated quadratic
numbers and $0\in\overline{\Omega}$ then $\Sigma_{\varepsilon,\eta}(\Omega)$ is
self-similar. For that we have to find a self-similarity factor $\gamma$, such that
$\gamma\Sigma_{\eta',\eta}(\Omega)\subset\Sigma_{\eta',\eta}(\Omega)$.

Let $\varepsilon,\eta$ be the roots of the equation $Mx^2=Kx+L$ for some integers $K$, $L$, $M$.
We look for $\gamma$ in the form $\gamma=a+Mb\eta$ for some integers $a,b$. Such $\gamma$ satisfies
$\gamma\in\Z[\eta]$ and $\gamma\eta=a\eta+Mb\eta^2=Lb+\eta(a+Kb)\in\Z[\eta]$. Therefore
$\gamma\Z[\eta]\subset\Z[\eta]$.

Since $\gamma\in\Z[\eta]\subset\Q(\eta)$, we determine the image of $\gamma$ under the Galois
automorphism $\gamma'=a+Mb\eta'$. Clearly, $\gamma\gamma'\in\Z$.
Since $\eta'$ is irrational, the set $\Z[\eta']=\Z+\Z\eta'$
is dense in $\R$ and thus there are infinitely many choices of $a,b\in\Z$ so that $\gamma'\in(0,1)$.
Together with the fact $0\in\overline{\Omega}$ it follows that $\gamma'\Omega\subset\Omega$. We use
the above to obtain
$$
\begin{aligned}
\gamma \Sigma_{\eta',\eta}(\Omega) &= \gamma \{x\in\Z[\eta] \mid x'\in\Omega\} =
\{\gamma x\in\gamma\Z[\eta]\mid \gamma'x'\in\gamma'\Omega\} \subset\\
&\subset\{\gamma x\in      \Z[\eta]\mid \gamma'x'\in       \Omega\} \subseteq
\{y\in\Z[\eta] \mid y'\in\Omega\} = \Sigma_{\eta',\eta}(\Omega)\,.
\end{aligned}
$$
Since $\gamma'\in(0,1)$ and $\gamma\gamma'\in\Z$, we have $|\gamma|>1$. If $\gamma>1$, it
is the desired self-similarity factor, in the opposite case we choose  $\gamma^2$ for the
self-similarity factor.

Let us prove the necessary condition for the self-similarity of a C\&P set. Let
$\gamma>1$ satisfy $\gamma\Sigma_{\varepsilon,\eta}(\Omega)\subset
\Sigma_{\varepsilon,\eta}(\Omega)$. For a chosen point
$x=a+b\eta\in\Sigma_{\varepsilon,\eta}(\Omega)$ we have $\gamma
x\in\Sigma_{\varepsilon,\eta}(\Omega) \subset\Z[\eta]$. Therefore there must exist
integers $\tilde{a}$, $\tilde{b}$ such that $\gamma
x=\gamma(a+b\eta)=\tilde{a}+\tilde{b}\eta$. This implies
$$
\gamma= \frac{\tilde{a}+\tilde{b}\eta}{a+b\eta}\qquad\hbox{and}\qquad
\eta= \frac{-\tilde{a}+a\gamma}{\tilde{b}-b\gamma}\,.
$$
Therefore $\Q(\gamma)=\Q(\eta)$.

Let $(x_n)_{n\in\Z}$ be the strictly increasing sequence such that
$\Sigma_{\varepsilon,\eta}(\Omega) =\{x_n\mid n\in\Z\}$. Recall that the distances
between neighbouring points of $\Sigma_{\varepsilon,\eta}(\Omega)$ take values
$x_{n+1}-x_n\in\{\Delta_1, \Delta_2, \Delta_1+\Delta_2\}$, where $\Delta_1$, $\Delta_2$
are positive numbers in $\Z[\eta]$ linearly independent over $\Q$. Take an index $n$ such
that $x_{n+1}-x_n=\Delta_1$. Since $\Sigma_{\varepsilon,\eta}(\Omega)$ is self-similar
with the factor $\gamma$, both $\gamma x_n$ and $\gamma x_{n+1}$ belong to
$\Sigma_{\varepsilon,\eta}(\Omega)$. Therefore the gap between the two points is filled
by distances $\Delta_1$, $\Delta_2$ and $\Delta_1+\Delta_2$. It follows that the distance
$\gamma x_{n+1} - \gamma x_n$ is an integer combination of $\Delta_1$, $\Delta_2$ with
positive coefficients,
$$
\gamma\Delta_1 = \gamma x_{n+1} - \gamma x_n = k_{11}\Delta_1 + k_{12}\Delta_2
$$
for some non-negative integers $k_{11}$, $k_{12}$. Analogously we obtain
$$
\gamma\Delta_2 = k_{21}\Delta_1 + k_{22}\Delta_2\,,\qquad k_{21},k_{22}\in\N_0\,.
$$
We denote by ${\mathbb K}$ the $2\times2$ matrix ${\mathbb K}=(k_{ij})$ and write the above
as
$$
{\mathbb K} \binom{\Delta_1}{\Delta_2} = \gamma  \binom{\Delta_1}{\Delta_2} \,.
$$
This means that $\gamma$ is an eigenvalue of the integer-valued $2\times2$ matrix ${\mathbb K}$
and as such is a root of a monic quadratic polynomial with integer coefficients. Since
$\Q(\eta)=\Q(\gamma)$ and $\eta$ is irrational, $\gamma$ is a quadratic integer.

The eigenvector corresponding to $\gamma$ is $\binom{\Delta_1}{\Delta_2}$.
As $\Delta_1$, $\Delta_2$ belong to the quadratic field
$\Q(\eta)=\Q(\gamma)$, we can apply the Galois automorphism to obtain the other
eigenvector and eigenvalue of the matrix ${\mathbb K}$,
$$
{\mathbb K} \binom{\Delta'_1}{\Delta'_2} = \gamma'  \binom{\Delta'_1}{\Delta'_2} \,.
$$
The Perron-Frobenius theorem for positive integer matrices implies that $|\gamma'|<\gamma$.

We now explain the relation between the Galois automorphism and the star map in the
cut-and-project scheme. Take any pair of points
$x,y\in\Sigma_{\varepsilon,\eta}(\Omega)$, $x<y$. Their distance $y-x$ belongs to
$\Sigma_{\varepsilon,\eta}(\Omega)-\Sigma_{\varepsilon,\eta}(\Omega)=
\Sigma_{\varepsilon,\eta}(\Omega-\Omega)$. From the definition of the C\&P set, we have
$(y-x)^\star\in\Omega-\Omega$. Now let $x=\gamma^m x_n$, $y=\gamma^m x_{n+1}$ for any
integer power $m$ and for some $n$ such that $x_{n+1}-x_n=\Delta_1$, or
$x_{n+1}-x_n=\Delta_2$ respectively. From the self-similarity of
$\Sigma_{\varepsilon,\eta}(\Omega)$, the points $x,y$ belong to
$\Sigma_{\varepsilon,\eta}(\Omega)$, and hence $(\gamma^m\Delta_1)^\star$,
$(\gamma^m\Delta_2)^\star\in\Omega-\Omega$. Therefore the sequence of vectors
\begin{equation}\label{e:vectors}
{\mathbb K}^m\binom{\Delta^\star_1}{\Delta^\star_2} = \left({\mathbb
K}^m\binom{\Delta_1}{\Delta_2}\right)^\star
=\left(\gamma^m\binom{\Delta_1}{\Delta_2}\right)^\star =
\binom{(\gamma^m\Delta_1)^\star}{ (\gamma^m\Delta_2)^\star}
\end{equation}
is bounded with $m\to\infty$. In the above we have used the property of the star map
$(kx)^\star=kx^\star$ for $x\in\Z[\eta]$ and any integer $k$. Since the eigenvectors of
the matrix ${\mathbb K}$ form a basis of $\R^2$, we can write
$$
\binom{\Delta^\star_1}{\Delta^\star_2} = \alpha_1\binom{\Delta_1}{\Delta_2}+ \alpha_2
\binom{\Delta'_1}{\Delta'_2}
$$
for some real coefficients $\alpha_1$, $\alpha_2$.
Substituting into~\eqref{e:vectors} we derive that the sequence of vectors
$$
\alpha_1\gamma^m\binom{\Delta_1}{\Delta_2}+
\alpha_2{\gamma'}^m \binom{\Delta'_1}{\Delta'_2}
$$
is bounded. Since $\gamma>1$, we have $\alpha_1=0$ and $|\gamma'|<1$. We can conclude
that $\gamma$ is a quadratic Pisot number and $\binom{\Delta^\star_1}{\Delta^\star_2} =
\alpha_2\binom{\Delta'_1}{\Delta'_2}$. The lengths $\Delta_1$, $\Delta_2$ belong to
$\Z[\eta]$ and hence can be written in the form $\Delta_1=a_1+b_1\eta$,
$\Delta_2=a_2+b_2\eta$ for some integers $a_1,a_2,b_1,b_2$. We have
$$
\alpha_2=\frac{a_1+b_1\varepsilon}{a_2+b_2\varepsilon}=\frac{a_1+b_1\eta'}{a_2+b_2\eta'}\,,
$$
which implies $(a_1b_2-a_2b_1)(\varepsilon-\eta')=0$. Since $\Delta_1$, $\Delta_2$ are
linearly independent over $\Q$, we have $a_1b_2-a_2b_1\neq0$ and thus $\varepsilon=\eta'$
as the theorem claims. The star map in such a cut-and-project scheme coincides with the
Galois automorphism on the quadratic field $\Q(\eta)=\Q(\varepsilon)=\Q(\gamma)$.

The last to be verified is that $\overline{\Omega}$ contains the origin.
Since $\gamma\Sigma_{\varepsilon,\eta}(\Omega)\subset\Sigma_{\varepsilon,\eta}(\Omega)$,
it follows easily that $\gamma'\Omega\subset\Omega$. This implies $0\in\overline{\Omega}$.
\pfk

From the proof of the above theorem it follows that if $\varepsilon$, $\eta$ are mutually
conjugated quadratic numbers and $0\in\overline{\Omega}$, (i.e.
$\Sigma_{\varepsilon,\eta}(\Omega)$ is self-similar) there exists infinitely many factors
$\gamma$ such that $\gamma\Sigma_{\varepsilon,\eta}(\Omega)\subset
\Sigma_{\varepsilon,\eta}(\Omega)$. It can be shown that all of these factors are
quadratic Pisot numbers in $\Z[\eta]$. The exact description of all self-similarity
factors of a given C\&P set is straightforward, but rather technical. For a
generalisation of self-similarity studied on the most common example $\eta=\tau$ we refer
to~\cite{infl}.

Finally, let us mention that first results about self-similar
Delone sets with Meyer property (which include C\&P sets) have
been obtained by Meyer in~\cite{meyer2}. He shows that the
self-similarity factor of such sets must be a Pisot or Salem
number, i.e.\ an algebraic integer $>1$ with all conjugates in the
unit disc. In a even more general setting, self-similarity of
Delone sets has been studied in~\cite{kenyon1,kenyon2}.

\section{Non-standard numeration systems and C\&P sequences}

Another example of self-similar sets are sequences formed by
$\beta$-integers. We show how they are related to C\&P sequences.
For the definition of $\beta$-integers we introduce the notion of
$\beta$-expansion, which has been first given by
R\'enyi~\cite{renyi}. The $\beta$-expansions are studied from the
arithmetical point of view for example
in~\cite{schmidt,parry,FruSo,bertrand}.

Let $\beta$ be a real number greater than $1$. For a non-negative $x\in\R$ we find a unique $k$
such that $\beta^k\leq x < \beta^{k+1}$ and put
$$
x_k:=\left[\frac{x}{\beta^k}\right]\,,\qquad r_k:=x-x_k\beta^k\,.
$$
The coefficients $x_i$, $i\in\Z$, $i\leq k-1$ we define recursively
$$
x_i:=\left[\beta r_{i+1}\right]\,,\qquad r_i:=\beta r_{i+1}-x_i\,.
$$
The described procedure is called the greedy algorithm. It ensures that
$$
x=\sum_{i=-\infty}^k x_i\beta^i\,.
$$
The above expression of $x$ using an infinite series is called the $\beta$-expansion of
$x$. For $\beta=2$ or $\beta=10$ we obtain the usual binary or decimal expansion of $x$.
The real numbers $x$ for which the coefficients $x_{-1}$, $x_{-2}$, $x_{-3}$, \dots in
the $\beta$-expansion of $|x|$ vanish, are called $\beta$-integers. They form the set
denoted by $\Z_\beta$,
$$
\Z_\beta = \left\{\pm\sum_{i=0}^k x_i\beta^i \,\biggm|\, \sum_{i=0}^k x_i\beta^i \hbox{ is a
$\beta$-expansion of an }x\geq 0\right\}\,.
$$
The greedy algorithm implies that if $x=\sum_{i=-\infty}^k x_i\beta^i$ is the
$\beta$-expansion of a number $x$, then $\sum_{i=-\infty}^{k+1} x_{i-1}\beta^{i}$ is the
$\beta$-expansion of $\beta x$. Therefore we trivially have for $\beta$-integers
$$
\beta\Z_\beta\subset\Z_\beta\,.
$$
Let us mention that $\Z_\beta$ has also many other self-similarity factors.

If $\beta$ is an integer greater than 1, the set of $\beta$-integers coincides with
rational integers, $\Z_\beta=\Z$. Drawn on the real line, the distances between adjacent
points of $\Z_\beta$ are all 1, and all integers $>1$ are self-similarity factors of
$\Z_\beta$. In this case the coefficients (digits) $x_i$ in a $\beta$-expansion take
values $0$, $1$, \dots, $\beta-1$, and every finite sequence formed by these digits is a
$\beta$-expansion of some number $x$.

The situation is very different if $\beta\notin\Z$. As a consequence of the greedy algorithm,
the digits in a $\beta$-expansion take values $0$, $1$, \dots, $\lceil\beta\rceil-1$. However,
not all strings of these digits correspond to a number $x$ as its $\beta$-expansion. Which sequences
of digits are permissible in $\beta$-expansions and what are the distances between adjacent points
in $\Z_\beta$ depends on the so-called R\'enyi development of $1$.  We define a mapping
$$
T_\beta (x)= \beta x-[\beta x]\,,\qquad \hbox{ for }\ x\in[0,1]\,.
$$
Put $t_i:=\bigl[\beta T_\beta^{i-1}(1)\bigr]$ for $i=1,2,3,\dots$. The sequence
$$
d_\beta(1)=t_1t_2t_3\cdots
$$
is called the R\'enyi development of $1$.

In order to decide, whether a series $\sum_{i=0}^{n}x_i\beta^i$ is a $\beta$-expansion, we use
the condition of Parry~\cite{parry}.

\begin{prop}
Let $\beta>1$. Then $\sum_{i=0}^{n}x_i\beta^i$ is a $\beta$-expansion of a number $x$
if and only if the word $x_ix_{i-1}\cdots x_0$ is lexicographically strictly smaller
than $t_1t_2t_3\cdots$ for all $0\leq i\leq n$.
\end{prop}

In~\cite{thurston} it is shown that the distances between neighbouring points in the set $\Z_\beta$
are of the form
$$
\sum_{k=1}^\infty \frac{t_{i+k}}{\beta^k}\,,\qquad\hbox{ for }\ i=0,1,2,\dots\,.
$$
A necessary condition in order that $\Z_\beta$ has only finitely many distances between neighbouring
points is that the R\'enyi development of $1$ is eventually periodic. The construction of $d_\beta(1)$
implies that
\begin{equation}\label{e:jednicka}
1=\sum_{i=1}^\infty\frac{t_i}{\beta^i}\,.
\end{equation}
If moreover $d_\beta(1)$ is eventually periodic, $\beta$ is a root of a monic polynomial with
integer coefficients. Such $\beta$ is an algebraic integer.

Our aim is to describe which parameters have to be chosen in order
that the sets $\Sigma_{\varepsilon,\eta}(\Omega)$ and $\Z_\beta$
coincide on the positive half-axis\footnote{Positive half-axis
$[0,+\infty)$ is denoted by $\R^+_0$.}, i.e. when
\begin{equation}\label{e:betaintegr}
\Sigma_{\varepsilon,\eta}(\Omega) \ \cap \ \R^+_0 \quad=\quad \Z_\beta \ \cap \ \R^+_0 \,.
\end{equation}

In~\cite{GaFr..} an example of such a relation is given together with the
parameters $\varepsilon,\eta,\Omega$. In particular, the authors study the
case of quadratic Pisot units. All quadratic Pisot units can be expressed
as the positive roots of a quadratic equation
$$
\beta^2=m\beta+1\,,\quad\hbox{ for }m\geq 1 \qquad\hbox{ or }\qquad
\beta^2=m\beta-1\,,\quad\hbox{ for }m\geq 3\,.
$$
It is shown that
\begin{eqnarray}
\Sigma_{\beta',\beta}\bigl[-1,-\frac{1}{\beta'}\bigr)
 \ \cap \ \R^+_0  &=& \Z_\beta  \cap  \R^+_0 \quad \hbox{for }\ \beta^2=m\beta+1,\ m\geq 1,
 \label{eq:1}\\
 \Sigma_{\beta',\beta}\bigl[\,0,\frac{1}{\beta'}\bigr)
 \ \cap \ \R^+_0  &=& \Z_\beta  \cap  \R^+_0 \quad\hbox{for }\ \beta^2=m\beta-1,\ m\geq 3.
 \label{eq:2}
\end{eqnarray}

The above equalities imply that the C\&P sequences with given windows have two distances
only between adjacent points. Thus from Theorem~\ref{thmx} they are geometrically similar
to C\&P sequences with unit acceptance interval, and therefore the infinite binary words
corresponding to $\beta$-integers are sturmian words.

In the following proposition we prove that a quadratic Pisot unit
$\beta$ is the only example of a basis for which positive
$\beta$-integers coincide with the restriction of a
cut-and-project set to its positive part.

\begin{prop}\label{betapisot}
Positive part of the set $\Z_\beta$ coincides with the positive part of a cut-and-project set
$\Sigma_{\varepsilon,\eta}(\Omega)$ if and only if $\beta$ is a quadratic Pisot unit.
\end{prop}

\pf One implication is obvious from~\eqref{eq:1} and~\eqref{eq:2}. Let us prove
$\Rightarrow$. Since $\Z_\beta$ is a self-similar set, we impose the requirement of
self-similarity also on the C\&P sets, which implies that $\varepsilon$, $\eta$ are
mutually conjugated quadratic numbers, i.e. $\varepsilon=\eta'$ and
$0\in\overline{\Omega}$. According to Theorem~\ref{thmselfs} the self-similarity factor
$\beta$ is a quadratic Pisot number. Thus it remains to show that $\beta$ is a unit.

All quadratic Pisot numbers can be expressed as the positive roots of a quadratic equation
$$
\beta^2=m\beta+n,\ \hbox{ for }1\leq n\leq m, \quad\hbox{ or }\quad
\beta^2=m\beta-n,\ \hbox{ for }1\leq n\leq m-2.
$$
Our considerations can thus be divided into two cases.

\begin{trivlist}
\item1.\
Let $\beta^2=m\beta+n$, $m\geq n\geq 1$.
Then the conjugated root to $\beta$ is the number $\beta'\in(-1,0)$. The R\'enyi
development of $1$ has the form
$$
d_\beta(1)=mn
$$
and the distances between $\beta$-integers are
$$
\sum_{i=1}^\infty\frac{t_i}{\beta^i}=1 \qquad\hbox{ and }\qquad
\sum_{i=1}^\infty\frac{t_{i+1}}{\beta^i}=\beta-t_1 = \beta-m=\frac{n}\beta\,.
$$
In this case a series $\sum_{i=0}^k x_i\beta^i$ with non-negative integer coefficients
$x_i$ is a $\beta$-expansion if $x_ix_{i-1}$ is strictly lexicographically smaller than
$mn$, i.e. $x_ix_{i-1}\prec mn$ for all $1\leq i\leq k$, which means that
$x_i\in\{0,1,\dots,m\}$ and every digit $x_i=m$ in the string $x_kx_{k-1}\cdots x_1x_0$
is followed by a digit $x_{i-1}\leq n-1$.

Since in a self-similar C\&P set the star map and the Galois automorphism coincide, we
can find a candidate for the acceptance interval $\Omega$ in order
that~\eqref{e:betaintegr} be satisfied. In the following estimations we use
$\beta'\in(-1,0)$. For $x=\sum_{i=0}^kx_i\beta^i\in\Z_\beta$ we have
$$
x'=\sum_{i=0}^kx_i{\beta'}^i < m+ m{\beta'}^2 + m{\beta'}^4 + \cdots = \frac{m}{1-{\beta'}^2}\,.
$$
Similarly,
$$
x'=\sum_{i=0}^kx_i{\beta'}^i > m\beta'+ m{\beta'}^3 + m{\beta'}^5 + \cdots =
\frac{m\beta'}{1-{\beta'}^2}\,.
$$
Clearly, the only candidate for $\Omega$ is the interval
$\bigl[\frac{m\beta'}{1-{\beta'}^2},\frac{m}{1-{\beta'}^2}\bigr)$. It is obvious that for
such a window one inclusion of~\eqref{e:betaintegr} is verified,
$$
\Sigma_{\eta',\eta}(\Omega) \ \cap \ \R^+_0 \quad\supseteq\quad \Z_\beta \ \cap \ \R^+_0 \,.
$$
Since $\Z_\beta$ has only two possible distances between neighbouring elements,
name\-ly 1 and $\frac{n}{\beta}$, the equality in the above inclusion is reached according to 8.
of Remark~\ref{poznamka}
if $|\Omega|=\Delta'_1-\Delta'_2$, i.e. if
$$
\frac{m}{1-{\beta'}^2}-\frac{m\beta'}{1-{\beta'}^2} = \frac{m}{1+{\beta'}} = 1-\frac{n}{\beta}\,.
$$
Using the quadratic equation $\beta'^2=m\beta'+n$ we obtain the condition
$(1-n)\beta'=0$ which implies $n=1$. Thus $\beta$ is a unit.

\item2.\
Let us study the case $\beta^2=m\beta-n$, $m-2\geq n\geq 1$. Here the conjugated root
$\beta'$ belongs to the interval $(0,1)$. The R\'enyi development of $1$ has coefficients
$t_1=m-1$ and $t_i=m-n-1$ for $i\geq 2$, i.e.
$$
d_\beta(1)=(m-1)(m-n-1)(m-n-1)\cdots = (m-1)(m-n-1)^\omega\,.
$$
In this case a $\beta$-expansion of a number $x$ has digits in the set
$\{0,1,2,\dots,m-1\}$ and forbidden are the strings of digits equal or lexicographically
greater than $(m-1)(m-n-1)^{s}(m-n)$ for arbitrary non-negative integer $s$. The
distances between neighbouring $\beta$-integers are
$$
\sum_{i=1}^\infty\frac{t_i}{\beta^i}=1 \qquad\hbox{ and }\qquad
\sum_{i=1}^\infty\frac{t_{i+1}}{\beta^i}=\beta-t_1 = \beta-(m-1)=1-\frac{n}\beta\,.
$$
Again, using the Galois conjugation of $x=\sum_{i=0}^kx_i\beta^i\in\Z_\beta$ we find a candidate
on the acceptance interval $\Omega$ for~\eqref{e:betaintegr},
$$
0\leq x'=\sum_{i=0}^kx_i\beta'^i < m-1 + (m-2)\beta' + (m-2)\beta'^2 + \cdots =
1+\frac{m-2}{1-\beta'}\,.
$$
Let us therefore set $\Omega=\bigl[0,1+\frac{m-2}{1-\beta'}\bigr)$. In order that a C\&P
set with such an acceptance window have only two distances between neighbours, we must
have
$$
|\Omega| = 1+\frac{m-2}{1-\beta'} = \Delta'_1-\Delta'_2 = 1-1+\frac{n}{\beta'} = \frac{n}{\beta'}\,.
$$
After manipulations we obtain $(n-1)\beta'=0$ which implies $n=1$. This completes the proof.
\end{trivlist}
\vskip-0.8cm
\pfk

\section{C\&P sequences and substitutions}\label{s:subst}

Construction of an arbitrarily long segment of a C\&P sequence directly from the
definition of $\Sigma_{\varepsilon,\eta}(\Omega)$ is numerically very demanding, since
precise computation with irrational numbers requires a special arithmetics dependent on
the form in which the irrational numbers $\varepsilon,\eta$ are given. For a class of
self-similar C\&P sequences the sequence of distances between adjacent points (i.e. the
infinite word $u_{\varepsilon,\eta}(\Omega)$) can be generated effectively using
substitution rules. For this purpose we introduce the following notions.

The set ${\mathcal A}^*$ of finite words on an alphabet ${\mathcal A}$ equipped with the empty
word $\epsilon$ and the operation of concatenation is a free monoid.
A morphism on the monoid $\A^*$ is a map $\varphi:\A^*\to\A^*$ satisfying
$\varphi(wz)=\varphi(w)\varphi(z)$ for any pair of words $w,z\in\A^*$. Clearly, the morphism $\varphi$
is determined by $\varphi(a)$ for all $a\in\A$. The action of a morphism $\varphi$ can be easily
extended to one-directional infinite words $u=u_0u_1u_2\cdots $ over $\A$ by the prescription
$$
\varphi(u)=\varphi(u_0u_1u_2\cdots)=\varphi(u_0)\varphi(u_1)\varphi(u_2)\cdots
$$
Let $u=u_0u_1u_2\cdots$ be an infinite word over an alphabet $\A$
and let $\varphi$ be a morphism on $\A^*$ satisfying
$|\varphi(a)|\geq1$ for all $a\in\A$ and $|\varphi(u_0)|>1$. We
say that the word $u$ is invariant under the substitution
$\varphi$ if $u$ is its fixed point, i.e. $\varphi(u)=u$.

The $\varphi$-invariance of $u$ implies that $\varphi(u_0)$ has the form
$\varphi(u_0)=u_0u'$ for some non-empty word $u'\in\A^*$ and that $\varphi^n(u)=u$ for
every $n\in\N$. The word $\varphi^n(u_0)$ is a prefix of the fixed point $u$ and its
length grows to infinity with $n$, therefore we can formally write
$u=\lim_{n\to\infty}\varphi^n(u_0)$. The substitution under which an infinite word $u$ is
invariant allows one to generate $u$ starting from the initial letter $u_0$ repeating the
rewriting rules infinitely many times.

As an example of a substitution invariant C\&P sequence let us recall the
$\beta$-integers, as presented in the previous section. Consider first $\Z_\beta$ for
$\beta^2=m\beta+1$, where the distances are $\Delta_1=1$ and $\Delta_2=\frac1{\beta}$.
Associating the letter $A$ to the distance $1$ and the letter $B$ to the distance
$\frac1\beta$ we create a one-directional infinite word $u$ in the alphabet $\{A,B\}$. It
can be easily seen from the properties of $\beta$-expansions and from the Parry condition
that, if $x,y$ are neighbours in $\Z_\beta$ such that $y-x=1$, then, between the points
$\beta x$ and $\beta y$, there is $m$ times the distance 1 followed by one distance
$\frac1\beta$. Similarly, if $x,y$ are neighbours in $\Z_\beta$ such that
$y-x=\frac1\beta$, then the points $\beta x$, $\beta y$ are also neighbours and have
distance $1$. The above considerations imply that the infinite word $u$ is not changed,
if every letter $A$ is replaced by the finite word $A^mB$, and every letter $B$ is
replaced by $A$. We say, that the word $u$ corresponding to $\Z_\beta$ is invariant under
the substitution $\varphi$ given by
$$
\varphi(A)=A^mB,\qquad
\varphi(B)=A.
$$

Similarly we can derive that for $\beta$-integers where $\beta^2=m\beta-1$, the infinite word
$u$ corresponding to $\Z_\beta$ is invariant under the substitution
$$
\varphi(A)=A^{m-1}B,\qquad
\varphi(B)=A^{m-2}B.
$$

We have presented the substitutions only for those $\beta$-integers that correspond to C\&P sequences.
However in general, every $\Z_\beta$ which has a finite number of distances between neighbours
is invariant under a non-trivial substitution~\cite{fabre}.

To every substitution $\varphi$ on the alphabet
$\A=\{a_1,\dots,a_k\}$ one associates naturally the substitution
matrix $M\in M_k(\N_0)$, where
$$
M_{ij} = \hbox{the number of letters $a_j$ in the word $\varphi(a_i)$}\,.
$$
If all letters $a_i$ of the alphabet $\A$ have a well defined
density $\varrho_i$ in the infinite word $u$ invariant under the
substitution $\varphi$, then the vector
$(\varrho_1,\varrho_2,\dots,\varrho_k)$ is a left eigenvector of
the matrix $M$. If the matrix $M$ is primitive, i.e.\ it has a
positive power, then the substitution is called {\em primitive}. A
fixed point of a primitive substitution can be represented
geometrically as a self-similar sequence in the following way.

According to the Perron-Frobenius theorem, the matrix $M$ has an up to a scalar factor
unique posi\-tive right eigenvector
$(y_1,y_2,\dots,y_k)^T$  corresponding to the dominant eigen\-value, say $\lambda$. To the infinite
word $u=u_0u_1u_2\cdots$ we associate the sequence $(z_n)_{n\in\N_0}$ such that
\begin{equation}\label{e:gr}
z_0=0\qquad\hbox{ and }\qquad z_{n+1}-z_n = y_i \quad\hbox{ if }\ u_n=a_i\,.
\end{equation}
The sequence $(z_n)_{n\in\N_0}$ is self-similar, since we have
$$
\lambda\{ z_n \mid n\in\N_0\} \ \subset \ \{ z_n \mid n\in\N_0\}\,.
$$

\subsection{Substitution invariance}

Infinite words corresponding to C\&P sequences are bidirectional. From the property 2~of
Remark~\ref{poznamka} it follows that, without loss of generality, we can consider only
those C\&P sequences which have $0\in\Omega$, i.e.
$0\in\Sigma_{\varepsilon,\eta}(\Omega)$. We define a pointed bidirectional infinite word
$u_{\varepsilon,\eta}(\Omega)=\cdots u_{-2}u_{-1}|u_0u_1u_2\cdots$ such that
$u_0u_1u_2\cdots$ corresponds to the order of distances between adjacent points of
$\Sigma_{\varepsilon,\eta}(\Omega)$ on the right of 0, and $\cdots u_{-3}u_{-2}u_{-1}$
corresponds to the order of distances between adjacent points of
$\Sigma_{\varepsilon,\eta}(\Omega)$ on the left of 0. This word is ternary or binary.

Let $u=\cdots u_{-2}u_{-1}|u_0u_1u_2\cdots$ be a pointed bidirectional infinite word over an alphabet
$\A$. Let $\varphi$ be a morphism on $\A^*$ such that $|\varphi(u_{-1})|>1$ and $|\varphi(u_0)|>1$.
We say that the word $u$ is invariant under the substitution $\varphi$, if it satisfies
$$
u=\cdots u_{-2}u_{-1}|u_0u_1u_2\cdots =
\cdots \varphi(u_{-2})\varphi(u_{-1})|\varphi(u_0)\varphi(u_1)\varphi(u_2)\cdots = \varphi(u)\,.
$$
In this case we formally write
$$
u=\lim_{n\to\infty} \varphi^n(u_{-1})\mid\varphi^n(u_0)\,.
$$

Let us mention what is known about the substitution invariance of infinite words
associated to C\&P sequences. First consider the binary words. As explained in
Example~\ref{ex}, all such words coincide with lower and upper mechanical words
$\underline{s}_{\alpha,\beta}$, $\overline{s}_{\alpha,\beta}$ for irrational
$\alpha\in(0,1)$ and any real $\beta\in[0,1)$. The question about substitution invariance
of mechanical words has been solved independently by different
authors~\cite{yasutomi,berthe,sturmian}. In order to state the result we need to
introduce the notion of a Sturm number.

\begin{de}
A quadratic irrational number $\alpha\in(0,1)$ whose algebraic conjugate $\alpha'$ satisfies
$\alpha'\notin(0,1)$ is called a Sturm number.
\end{de}

Let us mention that originally Sturm numbers were defined by a
special form of their continued fraction. The characterization
presented here is due to~\cite{allauzen}. The necessary and
sufficient condition for  substitution invariance of mechanical
words is given by the following theorem~\cite{sturmian}.

\begin{thm}
The mechanical word $\underline{s}_{\alpha,\beta}$, resp.
$\overline{s}_{\alpha,\beta}$, for irrational $\alpha\in(0,1)$ and
real $\beta\in[0,1)$ is invariant under a substitution if and only
if
\begin{itemize}
\item[(i)] $\alpha$ is a Sturm number,
\item[(ii)] $\beta\in\Q(\alpha)$,
\item[(iii)] $\alpha'\leq \beta'\leq 1-\alpha'$ or $1-\alpha'\leq \beta'\leq\alpha'$.
\end{itemize}
\end{thm}

If we represent the substitution invariant lower mechanical word
$\underline{s}_{\alpha,\beta}$ geometrically, as described in~\eqref{e:gr}, we find that
this geometrical representation coincides with the C\&P sequence
$\Sigma_{\varepsilon,\varepsilon'}(\beta-1,\beta]$, where $\alpha=-\varepsilon$. Similar
statement is valid for the substitution invariant upper mechanical word
$\overline{s}_{\alpha,\beta}$. This implies that substitution invariance of a binary word
$u$ associated to a C\&P sequence forces existence of a self-similar C\&P sequence
$\Sigma_{\varepsilon,\varepsilon'}(\Omega)$ such that
$u=u_{\varepsilon,\varepsilon'}(\Omega)$.

Substitution invariance of ternary words corresponding to C\&P sequences has not yet been
solved completely. The authors however conjecture that, even in this case, substitution
invariance forces self-similarity of the corresponding C\&P sequence.

\subsection{Substitutivity}\label{subst}

The original aim for studying substitution properties of C\&P
sequences  was the possibility of symbolic generation of
$u_{\varepsilon,\eta}(\Omega)$. For our present purpose, it is
enough to consider a property weaker than substitution invariance,
namely the substitutivity. We take the formulation of
Durand~\cite{durand}.

\begin{de}
We say that the infinite word $u$ over an alphabet $\A$ is
substitutive if there exist an infinite word $v$ over an alphabet
${\mathcal B}$ and a letter projection $\psi:{\mathcal B}\to\A$
such that $v$ is invariant under a substitution $\varphi$ on
$\B^*$ and
$$
\cdots\psi(v_{-2})\psi(v_{-2})|\psi(v_0)\psi(v_1)\psi(v_2)\cdots
=\cdots u_{-2}u_{-1}|u_0u_1u_2\cdots
$$
If moreover $\varphi$ is a primitive substitution, then the
infinite word $u$ is said to be primitive substitutive.
\end{de}

If an infinite word $u$ is substitutive, it can be constructed in such a way that
generating by substitution the word $v$ and using the projection $\psi$ allows us to
obtain $u$.

Using Theorem~\ref{thmx} we can without loss of generality consider the ternary words associated to
C\&P sequences $\Sigma_{\varepsilon,\eta}[c,c+\ell)$, where
\begin{equation}\label{e:adam}
\varepsilon\in(-1,0),\quad \eta>0,\quad c\leq0<c+\ell\quad\hbox{ and }\quad
\max(-\varepsilon,1+\varepsilon)<\ell<1\,.
\end{equation}

The description of infinite words associated to C\&P sequences which are substitutive can be
derived from the paper of Adamczewski~\cite{adam}.

\begin{thm}
Let $\varepsilon,\eta,c,\ell$ satisfy~\eqref{e:adam}. The infinite
word $u_{\varepsilon,\eta}[c,c+\ell)$ is primitive substitutive if
and only if $\varepsilon$ is a quadratic irrational number and
$c,\ell\in\Q(\varepsilon)$.
\end{thm}

From a practical point of view it is important to know the procedure which, given a
substitutive word $u$ over a ternary alphabet $\{A,B,C\}$, allows to determine an
alphabet ${\mathcal B}$, a substitution invariant word $v$ over ${\mathcal B}$ and a
projection $\psi:{\mathcal B}\to\A$ such that $\psi(v)=u$. The bidirectional pointed word
$v=\cdots v_{-2}v_{-1}|v_0v_1v_2$ is in fact given by the initial letters $v_{-1}|v_0$
and the substitution $\varphi$ under which it is invariant, since we have
$v=\lim_{n\to\infty}\varphi^n(v_{-1})|\varphi^n(v_0)$.

In the rest of this section we describe the algorithm for solving this problem in case
that the parameters satisfy besides the necessary conditions~\eqref{e:adam}, an
additional condition that $-\varepsilon$ is a Sturm number, i.e.\ the algebraic conjugate
$\varepsilon'$ of $\varepsilon$ satisfies $\varepsilon'<-1$ or $\varepsilon'>0$. Using
the transformations~\eqref{e:trans1} and~\eqref{e:trans2} we have
$$
\Sigma_{\varepsilon,\eta}(\Omega) = \Sigma_{-1-\varepsilon,-1-\eta}(-\Omega)\,,
$$
and thus we can without loss of generality consider only $\varepsilon'>0$. As for the
parameter $\eta$, we know that, for fixed $\varepsilon,c,\ell$ satisfying~\eqref{e:adam},
the words $u_{\varepsilon,\eta}[c,c+\ell)$ coincide for all choices of $\eta>0$. In case
that $\varepsilon'>0$, it is suitable to put $\eta=\varepsilon'$. In this case
$\Sigma_{\varepsilon,\varepsilon'}[c,c+\ell)$ is a self-similar set. This is a crucial
property for proving the correctness of the algorithm presented below.

\begin{pozn}
The sequence $\Sigma_{\varepsilon,\varepsilon'}[c,c+\ell)$ with
parameters satisfying~\eqref{e:adam} has according to
Theorem~\ref{thmx} three types of distances between adjacent
points, namely $\varepsilon',1+\varepsilon',1+2\varepsilon'$.
According to Theorem~\ref{thmselfs} it is a self-similar set. The
stepping function $f$ on the acceptance interval
$\Omega=[c,c+\ell)$ has in this case the form
$$
f(x)=\left\{\begin{array}{clcr}
x+1+\varepsilon &\hbox{ for }& x\in[c,c+\ell-1-\varepsilon)&=:\Omega_A\,,\\
x+1+2\varepsilon &\hbox{ for }& x\in[c+\ell-1-\varepsilon,c-\varepsilon)&=:\Omega_B\,,\\
x+\varepsilon &\hbox{ for }& x\in[c-\varepsilon,c+\ell)&=:\Omega_C\,.\\
\end{array}\right.
$$
This function is a bijection on the acceptance interval $\Omega$, i.e. is invertible.
\end{pozn}

\bigskip
\noindent \subsubsection*{Algorithm:}~

\smallskip
\noindent
{\it Input:} quadratic $\varepsilon\in(-1,0)$ with
$\varepsilon'>0$, $c,\ell\in\Q(\varepsilon)$, such that $c\leq 0<c+\ell$,
$\max(-\varepsilon,1+\varepsilon)<\ell\leq1$.

\smallskip
\noindent
{\it Output:} alphabet ${\mathcal B}$, letters $v_{-1},v_0\in{\mathcal B}$,
morphism $\varphi$ on ${\mathcal B}^*$, projection $\psi:{\mathcal B}\to\A$.

\begin{itemize}

\item[\underline{Step 1}] Find a quadratic unit $\gamma\in(0,1)$ such that
$\gamma\Z[\varepsilon]=\Z[\varepsilon]$ and its conjugate $\gamma'>1$. It results in
solving a Diophantine equation (more precisely the so-called Pell equation) which has
always a solution.

\item[\underline{Step 2}] For $x\in\Omega$ we define
\begin{equation}\label{e:defg}
g_\gamma(x) = \frac1\gamma f^{-{\rm ind}(x)}(x),\quad\hbox{ where }\quad
{\rm ind}(x) = \min \{ i \in\N_0 \mid f^{-i}(x) \in\gamma\Omega\}.
\end{equation}
Find the minimal set $S\subset\Omega$ such that
$$
\{c,c+\ell-1-\varepsilon,c-\varepsilon\} \subseteq S\qquad\hbox{ and }\qquad
g_{\gamma}(S) \subseteq S\,.
$$
Such a set is finite, let us denote its elements by $S=\{c_0,c_1,\dots,c_k\}$, where
$c=c_0<c_1<\cdots < c_k$, and denote $c_{k+1}:=c+\ell$. Note that the elements of the set
$S$ divide the acceptance window into small subintervals
$$
\Omega=\bigcup_{i=0}^k[c_i,c_{i+1})\,.
$$

\item[\underline{Step 3}] Define the alphabet ${\mathcal B}:=\{0,1,\dots,k\}$ and to every letter
$i\in{\mathcal B}$ associate the number
$$
j_i=\min\{j\in\N\mid f^j(\gamma c_i)\in\gamma\Omega\}
$$
and the word $\varphi(i)=w^{(i)}_0w^{(i)}_1\cdots w^{(i)}_{j_i-1}$ by the prescription
$$
w_j^{(i)}:= m\in{\mathcal B}\qquad\hbox{ if }\quad f^j(\gamma
c_i)\in[c_m,c_{m+1})\,.
$$

\item[\underline{Step 4}] Define the initial letters of the infinite word $v=\cdots v_{-2}v_{-1}|v_0v_1v_2\cdots$
over the alphabet ${\mathcal B}$ as
$$
\begin{array}{cclccl}
v_0&=&m\in{\mathcal B}\qquad&\hbox{ if }\ &0&\in \ [c_m,c_{m+1})\,,\\[2mm]
v_{-1}&=&m\in{\mathcal B}\qquad&\hbox{ if }\ &f^{-1}(0)&\in \ [c_m,c_{m+1})\,.
\end{array}
$$

\item[\underline{Step 5}] Define the projection $\psi:\B\to\A=\{A,B,C\}$ by
$$
\psi(i)=\left\{\begin{array}{l}
A\hbox{ if } c_i\in\Omega_A\,,\\
B\hbox{ if } c_i\in\Omega_B\,,\\
C\hbox{ if } c_i\in\Omega_C\,.
\end{array}\right.
$$
\end{itemize}

\begin{thm}\label{algor}
Let parameters $\varepsilon$, $\eta$, $c$, $\ell$ satisfy~\eqref{e:adam}. Let moreover
$\varepsilon$ be a quadratic irrational number, such that $\varepsilon'>0$, and let
$c,\ell\in\Q(\varepsilon)$. Then the alphabet ${\mathcal B}$, letters
$v_{-1},v_0\in{\mathcal B}$, morphism $\varphi$ on ${\mathcal B}^*$, and projection
$\psi:{\mathcal B}\to\A$, defined in the above algorithm, satisfy
$$
u_{\varepsilon,\eta}[c,c+\ell)=\psi(v)\,,\quad\hbox{ where }\
v=\lim_{n\to\infty}\varphi^n(v_{-1})|\varphi^n(v_{0})\,.
$$
\end{thm}

The proof of the theorem follows the same ideas as
in~\cite{subst}, where the correctness of the algorithm for
$\varepsilon=-\frac1\tau$ is shown. Note that the crucial point in
the algorithm is to ensure that the set $S$ of Step 2 is finite.

\begin{pozn}~

1. Given the infinite word $u_{\varepsilon,\eta}(\Omega)$, the
substitution $\varphi$ is not given uniquely. Indeed, the
ambiguity is found in the choice of the unit $\gamma$ in Step 1 of
the algorithm. Note that if $\gamma$ has required properties, then
so does any power $\gamma^j$, $j\in\N$.

2. The cardinality of the alphabet $\B$ is given by the cardinality of the set $S$, which
depends on the choice of $\gamma$. Taking a power of $\gamma$ as the unit factor may
reduce the number of letters in the alphabet.

3. In case that the word $u_{\varepsilon,\eta}(\Omega)$ is not
only substitutive but is also a fixed point of a substitution,
then suitable choice of $\gamma$ (sufficiently high power of
minimal factor satisfying Step 1) in the algorithm yields the
substitution under which $u_{\varepsilon,\eta}(\Omega)$ is
invariant.
\end{pozn}

Let us illustrate the algorithm for finding the substitution on an
example.

\begin{ex}
Consider the C\&P sequence $\Sigma_{\varepsilon,\eta}[c,c+\ell)$
with parameters
$$
\varepsilon=-\frac1{\sqrt2}\,,\qquad
\eta=\varepsilon'=\frac1{\sqrt2}\,,\qquad c=0\,,\qquad
\ell=-2+2\sqrt2=-2-4\varepsilon\,.
$$
Such parameters clearly satisfy the assumptions of the algorithm.
The distances between adjacent points of $\Sigma_{\varepsilon,\eta}[c,c+\ell)$
are $\Delta_1=1+\eta=1+\frac1{\sqrt2}$, $\Delta_2=\eta=\frac1{\sqrt2}$, and
$\Delta_1+\Delta_2=1+2\eta=1+{\sqrt2}$,
and therefore the explicit expression of the stepping function in
our case is
$$
f(x)=\left\{\begin{array}{cllr}
x+1+\varepsilon &\hbox{ for }&
x\in[0,-3-5\varepsilon)&=:\Omega_A\,,\\[1mm]
x+1+2\varepsilon &\hbox{ for }&
x\in[-3-5\varepsilon,-\varepsilon)&=:\Omega_B\,,\\[1mm]
x+\varepsilon &\hbox{ for }&
x\in[-\varepsilon,-2-4\varepsilon)&=:\Omega_C\,.
\end{array}\right.
$$
From that we derive the formula for the inverse function
$$
f^{-1}(x)=\left\{\begin{array}{cll} x-\varepsilon &\hbox{ for }&
x\in[0,-2-3\varepsilon)\,,\\[1mm]
x-1-2\varepsilon &\hbox{ for }&
x\in[-2-3\varepsilon,1+\varepsilon)\,,\\[1mm]
x-1-\varepsilon &\hbox{ for }&
x\in[-1-\varepsilon,-2-4\varepsilon)\,.
\end{array}\right.
$$
We know that $0\in\Sigma_{\varepsilon,\eta}[c,c+\ell)$. We can generate other elements of the
C\&P sequences on the right from 0 using the stepping function $f$. The elements on the left are generated
using $f^{-1}$. We have
$$
\begin{aligned}
\Sigma_{\varepsilon,\eta}[c,c+\ell) = \{\dots,&-5-8\eta,-4-6\eta,-3-5\eta,-2-3\eta,-1-2\eta,-\eta,\\
&\quad 0,\ 1+\eta,\ 2+2\eta,\ 3+4\eta,\ 4+5\eta,\ 5+6\eta,\ 5+7\eta,\dots \}\,,
\end{aligned}
$$
and graphically,

\setlength{\unitlength}{0.3mm}
\begin{picture}(370,38)
\put(-5,15){\line(1,0){380}}
\put(3,15){\circle*{4}}
\put(15,20){\mbox{\small $\Delta_1$}}
\put(37,15){\circle*{4}}
\put(44,20){\mbox{\small $\Delta_1+\Delta_2$}}
\put(85,15){\circle*{4}}
\put(97,20){\mbox{\small $\Delta_1$}}
\put(119,15){\circle*{4}}
\put(131,20){\mbox{\small $\Delta_1$}}
\put(153,15){\circle*{4}}
\put(155,20){\mbox{\small $\Delta_2$}}
\put(167,8){\line(0,1){7}}
\put(165,-2){\mbox{\small 0}}
\put(167,15){\circle*{4}}
\put(179,20){\mbox{\small $\Delta_1$}}
\put(201,15){\circle*{4}}
\put(213,20){\mbox{\small $\Delta_1$}}
\put(235,15){\circle*{4}}
\put(242,20){\mbox{\small $\Delta_1+\Delta_2$}}
\put(283,15){\circle*{4}}
\put(295,20){\mbox{\small $\Delta_1$}}
\put(317,15){\circle*{4}}
\put(329,20){\mbox{\small $\Delta_1$}}
\put(351,15){\circle*{4}}
\put(353,20){\mbox{\small $\Delta_2$}}
\put(365,15){\circle*{4}}
\end{picture}

\bigskip
The corresponding bidirectional infinite word $u_{\varepsilon,\eta}[c,c+\ell)$ is obtained
by replacing $\Delta_1$ with the letter $A$, $\Delta_2$ with the letter $C$ and
$\Delta_1+\Delta_2$ with the letter $B$, i.e.
$$
u_{\varepsilon,\eta}[c,c+\ell) = \cdots BABAAC|AABAAC\cdots
$$

Let us now use the algorithm described above to derive the substitution generating the word
$u_{\varepsilon,\eta}[c,c+\ell)$. We proceed according to the steps of the algorithm.

\begin{itemize}

\item[\underline{Step 1}] Put $\gamma=3+4\varepsilon=3-2\sqrt2$.
It is obvious that $\gamma$ is a unit in $\Z[\varepsilon]\cap
(0,1)$ and its algebraic conjugate $\gamma'$ satisfies
$\gamma'=\frac1\gamma=3-4\varepsilon=3+2\sqrt2>1$. We have yet to
verify that $\gamma\Z[\varepsilon]=\Z[\varepsilon]$. For that it
suffices to show that $\gamma\varepsilon\in\Z[\varepsilon]$ and
$\gamma^{-1}\varepsilon\in\Z[\varepsilon]$. We have
$$
\begin{array}{ccccrcl}
\gamma\varepsilon &=& \varepsilon(3+4\varepsilon) &=&
2+3\varepsilon &\in&\Z[\varepsilon]\,,\\[1mm]
\gamma^{-1}\varepsilon &=& \varepsilon(3-4\varepsilon) &=&
-2+3\varepsilon &\in&\Z[\varepsilon]\,,
\end{array}
$$
where we have used that $\varepsilon^2=\frac12$.

\item[\underline{Step 2}] We need to find the minimal set
$S\subset\Omega$ closed under the action of $g_\gamma$, containing
the points
$$
c=0\,,\qquad
c+\ell-1-\varepsilon=-3-5\varepsilon\,,\qquad
c-\varepsilon=-\varepsilon\,.
$$
Thus we search for values of all iterations $g_\gamma^j(x)$, $j\in\N$, of the above points.
Important for the definition~\eqref{e:defg} of the function $g_\gamma$ is the index
of a point $x$, i.e. the first exponent $i\in\N_0$ such that $f^{-i}(x)$ belongs to the interval
$$
\gamma\Omega = (3+4\varepsilon)\bigl[0,-2-4\varepsilon\bigr) =
\bigl[0,-14-20\varepsilon\bigr)\,.
$$
Let us find the image under $g_\gamma$ of the point $x=0$.
Clearly, $0\in\gamma\Omega$, thus ${\rm ind}(0)=0$. Therefore $g_\gamma(0)=\frac1\gamma f^0(0) =0$.
$g_{\gamma}^j(0)=0$ for all $j\in\N$.

Let us find the image under $g_\gamma$ of the point $x=-\varepsilon$. We have
$$
f^{-2}(-\varepsilon)=f^{-1}(-1-2\varepsilon)=-2-3\varepsilon \in\gamma\Omega\,,
$$
which implies
$$
{\rm ind}(-\varepsilon)=2\quad\hbox{ and }\quad
g_\gamma(-\varepsilon)=\frac1\gamma f^{-2}(-\varepsilon) =
(3-4\varepsilon)(-2-3\varepsilon)=-\varepsilon\,.
$$
Thus $g_{\gamma}^j(-\varepsilon)=-\varepsilon$ for all $j\in\N$.

Let us now find the image under $g_\gamma$ of the point $x=-3-5\varepsilon$.
We have
$$
\begin{aligned}
f^{-4}(-3-5\varepsilon)&=f^{-3}(-4-6\varepsilon)=f^{-2}(-5-8\varepsilon)
=\\
&=f^{-1}(-6-9\varepsilon)
=-7-10\varepsilon \in\gamma\Omega\,,
\end{aligned}
$$
which implies
$$
\begin{aligned}
{\rm ind}(-3-5\varepsilon)=4\quad\hbox{ and }\quad
g_\gamma(-\varepsilon)&=\frac1\gamma f^{-4}(-3-5\varepsilon)=\\
&=(3-4\varepsilon)(-7-10\varepsilon)=-1-2\varepsilon\,.
\end{aligned}
$$
Therefore the set $S$ must contain the point $-1-2\varepsilon$. In order to
find further iterations of $g_\gamma$ on the point $-3-5\varepsilon$, we determine
$g_\gamma(-1-2\varepsilon)$. We have seen that
$f^{-1}(-1-2\varepsilon)=-2-3\varepsilon \in\gamma\Omega$. Thus
$$
{\rm ind}(-1-2\varepsilon)=1\quad\hbox{ and }\quad
g_\gamma(-1-2\varepsilon)=\frac1\gamma f^{-1}(-1-2\varepsilon)=-\varepsilon\,.
$$

Altogether, we obtain
$$
\begin{array}{rccl}
 g_{\gamma}^j(0)&=&0, &\hbox{ for all }\ j\in\N\,,\\
 g_{\gamma}^j(-\varepsilon),&=&-\varepsilon,&\hbox{ for all }\ j\in\N\,,\\
 g_{\gamma}(-3-5\varepsilon)&=&-1-2\varepsilon,\\
 g_{\gamma}^j(-3-5\varepsilon)&=&-\varepsilon,&\hbox{ for all }\ j\in\N\,,\ j\geq2\,.
\end{array}
$$
We can conclude that $S$ contains four elements,
$$
c_0=c=0\,,\quad c_1=-1-2\varepsilon\,,\quad
c_2=-3-5\varepsilon\,,\quad c_3=-\varepsilon\,,
$$
and we put $c_4=-2-4\varepsilon$.

Note that the elements of the set $S$ divide the acceptance window into small subintervals
$$
\begin{aligned}
\Omega&=\bigcup_{i=0}^3[c_i,c_{i+1}) \ = \\
&= [0,-1-2\varepsilon)  \cup
[-1-2\varepsilon,-3-5\varepsilon)  \cup
[-3-5\varepsilon,-\varepsilon)  \cup  [-\varepsilon,-2-4\varepsilon),
\end{aligned}
$$
as it is illustrated on the following figure.

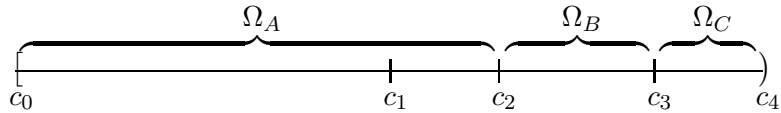
\begin{center}
\setlength{\unitlength}{0.3mm}
\begin{picture}(331,47)
\put(0,15){\line(1,0){331}}
\put(-2,12.5){\mbox{$\bigl[$}}
\put(-3,0){\mbox{$c_0$}}
\put(2,23){\mbox{$\overbrace{\hspace*{179pt}}^{\hbox{\normalsize $\ \Omega_A$}}$}}
\put(166,10){\line(0,1){10}}
\put(163,0){\mbox{$c_1$}}
\put(214,10){\line(0,1){10}}
\put(211,0){\mbox{$c_2$}}
\put(217,23){\mbox{$\overbrace{\hspace*{54pt}}^{\hbox{\normalsize $\ \Omega_B$}}$}}
\put(283,10){\line(0,1){10}}
\put(280,0){\mbox{$c_3$}}
\put(285.5,23){\mbox{$\overbrace{\hspace*{36pt}}^{\hbox{\normalsize $\ \Omega_C$}}$}}
\put(329,12.5){\mbox{$\bigr)$}}
\put(328,0){\mbox{$c_4$}}
\end{picture}
\end{center}

\item[\underline{Step 3}] Since $S$ has four elements, we have the alphabet ${\mathcal
B}:=\{0,1,2,3\}$ on four letters. In order to define the substitution we compute
the iterations $f^{j}(\gamma c_i)$ and we stop when $f^{j}(\gamma c_i)\in\gamma\Omega$.
First take $i=0$. We have $\gamma c_0 = 0$ and the iterations
$$
\begin{array}{rcccl}
 f^{0}(0)&=&0&\in&[c_0,c_1),\\
 f^{1}(0)&=&1+\varepsilon&\in&[c_0,c_1),\\
 f^{2}(0)&=&2+2\varepsilon&\in&[c_2,c_3),\\
 f^{3}(0)&=&3+4\varepsilon&\in&[c_0,c_1),\\
 f^{4}(0)&=&4+5\varepsilon&\in&[c_1,c_2),\\
 f^{5}(0)&=&5+6\varepsilon&\in&[c_3,c_4),\\
 f^{6}(0)&=&5+7\varepsilon&\in&\gamma\Omega,
\end{array}
\quad
\begin{array}{ll}
\hbox{ thus } & j_0=6 \ \hbox{ and }\\[1mm] \
&\varphi(0)=002013.
\end{array}
$$
Note that the word $\varphi(0)=002013$ is formed by the indices $m$ of the left-end-points of
the intervals $[c_m,c_{m+1})$, read in the column.
Similarly for $i=1$, we have $\gamma c_1 = \gamma (-1-2\varepsilon) = -7-10\varepsilon$. Thus
$$
\begin{array}{rcccl}
 f^{0}(-7-10\varepsilon)&=&-7-10\varepsilon&\in&[c_0,c_1),\\
 f^{1}(-7-10\varepsilon)&=&-6-9\varepsilon&\in&[c_0,c_1),\\
 f^{2}(-7-10\varepsilon)&=&-5-8\varepsilon&\in&[c_2,c_3),\\
 f^{3}(-7-10\varepsilon)&=&-4-6\varepsilon&\in&[c_0,c_1),\\
 f^{4}(-7-10\varepsilon)&=&-3-5\varepsilon&\in&[c_2,c_3),\\
 f^{5}(-7-10\varepsilon)&=&-2-3\varepsilon&\in&\gamma\Omega,
\end{array}
\quad
\begin{array}{ll}
\hbox{ hence } & j_1=5 \ \hbox{ and }\\[1mm] \
&\varphi(1)=00202.
\end{array}
$$
For $i=2$, we have $\gamma c_2 = \gamma (-3-5\varepsilon) = -19-27\varepsilon$. Therefore
$$
\begin{array}{rcccl}
 f^{0}(-19-27\varepsilon)&=&-19-27\varepsilon&\in&[c_0,c_1),\\
 f^{1}(-19-27\varepsilon)&=&-18-26\varepsilon&\in&[c_0,c_1),\\
 f^{2}(-19-27\varepsilon)&=&-17-25\varepsilon&\in&[c_2,c_3),\\
 f^{3}(-19-27\varepsilon)&=&-16-23\varepsilon&\in&[c_0,c_1),\\
 f^{4}(-19-27\varepsilon)&=&-15-22\varepsilon&\in&[c_2,c_3),\\
 f^{5}(-19-27\varepsilon)&=&-14-20\varepsilon&\in&[c_0,c_1),\\
 f^{6}(-19-27\varepsilon)&=&-13-19\varepsilon&\in&[c_1,c_2),\\
 f^{7}(-19-27\varepsilon)&=&-12-18\varepsilon&\in&[c_3,c_4),\\
 f^{8}(-19-27\varepsilon)&=&-12-17\varepsilon&\in&\gamma\Omega,
\end{array}
\quad
\begin{array}{l}
\hbox{ hence } j_2=8 \ \hbox{ and }\\[1mm]
\quad \varphi(2)=00202013.
\end{array}
$$
Last, for $i=3$, we have $\gamma c_3 = \gamma (-\varepsilon) = -2-3\varepsilon$. Thus
$$
\begin{array}{rcccc}
 f^{0}(-2-3\varepsilon)&=&-2-3\varepsilon&\in&[c_0,c_1),\\
 f^{1}(-2-3\varepsilon)&=&-1-2\varepsilon&\in&[c_1,c_2),\\
 f^{2}(-2-3\varepsilon)&=&-\varepsilon&\in&[c_3,c_4),\\
 f^{3}(-2-3\varepsilon)&=&0&\in&\gamma\Omega\,,
\end{array}
\quad
\begin{array}{ll}
\hbox{ hence } & j_3=3 \ \hbox{ and }\\[1mm] \
&\varphi(3)=013.
\end{array}
$$
Altogether, we have the substitution
$$
\begin{array}{rcl}
 \varphi(0)&=&002013\,,\\
 \varphi(1)&=&00202\,,\\
 \varphi(2)&=&00202013\,,\\
 \varphi(3)&=&013\,.
\end{array}
$$

\item[\underline{Step 4}] Since $0\in[c_0,c_1)$ and
$f^{-1}(0)=-\varepsilon\in[c_3,c_4)$, we put as the initial
letters $v_0=0$, $v_{-1}=3$. Note that 0 is a prefix of
$\varphi(0)$ and 3 is a suffix of $\varphi(3)$, thus the word
$v=\lim_{n\to\infty}\varphi^n(v_{-1})|\varphi^n(v_{0}) =
\lim_{n\to\infty}\varphi^n(3)|\varphi^n(0)$ is well defined.

\item[\underline{Step 5}] Since $c_0,c_1\subset\Omega_A$,
$c_2\subset\Omega_B$, $c_3\subset\Omega_C$, we have the projection
$\psi:\{0,1,2,3\}\to\A=\{A,B,C\}$ by
$$
\psi(0)=\psi(1)=A\,,\qquad \psi(2)=B\,,\qquad \psi(3)=C\,.
$$
\end{itemize}

Let us write the subsequent iterations of the substitution $\varphi$ on the pair of initial letters
$3|0$, i.e. $\varphi^n(3)|\varphi^n(0)$. We have for $n=0,1,2$,
$$
\begin{array}{rcl}
3&\hspace*{-3mm}|\hspace*{-3mm}&0\\
013&\hspace*{-3mm}|\hspace*{-3mm}&002013\\
00201300202013&\hspace*{-3mm}|\hspace*{-3mm}&0020130020130020201300201300202013\\
&\hspace*{-3mm}\vdots\hspace*{-3mm}&
\end{array}
$$
Since each row is a factor of the next one, in the limit we obtain the infinite word
$$
v=\cdots 00201300202013|0020130020130020201300201300202013\cdots\,.
$$
Now we apply the letter projection $\psi$, which collapses the letters $0$ and $1$. We
have {\small
$$
\begin{array}{r@{\ }c@{}c@{}c@{}c@{}c@{}c@{}c@{}c@{}c@{}c@{}c@{}c@{}c@{}
c@{}c@{}c@{}c@{}c@{}c@{}c@{}c@{}c@{}c@{}c@{}c@{}c@{}c@{}c@{}
c@{}c@{}c@{}c@{}c@{}c@{}c}
v=&\cdots&1&3&0&0&2&0&2&0&1&3&|&0&0&2&0&1&3&0&0&2&0&1&3&0&0&2&0&2&0&1&3&0&0&\cdots\\
&&\downarrow&\downarrow&\downarrow&\downarrow&\downarrow&\downarrow&\downarrow&\downarrow&\downarrow&
\downarrow&&\downarrow&\downarrow&\downarrow&\downarrow&\downarrow&\downarrow&\downarrow&
\downarrow&\downarrow&\downarrow&\downarrow&\downarrow&\downarrow&\downarrow&\downarrow&\downarrow&
\downarrow&\downarrow&\downarrow&\downarrow&\downarrow&\downarrow\\
u_{\varepsilon\!,\eta}[c,c\!+\!\ell)=
&\cdots&A&C&A&A&B&A&B&A&A&C&|&A&A&B&A&A&C&A&A&B&A&A&C&A&A&B&A&B&A&A&C&A&A&\cdots
\end{array}
$$
}

Let us study the second iteration of the above substitution $\varphi$. We obtain
 $$
\begin{array}{rcl}
 \varphi^2(0)&=&\varphi(002013)\ = \\
 &=&0020130020130020201300201300202013\,,\\
 \varphi^2(1)&=&\varphi(00202)\ =\\
 &=&0020130020130020201300201300202013\,,\\
 \varphi^2(2)&=&\varphi(00202013) =\\
 &=&002013002013002020130020130020201300201300202013\,,\\
 \varphi^2(3)&=&\varphi(013)\ =\\
 &=&00201300202013\,.
\end{array}
$$
 Note that $\varphi^2(0)=\varphi^2(1)$. Therefore we can consider the letters $0$ and
$1$ as identic. It enables us to define a new substitution
$\tilde{\varphi}:\{A,B,C\}^*\to\{A,B,C\}^*$ by
$$
\begin{array}{rcl}
 \tilde{\varphi}(A)&=&AABAACAABAACAABABAACAABAACAABABAAC\,,\\
 \tilde{\varphi}(B)&=&AABAACAABAACAABABAACAABAACAABABAAC\,\circ\\
  &&\circ\, AABAACAABABAAC\,,\\
 \tilde{\varphi}(C)&=&AABAACAABABAAC\,,
\end{array}
$$
under which the word $u_{\varepsilon,\eta}[c,c+\ell)$ is invariant, we namely have
$$
u_{\varepsilon,\eta}[c,c+\ell) = \lim_{n\to\infty}\tilde{\varphi}^n(C)|\tilde{\varphi}^n(A)\,.
$$
Note that the symbol $\circ$ in the formula for the substitution stands for
concatenation.
\end{ex}

\section{Conclusions}

In this paper, we have attempted to give a unifying view of the
one-dimensional cut-and-project point sets obtained from the
square lattice in the plane. A part of the work is a review of
former results which had to be recalled for the sake of clarity.
Let us now indicate some possible continuations or applications of
our results.

First, it would be   interesting to extend the work~\cite{anbuga} to the case of splines
of larger regularity for
  other  cut-and-project sets, with quadratic self-similarity, or
without self-similarity at all,
  especially having in view
the  relation between scaling equations and substitution properties of the considered
discretizations of $\R$.

Mathematical diffraction  of such aperiodic sets obtained by cut and projection should be
also envisaged in a systematic way, in relation with the existence of those
multiresolution analysis and related wavelets. Concerning diffraction, one can find in
the literature on quasicrystals many works devoted to this important subject, in which
substitutional properties, or self-similarity, or cut and projection from
higher-dimensional lattices, play a central role in the elaboration of rigorous results
(see for instance~\cite{bombieri,hof,luck} and~\cite{Lagarias} for a recent review on
these questions). The analysis of diffraction spectra by using adapted wavelets, i.e.\
wavelets ``living'' on the diffracting aperiodic structure, is a project which remains to
be really developed.

Another nice application of the results of this paper can be envisaged in the
construction of a new type of pseudo-random number generators. First step in this
direction has been made in~\cite{aprng}, where the authors use sturmian sequences to
combine classical periodic pseudo-random sequences to produce an aperiodic pseudo-random
sequence. These aperiodic pseudo-random number generators (APRNG) have been tested using
the DIEHARD test suite and using the Maurer test and it turned out that statistical
properties of these APRNG's are significantly better than of the original periodic
sequences. Moreover, the authors prove that the APRNG passes the spectral test. It would
be very interesting to pursue the study of the APRNG's extending the definition to
generic (i.e. ternary) cut-and-project sequences.

\section*{Acknowledgements}

J.P.G.,  Z.M. and E.P.
acknowledge partial support by Czech Science Foundation GA \v CR 201/05/0169.


\end{document}